%% file: main.tex
\documentclass[twocolumn]{aastex63}%,trackchanges
\usepackage{amsmath,color,textcomp,url,graphicx}
\usepackage{subfigure}
\usepackage{listings}
\usepackage{comment}
\usepackage{lineno}

\newcommand{\kms}{\hbox{km\,s$^{-1}$}}
\newcommand{\ncrius}{241}
\newcommand{\nselectedcrius}{84}
\newcommand{\newgroups}{32}
\newcommand{\newcoronae}{8}
\newcommand{\newtheiaextensions}{9}
\newcommand{\nearestvca}{22}

\received{2022 March 24}
\revised{2022 May 17}
\accepted{2022 August 16}
\submitjournal{ApJ}

\shorttitle{New coronae and stellar associations in the Solar neighborhood}
\shortauthors{Moranta et al.}

\begin{document}

\title{NEW CORONAE AND STELLAR ASSOCIATIONS REVEALED BY A CLUSTERING ANALYSIS OF THE SOLAR NEIGHBORHOOD}

\author[0000-0001-7171-5538]{Leslie Moranta}
\affiliation{Plan\'etarium Rio Tinto Alcan, Espace pour la Vie, 4801 av. Pierre-de Coubertin, Montr\'eal, Qu\'ebec, Canada}
\affiliation{Institute for Research on Exoplanets, Universit\'e de Montr\'eal, D\'epartement de Physique, C.P.~6128 Succ. Centre-ville, Montr\'eal, QC H3C~3J7, Canada}
\email{leslie.moranta@umontreal.ca}
\author[0000-0002-2592-9612]{Jonathan Gagn\'e}
\affiliation{Plan\'etarium Rio Tinto Alcan, Espace pour la Vie, 4801 av. Pierre-de Coubertin, Montr\'eal, Qu\'ebec, Canada}
\affiliation{Institute for Research on Exoplanets, Universit\'e de Montr\'eal, D\'epartement de Physique, C.P.~6128 Succ. Centre-ville, Montr\'eal, QC H3C~3J7, Canada}
\author{Dominic Couture}
\affiliation{Institute for Research on Exoplanets, Universit\'e de Montr\'eal, D\'epartement de Physique, C.P.~6128 Succ. Centre-ville, Montr\'eal, QC H3C~3J7, Canada}
\author[0000-0001-6251-0573]{Jacqueline K. Faherty}
\affiliation{Department of Astrophysics, American Museum of Natural History, Central Park West at 79th St., New York, NY 10024, USA}

\begin{abstract}

We present the results of a density-based clustering analysis of the 6-dimensional $XYZ$ Galactic positions and $UVW$ space velocities of nearby ($\leq 200$\,pc) Gaia~EDR3 stars with radial velocities using HDBSCAN, in opposition to previous studies \citep{2019AJ....158..122K, 2021AA...645A..84M} that only included positions and tangential velocities. Among the 241 recovered clusters, we identify more than 50 known associations, 32 new candidate stellar streams aged 100\,Myr--3\,Gyr, 9 extensions of known Theia groups uncovered by \cite{2019AJ....158..122K}, and 8 newly recognized coronae around nearby open clusters. Three confirmed exoplanet-hosting stars and three more TESS transiting exoplanet candidates are part of the new groups discovered here, including TOI--1807 and TOI--2076 from \cite{2021AJ....162...54H} that were suspected to belong to a yet unidentified moving group. The new groups presented here were not previously recognized because of their older ages, low spatial density, and projection effects that spread out the tangential velocities of their nearby co-moving members. Several newly identified structures reach distances within 60\,pc of the Sun, providing new grounds for the identification of isolated planetary-mass objects. The nearest member of the newly recognized corona of Volans-Carina is V419~Hya, a known young debris disk star at a distance of 22\,pc. This study outlines the importance of further characterization of young associations in the immediate Solar neighborhood, which will provide new laboratories for the precise age calibration of nearby stars, exoplanets and substellar objects.

\end{abstract}


\section{INTRODUCTION}\label{sec:intro}

Stars tend to be formed within a clustered environment caused by the collapse of a molecular cloud, dense and extremely cold structures primarily made of hydrogen gas and dust. As a consequence, young stars are rarely found in an isolated environment, but rather in co-moving groups or stellar associations which dissipate over time \citep{2003ARAA..41...57L}. The study of such young stellar associations provides unique laboratories where the ages of stars are precisely constrained, to an extent that can rarely be achieved with an isolated star. These young associations can generally be recognized from their coherent Galactic space velocities for several hundreds of millions of years and up to billions of years for denser open clusters (e.g., see \citealp{2003MNRAS.338..717B}). Over time, they become gradually tidally disrupted, which can cause elongated structures to detach from the core \citep{2019AA...621L...3M}, up to the point where they cannot be distinguished from unrelated field stars. This process usually happens over a time scale of $\approx$\,100\,Myr or more \citep{2020AA...640A..84D}, but recent discoveries of diffuse structures around younger associations (e.g., \citealp{2019MNRAS.489.4418J,2021ApJS..254...20L,2021ApJ...915L..29G}) indicates that star-formation events may happen over regions spatially larger than previously thought, or that interactions or collisions between dense clusters may be common.

The Gaia Early Data Release 3 (Gaia~EDR3; \citealp{2021AA...649A...6G,2021AA...649A...5F}) data products include measurements of sky positions, parallaxes, and proper motions for 1.468 billion stars, obtained by the Gaia mission \citep{2016AA...595A...1G} with an unprecedented accuracy that reaches 0.02--0.03\,mas for the parallax component of bright ($G < 15$) stars \citep{2021AA...649A...6G}\footnote{See \url{https://www.cosmos.esa.int/web/gaia/earlydr3}}. A radial velocity spectrometer is also located on board of the spacecraft to complete the 3D radial velocities of the brightest stars \citep{2018AA...616A...5C,2018AA...616A...7S,2018AA...616A...6S}. There are no new radial velocity measurements in the Gaia~EDR3 data release, however, it includes those from Gaia~DR2 \citep{2018AA...616A...1G} that were constrained to the $\approx$ 7.1-million brighter stars ($4 < G_{\rm RVS} < 13$) with approximate spectral types in the range F2--M2 ($3550 < T_{\rm eff} < 6900$), and reached typical radial velocity precisions of 1--3\,\kms\ in the case of radial velocity-stable stars. Gaia DR2 radial velocities are based on a median of $\approx$ 2--20 distinct visits \citep{2019AA...622A.205K}, which favors measuring the stable heliocentric radial velocity component associated with the $UVW$ velocity of a star over the components due to binary orbits or other sources of variation. There are approximately 6.6\,million sources in Gaia~DR2 that benefit from a radial velocity measurement with a precision of 5\,\kms\ or better.

These data make it possible to calculate the $UVW$ space velocities of bright stars in the Solar neighborhood with a precision of 0.1--0.5\,\kms, comparable or smaller than the intrinsic velocity dispersions in open clusters and loose stellar associations (e.g., \citealp{2004ARAA..42..685Z}). This provides an unprecedented opportunity to discover and probe the structural features and dynamical evolution of stellar associations, open clusters and their more recently discovered “coronae” \citep{2019AA...621L...3M,2021AA...645A..84M,2019AA...627A...4R,2019ApJ...877...12T}, a term recently introduced to designate loose ensembles of stars that appear to be coeval and comoving with denser cores of open clusters. Such coronae probably correspond to tidal disruption tails \citep{2021AA...645A..84M}.

Recent studies demonstrated that the Hierarchical Density-Based Spatial Clustering of Applications with Noise (HDBSCAN) is efficient at recovering known and new stellar associations (e.g., see \citealp{2019AJ....158..122K,2021AA...645A..84M,2021ApJ...917...23K}). HDBSCAN is robust to non-typical cluster shapes or uneven cluster densities, and provides a computationally efficient method for detecting over-densities in $N$--dimensional space \citep{campello2013density, 2017arXiv170507321M}. However, the lack of heliocentric radial velocities for most Gaia stars encouraged past studies (e.g., \citealp{2019AJ....158..122K} and \citealp{2021AA...645A..84M}) to perform HDBSCAN clustering in 5D space ($XYZ$ Galactic positions and 2D tangential velocities). As a consequence of the projection effects where nearby stars can show large variations in their tangential velocities even while having similar $UVW$ space velocities, such previous studies have not recovered most of the nearest known stellar associations, within $\approx 100$\,pc of the Sun.

Young associations of stars which are located close to the Sun are especially interesting for the detailed characterization of young exoplanets and substellar objects, because their proximity allows us to achieve much higher signal-to-noise ratios even on the lowest mass members. In an effort to complete the census of such nearby young associations, we present a clustering analysis of nearby stars based on their full 6D $XYZ$ Galactic positions and $UVW$ space velocities calculated from Gaia~EDR3, and demonstrate that we can recover a significant number of known and new structures despite the restriction to stars with radial velocities reported in Gaia~DR2. This study allowed us to uncover spatial extensions to several Theia groups from \cite{2019AJ....158..122K} towards the Sun, \newcoronae\ previously unrecognized coronae around nearby open clusters, and \newgroups\ new co-moving associations with small dispersions in $Z$ and $UVW$ and color-magnitude diagrams consistent with ages above $\approx 100$\,Myr.

In Section~\ref{sec:sample_method}, we discuss the selection of our input sample that consists of nearby Gaia~EDR3 entries with good-quality kinematics, and provide a brief description of the HDBSCAN algorithm and how we apply it to uncover \ncrius\ clusters. We discuss these clusters in Section~\ref{sec:discussion} by comparing them with known associations in the literature, and identifying the clusters most likely to correspond to new coeval associations. We also discuss known exoplanets that are members of these newly discovered clusters, and we summarize our work in Section~\ref{sec:conclusion}.

\section{METHOD}\label{sec:sample_method}

\subsection{Sample}\label{sec:sample}

We used a subset of all Gaia~EDR3 entries within approximately 200\,pc of the Sun with good-quality 6-dimensional kinematics to identify nearby streams of stars. To achieve this, we selected stars with a Gaia~EDR3 parallax above 5\,mas with Renormalised Unit Weight Errors (RUWE\footnote{The RUWE is documented in the Gaia Data Release 2 Documentation at \url{https://gea.esac.esa.int/archive/documentation/GDR2/Gaia_archive/chap_datamodel/sec_dm_main_tables/ssec_dm_ruwe.html}.}) above 3 to exclude astrometric solutions strongly affected by unresolved companions or other systematic problems such as cross-matching errors. We have found that combining such a RUWE~$<$~3 criterion with additional quality criteria on parallax and proper motions was necessary to avoid an imprint of the distant Galactic plane into high-parallax (i.e., nearby) Gaia~EDR3 samples, because the crowded Galactic plane is prone to cross-matching errors that can mimick large parallax values. However, we have found that additional proper motion or parallax quality cuts were not necessary when working with the sample of Gaia~EDR3 stars with radial velocity measurements, because a simple RUWE~$<$~3 quality cut applied on stars with parallaxes 5\,mas or above resulted in high-quality parallax values with signal-to-noise values above 15, and similarly high-quality proper motion measurements with signal-to-noise values above 6. We note that the RUWE selection cut will bias our search against binary stars (most of which have RUWE~$>$~1.4; e.g., see \citealp{2021ApJ...907L..33S}), but will also greatly reduce sample contamination due to low-quality parallax or proper motion measurements.

We have calculated the individual $XYZ$ Galactic positions and $UVW$ space velocities of our sample with a Monte Carlo simulation. This $XYZUVW$ frame of reference is centered on the Solar position and motion, where the positive $X$/$U$ axes point towards the Galactic center, and the $Y$/$V$ axes are in the direction of the rotational motion of the Galaxy, such that $XYZ$ and $UVW$ form right-handed coordinate systems. For each star in our sample, we have taken 100 artificial measurements of proper motion, radial velocities and parallaxes centered on the Gaia~EDR3 values with standard deviations equal to their respective measurement errors. We have then transformed each of these artificial measurements on the $XYZUVW$ frame of reference, and taken the median value along each direction as the most likely $XYZUVW$ values for each individual star. We have also calculated approximate measurement errors by taking the standard deviations of the resulting distributions along each dimension, and further rejected any star with resulting errors above 3\,pc or 3\,\kms\ in the individual $XYZUVW$ dimensions. We elected to apply these rigid selection cuts to reduce systematic and measurement errors because the HDBSCAN algorithm, described in Section~\ref{sec:hdbscan}, does not account for measurement errors by construction.

We note that the Monte Carlo simulation used here would be equivalent to using the flat-prior distance estimations of \citep{2021AJ....161..147B} in the specific step where parallaxes are transformed to distances, but our method also similarly deals with nonlinear effects in the transformation of measurement densities from radial velocities and proper motions to the $XYZUVW$ space. While non-flat Bayesian priors may be useful to calculate the distance probability density for more distant stars, our sample is constituted of stars within 200\,pc of the Sun with parallaxes at a high signal-to-noise ratio (above 15), meaning that the effects of a non-flat Bayesian prior would be negligible, especially after having rejected stars with a projected measurement density above 3\,pc or 3\,\kms\ in $XYZUVW$ space.

We have selected a volume of 200\,pc around the Sun for this search because previous surveys have not efficiently recovered loose associations within 100\,pc in a systematic way (see Section~\ref{sec:projection} for more details), and the 100--200\,pc region will be useful to identify extensions of known large-scale structures such as those recovered by \cite{2019AJ....158..122K} based on their spatial overlap. We also note that considering a larger volume would require significant computer memory depending on the chosen clustering algorithm, with little benefits compared to previous clustering surveys.

\begin{figure*}
	\centering
	\subfigure[Color-magnitude diagram]{\includegraphics[width=0.49\textwidth]{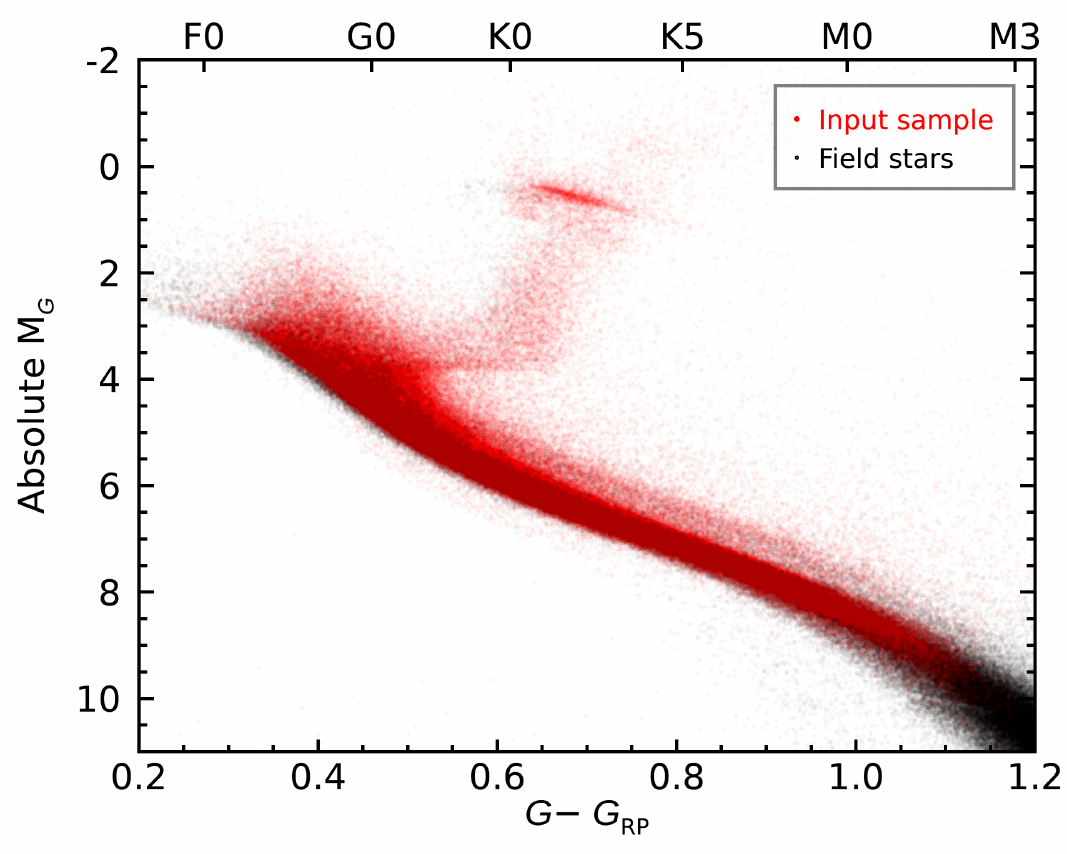}\label{fig:sample_cmd}}
	\subfigure[Color versus distance]{\includegraphics[width=0.49\textwidth]{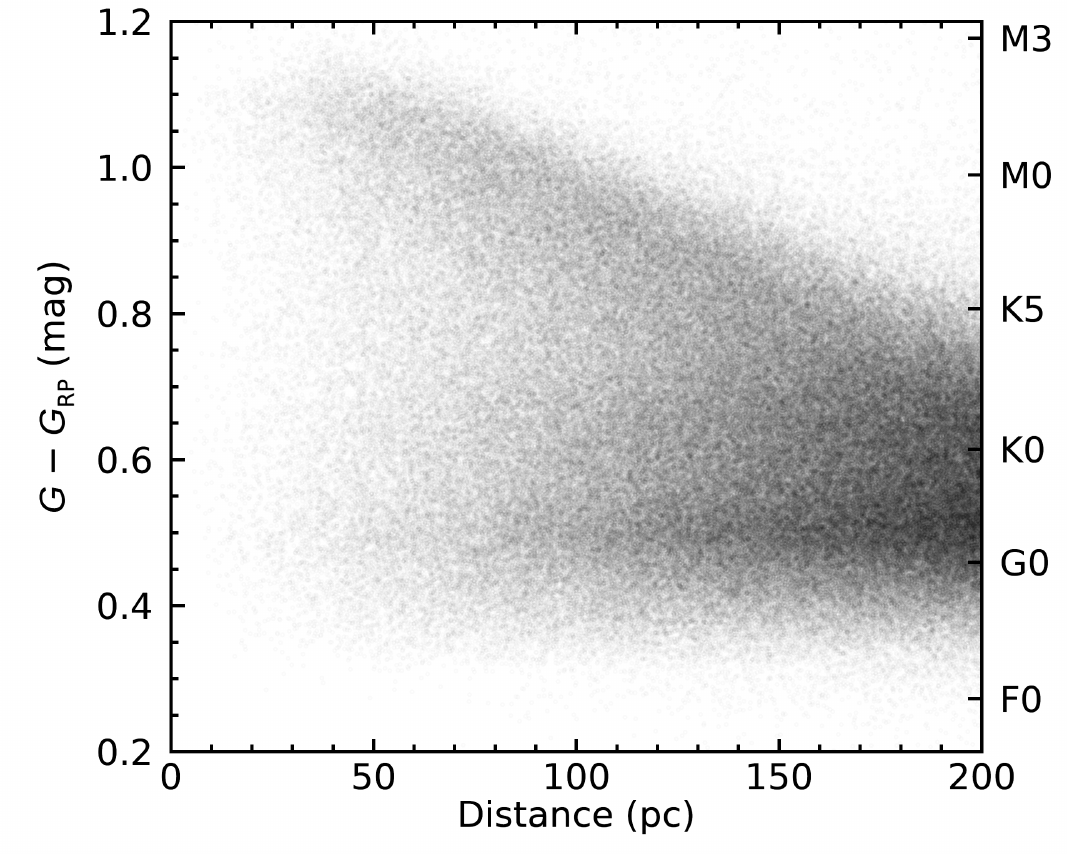}\label{fig:sample_dist_color}}
	\subfigure[]{\includegraphics[width=0.49\textwidth]{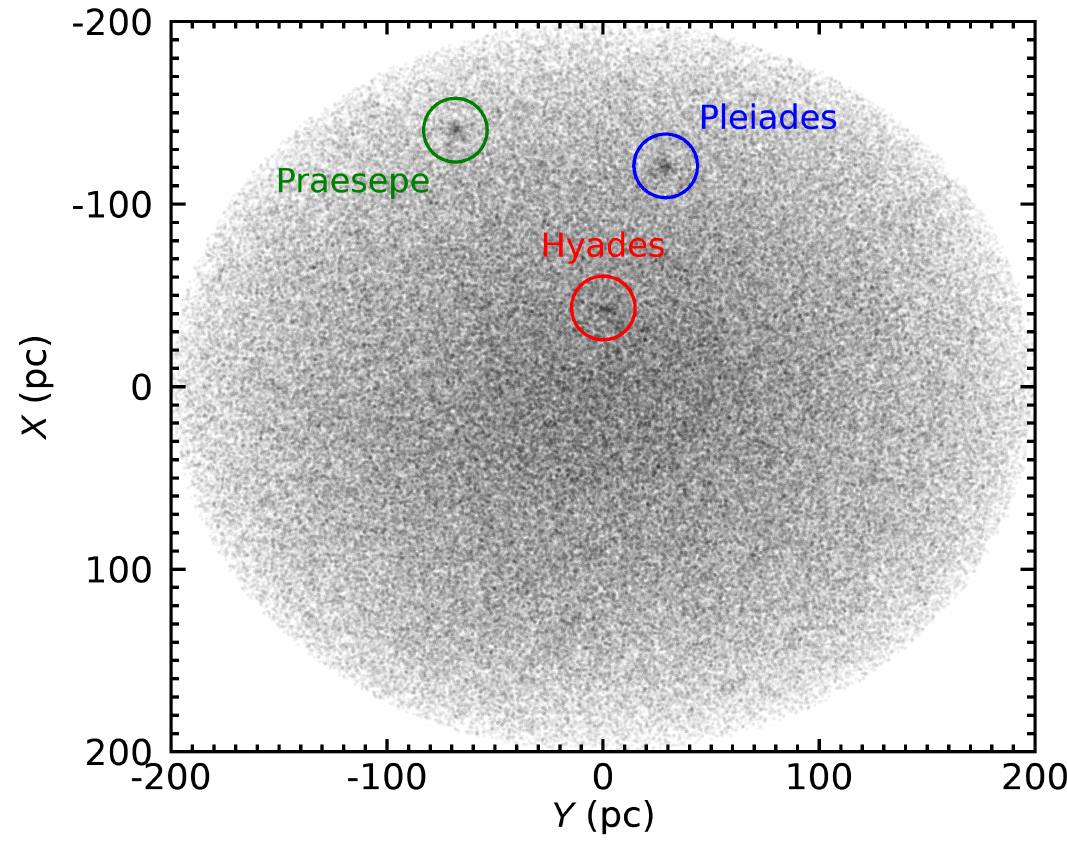}\label{fig:sample_xy}}
	\subfigure[]{\includegraphics[width=0.49\textwidth]{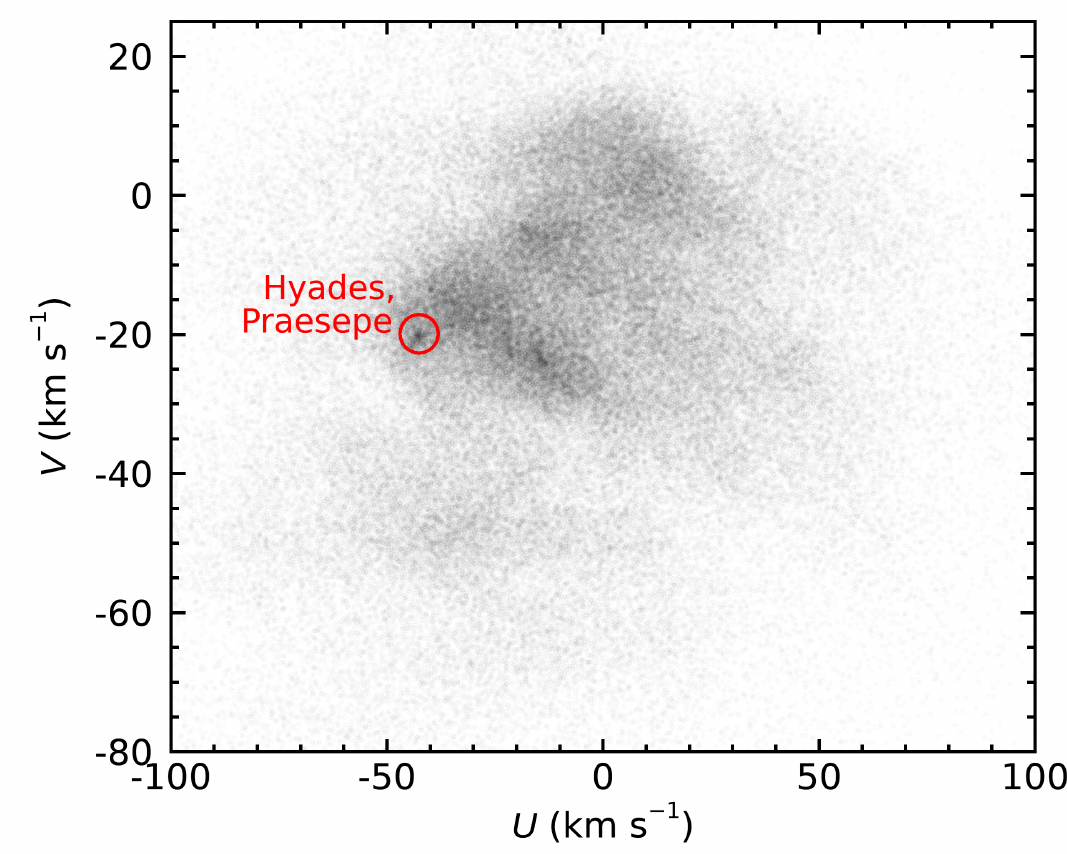}\label{fig:sample_uv}}
	\caption{Properties of the input sample considered here, constructed from Gaia~EDR3 entries with full kinematics. The input sample is identified with red dots in the upper-left color-magnitude diagram figure, and with black dots in all other panels. Our sample is heavily affected by the selection criteria for radial velocity measurements in Gaia~DR2 ($4 < G_{\rm RVS} < 13$, $3550 < T_{\rm eff} < 6900$, corresponding to spectral types F2--M2) because Gaia~EDR3 does not include additional radial velocity measurements. The distribution of colors versus distance reflects a combination of these selection cuts with the initial mass function that peaks at early M spectral types. The over-densities that are visible by eye in the $XY$ and $UV$ distributions were identified.}
	\label{fig:sample}
\end{figure*}

The general properties of our input sample are displayed in Figure~\ref{fig:sample}. The Galactic XYZ positions of our sample stars have typical measurement errors of 0.1--0.5\,pc, and the $UVW$ space velocities have typical measurement errors of 0.1--1.0\,\kms, with no significant differences in the distribution of measurement errors along the three spatial or kinematic axes. These typical measurement errors are in the range of typical velocity dispersions in star-forming regions and moving groups. Based on these combined selection criteria (parallax above 5\,mas, RUWE~$<$~3, $XYZ$ and $UVW$ errors below 3\,pc or 3\,\kms), we obtained an input sample of 303\,540 Gaia~EDR3 entries.

\subsection{HDBSCAN Clustering}\label{sec:hdbscan}

We used the Python \textit{scipy} implementation of the accelerated version of HDBSCAN \citep{2017arXiv170507321M}, which can take as an input a set of $N$--dimensional coordinates (with coherent spatial units) for $M$ objects, and offers several metrics to compute the mutual distances between individual objects (Euclidian distances are used by default). However, HDBSCAN can also take as an input an $N\times M$ matrix of pairwise distances that were pre-computed with any custom metric not offered in the library. Although it would be possible to pre-compute distances in $XYZUVW$ space that marginalize over the unknown heliocentric radial velocities of most Gaia~EDR3 entries, we found that such an approach required an unmanageable amount of computer memory even with much more restricted sample volumes. The distance metrics offered in the HDBSCAN library all respect the triangle inequality, allowing HDBSCAN to avoid computing most of the pairwise distances between all $M$ objects, and the resulting computational requirements scale approximately with $\mathcal{O}(M\log{}M)$ instead of $\mathcal{O}(M^2)$. However, distance metrics that rely on marginalized heliocentric radial velocities do not respect the triangle inequality -- because the best-matching radial velocity between a given pair of stars $AB$ may be different from the best-matching radial velocity when testing a different pair of stars $AC$. As a consequence, they cannot be implemented with an $\mathcal{O}(M\log{}M)$ scaling and require a pre-computation of a large matrix of pairwise distances.

We have therefore elected to use HDBSCAN with the default Euclidian distance metric, however, we have taken an approach similar to \cite{2021ApJ...917...23K} where the mutual distances in kinematic dimensions are given more importance than those in spatial dimensions because we expect coeval groups of stars to have formed with a narrow $UVW$ distribution (typically 1--3\,\kms; \citealp{2004ARAA..42..685Z}), but with much larger spatial extents over time -- up to $\approx 15$\,pc in $Z$, and hundreds of pc in $XY$, in some cases (e.g., see \citealp{2019AJ....158..122K}, \citealp{2019AA...622L..13M} and \citealp{2021AA...645A..84M}). Furthermore, the $XYZ$ spatial positions have a different physical dimension than $UVW$ velocities, and therefore it becomes important to consider a transformation factor which bears physical units for consistency. We have thus constructed a $6\times M$ array of $\left[X,Y,Z,c\cdot U,c\cdot V,c\cdot W \right]$ coordinates where $c = 12$\,pc$/$\kms. This value is larger than the $c = 6$\,pc$/$\kms\ used by \cite{2021ApJ...917...23K} because we aim to consider potentially older groups similar to \cite{2019AJ....158..122K}, whereas \cite{2021ApJ...917...23K} specifically investigated a sample of $< 50$\,Myr-old stars. In fact, the $\approx$\ 200 young associations, open clusters and tidal tails within 500\,pc of the Sun with ages $\lesssim$~1\,Gyr \citep{2018ApJ...856...23G,2021AA...647A..19T,2020AA...640A...1C} have typical spatial-to-kinematic ratios of their characteristic axes (defined as the cube root of their 1$\sigma$ spatial or kinematic volume) in the range 1--14, and 90\% of them have a ratio below 12. Those with lower ratios are usually the denser cores of open clusters, and those with larger ratios are either loose moving groups or open cluster coronae. Choosing a value of $c = 12$ will therefore ensure that our clustering step is able to recover associations that are spatially as large as curently known moving groups and open cluster coronae.

HDBSCAN offers two clustering approaches: “Excess of Mass” (EOM) and “leaf”, which directs how it separates sub-clusters within larger clusters of stars. In the EOM approach, clusters that are most stable under changes in detection threshold are preferred, whereas the leaf method will tend to split such clusters further apart when subclusters are detected. We elected to use the leaf clustering method to obtain a sample of individual substructures, even though some of them may be related to each other. For example, one might expect that extended structures like the Sco-Cen OB association would be preserved as a final cluster under the EOM clustering method, whereas the leaf method would further identify its sub-populations (e.g. Upper Scorpius, Lower Centaurus Crux and Upper Centaurus Lupus, see \citealp{1946PGro...52....1B}, \citealp{1999AJ....117..354D} and \citealp{2016MNRAS.461..794P} for more detail), which have similar but distinct ages. Our choice of using the “leaf” clustering method will provide us with a finer sampling of individual structures, but we may find upon further study that some of these structures are related to each other through a larger past star-forming complex, for example. We used the default values for the $\epsilon$ and $\alpha$ parameters\footnote{Described in more details at \url{https://hdbscan.readthedocs.io/en/latest/basic_hdbscan.html} and in \cite{2017arXiv170507321M}.}. We have chosen a minimum cluster size of $N=10$ with HDBSCAN to start investigating the most significant missing structures of the Solar neighborhood in this work. We have avoided values of $N<10$ to reduce the number of false clusters that may correspond to random collections of stars; choosing a much larger value, on the other hand, would prevent us from recovering some of the sparser young associations in the Solar neighborhood, especially because we only consider stars in the temperature and $G$-band magnitude ranges that benefit from Gaia~DR2 radial velocities. Currently known young associations within 100\,pc of the Sun have between 2--200 members with Gaia~DR2 radial velocity measurements, and 80\% of these associations have at least 10 such members (some exceptions are discussed in Section~\ref{sec:partial}).

\begin{figure}
    \centering
    \includegraphics[width=0.47\textwidth]{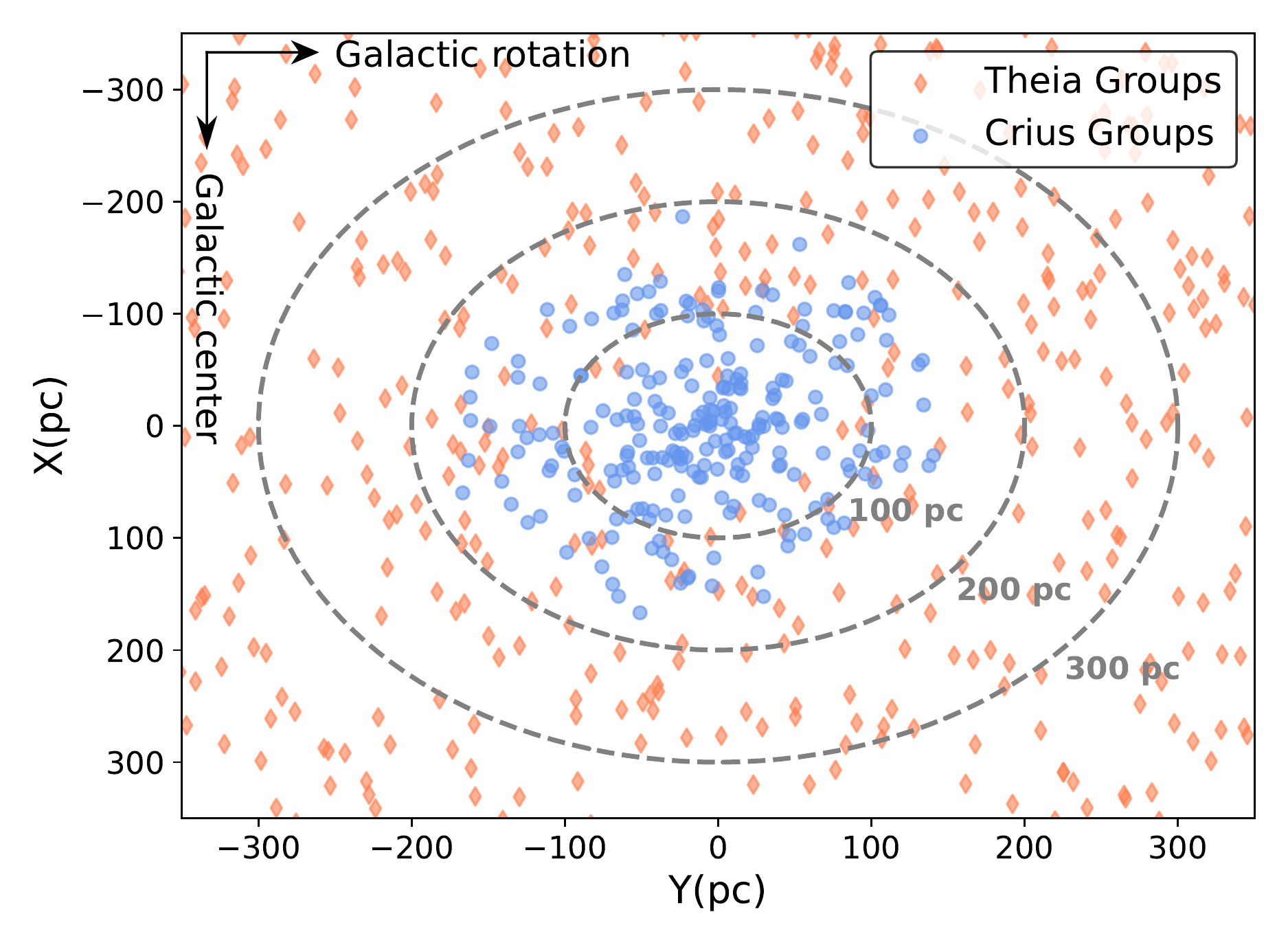}
    \caption{Median $XY$ Galactic positions of all \ncrius\ Crius groups recovered here, compared with the distribution of Theia groups from \cite{2019AJ....158..122K}. The clustering analysis presented here allowed us to recover groups much closer to the Sun, filling in the region within $\lesssim 100$\,pc from the Sun, because of its use of $UVW$ space velocities to prevent projection effects from artificially spreading out the members of nearby associations.}
    \label{fig:xy_dist}
\end{figure}

The application of HDBSCAN with this particular configuration to our sample described in Section~\ref{sec:sample} yielded a total of \ncrius\ clusters, which are listed in Table~\ref{tab:allstars} with their individual lists of Gaia~EDR3 source identifications. The median $XY$ Galactic positions of all Crius groups are shown in Figure~\ref{fig:xy_dist}, which shows how they fill the gap within $\approx 100$\,pc from the Sun where previous clustering studies were less efficient due to projection effects. Following the spirit of \cite{2019AJ....158..122K} in their nomenclature of Theia groups, we hereafter refer to our resulting \ncrius\ putative ensembles of co-moving stars as Crius~1 to Crius~241\footnote{The groups identified by \cite{2019AJ....158..122K} were named after Theia, a child of Gaia in the Greek mythology. We elected to follow this tradition by naming the groups identified here after Crius, another children of Gaia.}.

\section{DISCUSSION}\label{sec:discussion}

\begin{figure}
    \centering
    \includegraphics[width=0.52\textwidth]{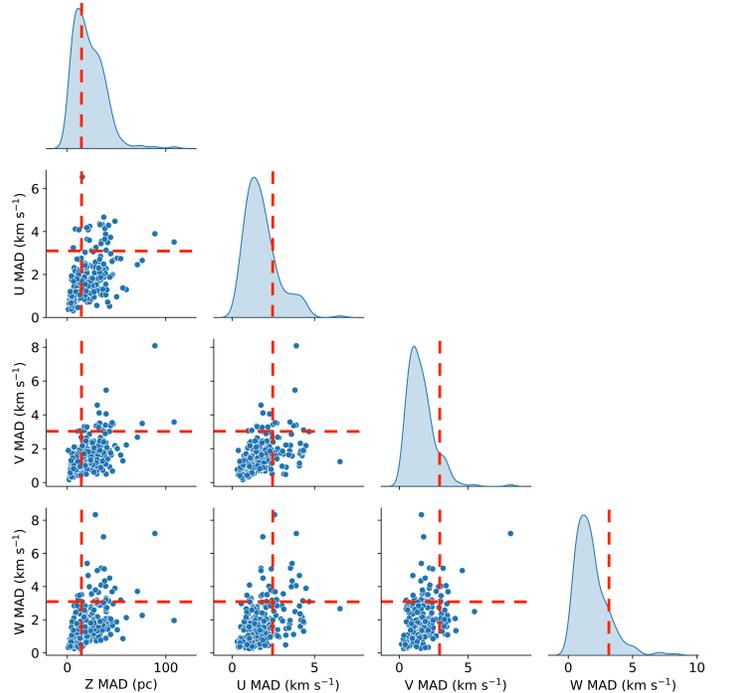}
    \caption{Corner plot of the median absolute deviations for the $Z$ Galactic coordinate and the $U$, $V$ and $W$ space velocities of the \ncrius\ Crius groups identified here. Red dashed lines correspond to the selection criteria described in Section~\ref{sec:discussion}.}
    \label{fig:zuvw_mad}
\end{figure}

A number of the \ncrius\ Crius groups identified in Section~\ref{sec:sample_method} may correspond to random over-densities of stars which are not coeval or did not form in a common birth region, much like some of the previously identified moving groups and streams of stars. Such unphysical groups tend to have larger $UVW$ velocity spreads compared with coeval groups (e.g., see \citealp{2016IAUS..314...21M}), however \cite{2019AJ....158..122K} also showed that older groups may display larger velocity spreads on average. Typically, coeval associations have intrinsic $UVW$ spreads of 1\,\kms\ or less \citep{2004ARAA..42..685Z,2016IAUS..314...21M}, but the observed spread of a coeval population can appear slightly larger (1--3\,\kms) when measurement errors contribute significantly to the $UVW$ standard deviation. A number of studies have also now demonstrated that coeval and co-moving groups of stars tend to be distributed along a relatively thin sheet in the $Z$ Galactic plane (e.g., see \citealp{2014AJ....147..146K} and \citealp{2021AA...645A..84M}). For these reasons, we have applied selection cuts on the Crius groups to focus on those that are most likely to consist of coeval stellar populations. We chose to further investigate only the Crius groups with individual $U$, $V$ and $W$ median absolute deviations below 3\,\kms, and $Z$ median absolute deviations below 15\,pc, which are in line with currently known young associations and allow for a modest contribution of radial velocity and parallax measurement errors. The distributions of $Z$ and $UVW$ median absolute deviations of all Crius groups are shown in Figure~\ref{fig:zuvw_mad}. We note that these rejected Crius groups may contain additional coeval structures, but they will likely also include a larger number of contaminated or unphysical clusters.

\begin{figure}
	\centering
	\includegraphics[width=0.47\textwidth]{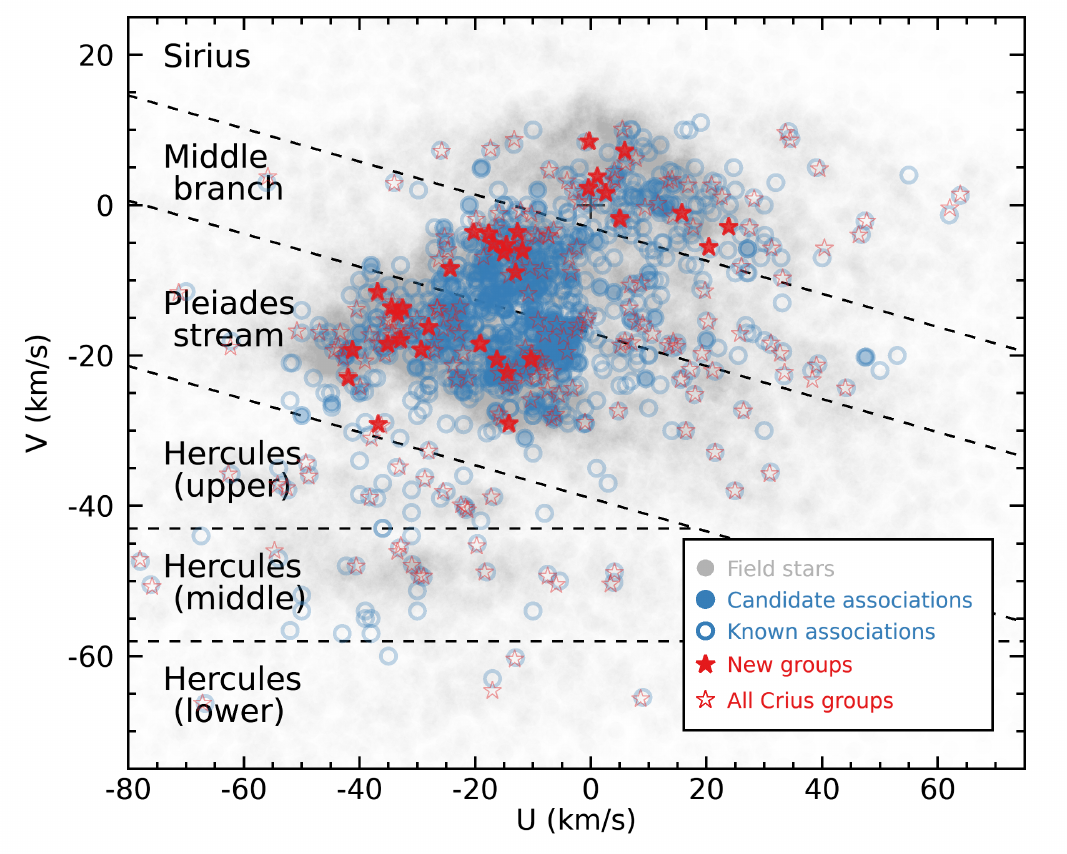}
	\caption{Median $UV$ space velocities of all Crius groups (open red stars), and new candidate associations presented here (filled red stars) compared with other known young associations in the literature (blue circles) and field stars within 100\,pc of the Sun in Gaia~EDR3 (grey circles). Open blue circles indicate “candidate” associations which have not yet been fully characterized, such as most of the \cite{2019AJ....158..122K} Theia groups, and we indicate the velocity regions introduced by \cite{1958JBAA...78...21E,1971PASP...83..251E}. We have excluded Crius groups within the Hercules regions as well as the higher-velocity groups at $V>40$\,\kms\ or $U < -60$\,\kms\ from this study \citep{2021AA...649A...6G}.}
	\label{fig:global_groups_uv}
\end{figure}

The distribution of $UV$ velocities of Crius groups is compared with other known associations and nearby field stars in Figure~\ref{fig:global_groups_uv}, along with the velocity regions defined by \cite{1958JBAA...78...21E} and \cite{1971PASP...83..251E}, and further discussed by \cite{2021AA...649A...6G}. Groups in the three Hercules “branches” are known to cover a wide range of stellar ages and are still not well understood; they may correspond to resonances in the Galactic orbits of relatively old stars caused by the Galactic arms or bar or past accretion events (e.g., see \citealp{2010LNEA....4...13B}). The Hercules streams are known to be substructured \citep{1998AJ....115.2384D}, in line with our analysis having retrieved clusters in this region of the $UV$ plane. We have excluded the Crius groups in the Hercules branches for the current study, and similarly exclude the groups with unusually high velocities ($V>40$\,\kms\ or $U < -60$\,\kms) that likely correspond to substructure in the older halo stars, but we note that these groups may be interesting for the study of older stellar populations.

\begin{figure}
	\centering
	\includegraphics[width=0.47\textwidth]{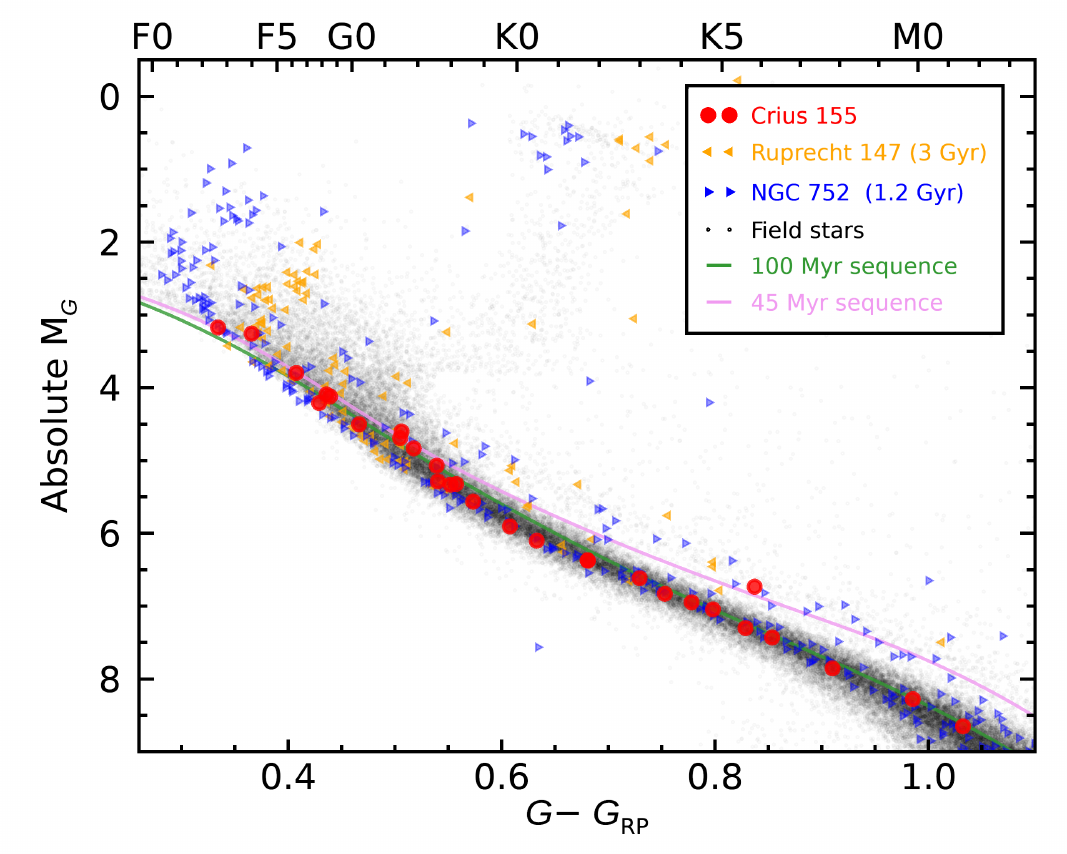}
	\caption{Gaia~EDR3 color-magnitude diagram for Crius~155 members, compared with the empirical sequences of \cite{2021ApJ...915L..29G} and the older open clusters NGC~752 and Ruprecht~147. The low-mass members of Crius~155 are consistent with the 100\,Myr-old sequence, but also with both older open clusters. The absence of a main-sequence turn-off of the most massive members of Crius~155 suggests a maximum age intermediate between NGC~752 and Ruprecht~147 ($1.2-3$\,Gyr; \citealp{2020AA...640A...1C}).}
	\label{fig:cr155_cmd}
\end{figure}

Because we have restricted our input sample to stars with heliocentric radial velocities in Gaia~DR2, the locus of our current Crius members have spectral types in the range FGK (see Figure~\ref{fig:cr155_cmd}), preventing an efficient determination of coevality with isochrone fitting in the range 100\,Myr--3\,Gyr. We have therefore not used color-diagram magnitudes in our quality cuts, and note that this will likely be feasible with the additional radial velocities in Gaia~DR3.

The selection cuts described above have yielded a subset of \nselectedcrius\ Crius groups listed in Table~\ref{tab:allgroups_pass} which we have investigated further to identify whether they correspond to known coeval associations. All Crius groups that did not pass these selection criteria are listed in Table~\ref{tab:allgroups_fail}.

\subsection{Known Associations}\label{sec:known}

We determined whether each of the Crius groups identified in the previous section corresponds to a known association, moving group or open cluster by comparing their list of Gaia~EDR3 names with membership lists of known associations in the Solar neighborhood. This set of memberships builds on those compiled by \cite{2013ApJ...762...88M,2017AJ....153...95R,2018ApJ...856...23G} and gathers more than 450 publications that includes but is not limited to membership lists on the CDS \citep{2000AAS..143...23O}. The full list, including those not discussed in this work, will be presented in the form of an online database in a future publication (J.~Gagn\'e et al., in preparation), and includes all known associations and open clusters discussed in the literature that we could identify with an average distance within 400\,pc of the Sun, including those that were discovered under more than one name, or are currently thought to be an unphysical collection of stars. These include recent surveys such as \cite{2021ApJ...917...23K}, \cite{2021ApJS..254...20L} and \cite{2021AA...647A..19T}. Every instance where a number of Crius Gaia~EDR3 entries match members of a known association were investigated in both $UVW$ space velocities and $XYZ$ Galactic positions using the 3D data visualization software Partiview \citep{2003IAUS..208..343L} to determine whether the overlap between the known association and the Crius group is significant.

Subsequently, all Crius groups for which no member matched a known association were also investigated in $UVW$ space velocities and $XYZ$ Galactic positions using Partiview, considering all plausible matching groups in the literature which bulk $UVW$ space velocities fell within a radius of $\approx 10$\,\kms\ of the median $UVW$ space velocity of Crius groups, and which bulk $XYZ$ Galactic positions fell within a radius of $\approx 50$\,pc in of the median $XYZ$ Galactic position of Crius groups.

These two methods allowed us to identify several Crius groups that likely correspond to known associations, listed in Table~\ref{tab:known} and discussed in Section~\ref{sec:known}. In most cases, the overlap was mostly complete, meaning that we have recovered the majority of the known members in the substructure, and the full spatial overlap of the Crius group is similar to that of the known young association.

In a few cases, a Crius group was recovered that only spans a fraction of the full spatial distribution of a known young association; these cases are discussed further in Section~\ref{sec:partial}. In other instances, we have recovered significant spatial extensions of known associations that share the same $UVW$ space velocities within 3\,\kms. A number of these extensions simply correspond to the forefront of known loose associations towards nearby distances or extensions in the tangential direction that prevented previous clustering algorithms from identifying them because they are based on 2D tangential velocities rather than 3D space velocities (as discussed in Section~\ref{sec:extensions}). However, a number of other Crius groups seem to correspond to the coronae of less-well-studied open clusters such as Group~23 of \citeauthor{2017AJ....153..257O} (\citeyear{2017AJ....153..257O}, sometimes also named Oh~23) and Volans-Carina \citep{2018ApJ...865..136G}, in line with the recent discoveries of a large number of similar structures around known open cluster (see e.g. \citealp{2021AA...645A..84M}). These new candidate coronae are listed in Table~\ref{tab:coronae} and discussed further in Section~\ref{sec:coronae}. The remaining Crius groups that do not seem to correspond to known structures are listed in Table~\ref{tab:new} and discussed further in Section~\ref{sec:new_crius}.

\subsection{Nearby Moving Groups not Recovered}\label{sec:partial}

There are a few relatively well-studied nearby moving groups that were notably not recovered in our clustering analysis of the Solar neighborhood. These groups are discussed further below.

\subsubsection{TW~Hya Association}

The TW~Hya association \citep{1997Sci...277...67K,1989ApJ...343L..61D} was likely not recovered because it lacks a significant number of members with F and G spectral types -- in fact, no members of this association are currently known with spectral types in the range A2--K5 \citep{2017ApJS..228...18G}. Given that the $G$-band magnitude limit of Gaia~DR2 radial velocity measurements is $G \approx 13$ and the distance distribution of known TWA members is 50--80\,pc \citep{2017ApJS..228...18G}, we can expect to have Gaia~DR2 radial velocities for TWA members with absolute $G$-band magnitudes in the range 8.5--9.5, corresponding to spectral types M0--M2\footnote{See \url{https://www.pas.rochester.edu/~emamajek/EEM_dwarf_UBVIJHK_colors_Teff.txt}.}. Only 5 known members of TWA have radial velocity measurements in Gaia~DR2, the latest-type of which is CD--29~8887 (M1.5), consistent with the magnitude constraint for Gaia~DR2 radial velocities. As a consequence of this fact, only 4 known members of the TW~Hya association were included in our input sample because other members lack Gaia~DR2 radial velocity measurements, and one member has a large RUWE value of 8.8.

\subsubsection{Carina-Near Moving Group}
Although the Carina-Near moving group \citep{2006ApJ...649L.115Z} did not appear to have been recovered in our analysis, Crius~209, which was not studied further because of its large median absolute deviations in $Z$ (36.9\,pc) and $W$ (3.0 \kms), seems at least partially related to Carina-Near. Four members of Crius~209 are also known members of Carina-Near, however 9 other members are in common with the “Greater Sco-Cen” region defined by \cite{2021ApJ...917...23K}. A more detailed follow-up of Crius~209 will be required to determine whether it was simply subject to contamination by Sco-Cen, which caused it to be rejected in our analysis.

\subsubsection{Argus Association}
The physical nature of the Argus association \citep{2008hsf2.book..757T} was called into question based on an isochrone analysis by \cite{2015MNRAS.454..593B}, suggesting that it was subject to a high level of contamination or constituted a random collection of stars that happened to share similar $UVW$ space velocities. More recently, \cite{2018AAS...23132603Z} provided further evidence that Argus may be a coeval association by assembling a cleaner list of members that displayed a more consistent population in the color-magnitude diagram as well as several debris disk host stars, and provided an age estimate of 40--50\,Myr. The Argus association has long been tentatively associated with the IC~2391 open cluster, and \cite{2021ApJ...915L..29G} hypothesized that Theia~114 and 115 of \cite{2019AJ....158..122K}, along with the Argus association, may constitute extended tidal tails around IC~2391. Although Crius~193 and Crius~196 seem related to Theia~115 and IC~2391 respectively, we find no significant spatial overlap between them and the \cite{2018AAS...23132603Z} list of Argus members. This is likely not a consequence of the selection cuts we have imposed on our input sample, given that it contains 29 members of the \cite{2018AAS...23132603Z} Argus sample.

\subsubsection{AB~Dor Moving Group}
The AB~Dor moving group (ABDMG; \citealp{2004ApJ...613L..65Z}) is a well-studied $133^{+15}_{-20}$\,Myr-old \citep{2015MNRAS.454..593B,2018ApJ...861L..13G} group whose members are distributed across all directions in a local bubble of $\approx 140$\,pc diameter around the Sun. \cite{2016IAUS..314...21M} noted that, while the nucleus of members at distances of 15--50\,pc from the Sun defined by \citep{2004ApJ...613L..65Z} is a convincing coeval and co-moving population, the status of other members are less clear due to their wider spread in Galactic positions and space velocities. Furthermore, \cite{2013ApJ...766....6B} found that only half of a 10--stars sample in the stream ABDMG members have coherent chemical composition, indicating that the stream members are highly contaminated at best.

More recently, \cite{2019AJ....158..122K} identified Theia~301, a very large structure that appears co-moving and coeval with the ABDMG, which \cite{2021ApJ...915L..29G} hypothesized may correspond to the still missing corona around the Pleiades open cluster, along with the ABDMG. It is unclear why most core members of the ABDMG were not recovered in this analysis, given that 22/35 of the core members and 89 further members are included in our input sample. Crius~231 (30 members) is clearly related to the ABDMG, with which it shares 18 members, but those are preferentially located in the vicinity of Theia~301. This indicates that perhaps current membership lists of the ABDMG are either contaminated by a large number of random interlopers, or it may consist of more than one distinct population as hypothesized by \cite{2016IAUS..314...21M}.

\subsubsection{Octans Association}
Although the relatively well-studied 30--40\,Myr-old Octans association \citep{2015MNRAS.447.1267M} was not recovered in this section, we found that Crius~164 (78 members), with a $Z$ median absolute deviation slightly above our selection criterion (15.1\,pc), shares common members and kinematics with the more distant half of the spatial distribution of Octans. It remains unclear why we have not recovered the other nearby half of Octans in this analysis.

\subsubsection{Pisces Moving Group}

\cite{2015MNRAS.452..173B,2018MNRAS.473.2465B} have recently identified a new candidate moving group which they named the “Pisces Moving Group” from a list of nearby young stars that are not members of other known young associations, and they determined an age of about 30--50\,Myr for the group of 14 stars, based on a color-magnitude diagram of its low-mass members. Recently, Gaia~EDR3 data provided full kinematics for all members of this putative group, and the wide resulting distributions in $UVW$ space velocities make its physical nature questionable as a single coeval association. The standard deviations in $UVW$ are larger than typical coeval groups of such a young age (5.3, 6.2, and 4.2\,\kms, respectively) and the median absolute deviations are also large (2.6, 4.3, and 4.5\,\kms), indicating that the standard deviations are not driven by a small number of outliers.

There are, however, two possible slight over-densities in the $UVW$ distribution of the original Pisces moving group members, with standard deviations typically near or below 1\,\kms. We will refer to these putative subgroups as “Subgroup~1” (with members HQ~Psc, GP~Psc, CH~Ari, and CF~Ari) grouped around $UVW = (-11.9, -5.8, -4.1)$\,\kms\ with standard deviations of $(1.3, 1.0, 1.9)$\,\kms; and “Subgroup~2” (with members TYC~5770--457--1, V395~Peg, TYC~584--343--1, and 2MASS~J23494539+3126272), grouped around $UVW = (-10.0, -0.4, -8.3)$\,\kms\ (hence 7.7\,\kms\ apart from Subgroup~1), with standard deviations of $(0.8, 1.8, 0.7)$\,\kms. We note that these two subgroups are also clustered spatially, with all 4 stars of Subgroup~1 fall below the $Z = 0$ plane, and all stars of Subgroup~2 fall above it, further suggesting that they are likely distinct populations.

While a Gaia color-magnitude diagram of the three earlier-type (G8--K1) members of Subgroup~1 seems consistent with an age of $\approx$\,20\,Myr, the K3--type star GP~Psc falls significantly below the expected young sequence, closer to what could be expected for $\approx$\,110\,Myr-old Pleiades members. GP~Psc is also the most outlier member of Subgroup~1 in $UVW$ space, at 3.3\,\kms\ from the locus of the other three members. Taking GP~Psc out from the list of Subgroup~1 members shifts their median positions to $UVW = (-11.9, -5.8, -5.1)$\,\kms\ with updated standard deviations of $(0.4, 1.1, 0.6)$\,\kms.

The members of Subgroup~2 (G8--K2) are all consistent with color-magnitude sequences of known associations with ages in the range 45--110\,Myr, and therefore appear slightly older than Subgroup~1. We note, however, that both of these subgroups are made of extremely small samples, and they are therefore not immediately convincing cases for new coeval associations. For the current purpose of comparing the Pisces moving group members to Crius groups, however, we found it necessary to first construct sets of stars within the Pisces moving group with consistent space velocities to avoid comparisons with spurious members.

After this revision of the Pisces moving group members, we have verified whether any members of a Crius group fell within 50\,pc of each individual member of the Pisces Subgroups~1 and 2, and determined which of these individual stars fell closest to each Pisces star in $UVW$ space. In the case of Pisces Subgroup~1, we have found that 2/3 of its members fall within 3\,\kms\ and 50\,pc of a star in Crius~169. Indeed, the members of Subgroup~1 appear to be located on the edge of both the spatial and kinematic distributions of Crius~169 members, which we have associated with the 30--40\,Myr-old Greater Taurus Subgroup~4 of \cite{2021ApJ...917...23K}.

The case of the Pisces Subgroup~2 is a bit less straightforward; one of its members (TYC~5770--457--1) is located within the spatial-kinematic the locus of Crius~173 members, but we have rejected this Crius group from our detailed analysis because of its high dispersion in the $Z$ direction (about 18\,pc), but we note here that its color-magnitude diagram would indicate an age of at least 100\,Myr if it were a coeval group. Two additional members of Subgroup~2 appear consistent with the spatial and kinematic distributions of Crius~153, which corresponds to the $\approx$\,220\,Myr-old Theia~372 group of \cite{2019AJ....158..122K}.

Following these considerations, we consider that while some members of the putative Pisces moving group seem related to Crius~173, Theia~372, and Greater Taurus Subgroup~4, the Pisces moving group itself is likely a spurious compilation of unrelated relatively young stars that fall in a similar vicinity of the $XYZUVW$ volume.

\subsubsection{Ursa Major Cluster}
\begin{figure*}
	\centering
	\subfigure[]{\includegraphics[width=0.49\textwidth]{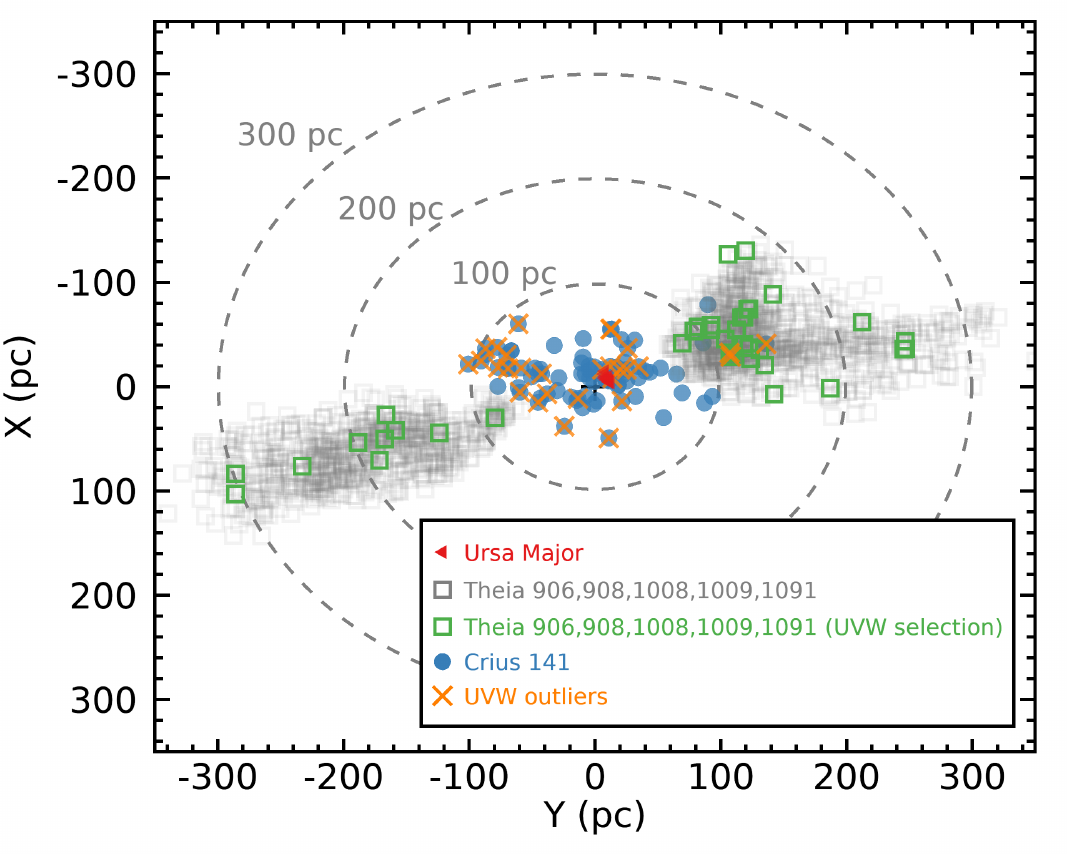}\label{fig:uma_xy}}
	\subfigure[]{\includegraphics[width=0.49\textwidth]{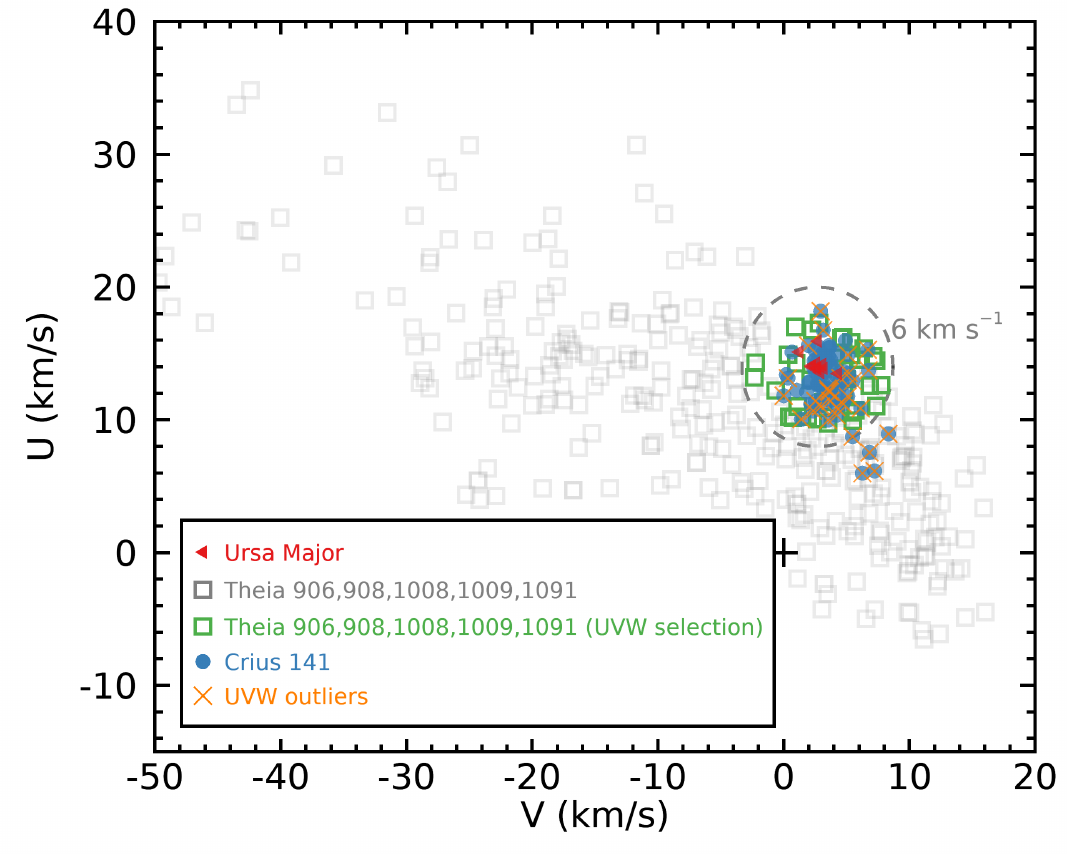}\label{fig:uma_uv}}
	\caption{Galactic positions and space velocities of Crius~141 members compared with the Ursa~Major core and the various \cite{2019AJ....158..122K} Theia groups tentatively identified as possible tidal tails around Ursa Major \citep{2020RNAAS...4...92G}. Crius~141 may be associated with the missing part of the tidal tails not recovered by \cite{2019AJ....158..122K} due to projection effects, and is likely associated with the Ursa~Major stream (or the “Sirius supercluster”) of \cite{1992AJ....104.1493E}. We identify stars with known 3D $UVW$ space velocities within 6\,\kms\ of the Ursa~Major core stars with a distinct color; we allow for a larger space velocity difference in this case because Crius~141 (and the other structures possibly associated with the Ursa~Major core) display a large spread in $UVW$ velocities.}
	\label{fig:uma}
\end{figure*}

We have further investigated whether one of the Crius groups recovered here may correspond to a candidate extension of the Ursa Major cluster (e.g., \citealp{2003AJ....125.1980K}), potentially extending it towards the more distant candidate tidal tails identified by \cite{2020RNAAS...4...92G} from an investigation of the Theia groups of \cite{2019AJ....158..122K}. We found that Crius~141 (84 members) appears to match the $UVW$ distribution of the core Ursa~Major members and seems more widely distributed spatially, potentially towards the five Theia groups that may constitute the tidal tails of Ursa Major (see Figure~\ref{fig:uma}), however, Crius~141 was not included in our detailed study of Crius groups because of its median absolute deviation of 16.5\,pc in $Z$, slightly above our 15\,pc selection criterion. In fact, Crius~141 shares 4 stars in common with the core list of \cite{2003AJ....125.1980K}, 4 with their list of Ursa Major “stream” members, and 2 more with the candidate tidal tails of \cite{2020RNAAS...4...92G}. Further characterization of its members will be required to determine whether Crius~141 is related to the Ursa~Major core.

\subsection{Candidate New Associations}\label{sec:new_crius}

\begin{figure}
	\centering
	\includegraphics[width=0.47\textwidth]{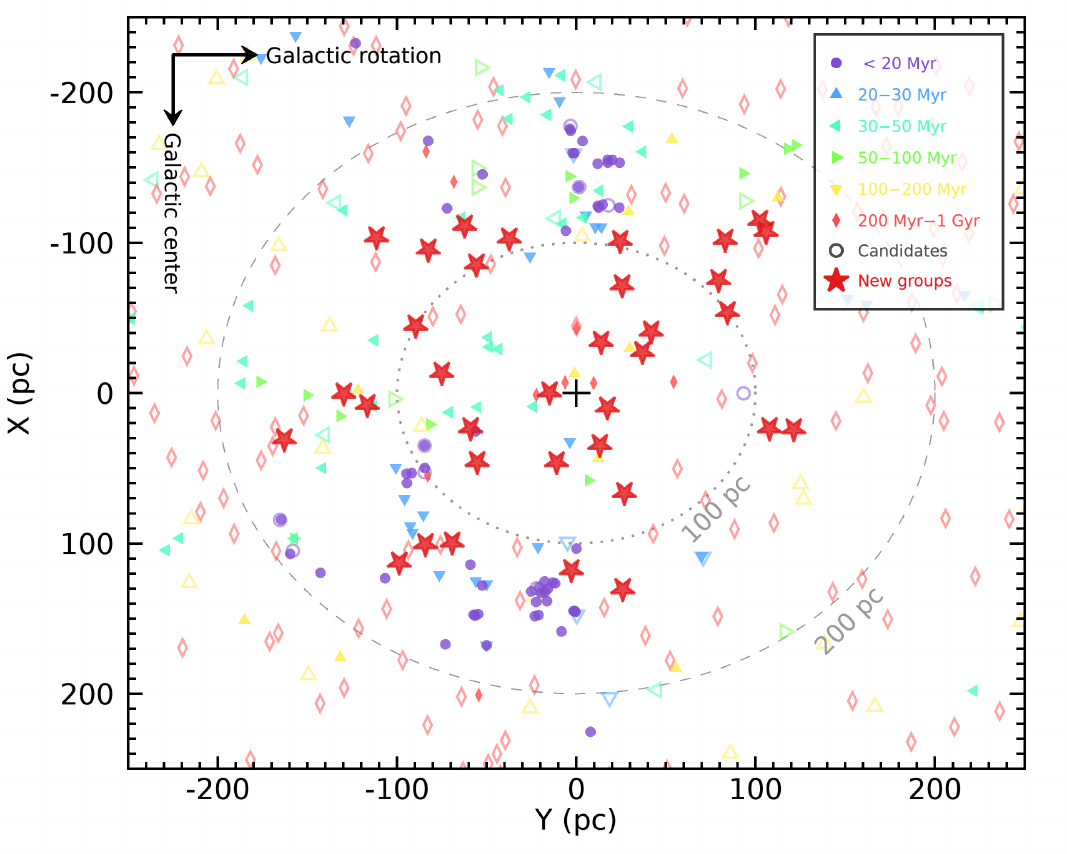}
	\caption{Median $XY$ Galactic positions of the new candidate associations presented in this work (large, red stars), compared with other known associations in the literature (color-coded by age). Open symbols indicate “candidate” associations which have not yet been fully characterized, such as most of the \cite{2019AJ....158..122K} Theia groups. All new candidate associations are older than $\approx 100$\,Myr, and some of then are located well within 100\,pc from the Sun.}
	\label{fig:global_groups_xy}
\end{figure}

A total of \newgroups\ Crius groups passed our quality selection criteria and do not appear to correspond to a known coeval association of stars. These groups, which may correspond to new loose coeval associations, are listed in Table~\ref{tab:new} and their median Galactic positions are shown in Figure~\ref{fig:global_groups_xy}. The typical number of stars in these new groups is in the range 10--28. When possible, we have assigned them with approximate age ranges based on their color-magnitude diagrams, which lower age range is based on the late-K or M stars compared with the empirical sequences of \cite{2021ApJ...915L..29G} constructed from members of nearby moving groups, and the higher age range is based on a main-sequence turnoff of their OBA stars compared with the old clusters NGC~752 ($1.2$\,Gyr; \citealp{2020AA...640A...1C}) and Ruprecht~147 ($3$\,Gyr; \citealp{2020AA...640A...1C}). An example is displayed in Figure~\ref{fig:cr155_cmd} for Crius~155. However, several of these putative new groups only have Gaia~DR2 heliocentric radial velocities for their FGK members, which prevents a more accurate age-dating.

The new candidate associations presented here still require detailed follow-up observations to determine their ages more precisely. The color-magnitude diagram positions of later-type members discovered either with a more detailed Bayesian analysis that does not require radial velocity measurements, or once Gaia~DR3 is released with about 33 million heliocentric radial velocities will aid in that analysis\footnote{See \url{https://www.cosmos.esa.int/web/gaia/dr3}.}. Lithium measurements and rotation periods will also be important to constrain the ages of these putative new associations and further test whether their members are coeval. The identification of white dwarf members of these young associations will also be an interesting avenue to provide an independent age calibration, however this will require an analysis that does not rely on heliocentric radial velocities given that they will not be available even in Gaia~DR3.

\subsection{Spatial Extensions of Known Associations}\label{sec:extensions}

Several groups identified by \cite{2019AJ....158..122K} show an elongated spatial structure, and we might expect that some of these would extend to the immediate Solar neighborhood ($\lesssim 100$\,pc), however, most of the nearest Theia groups stop short of this region, likely because of projection effects that start to impact the efficiency of clustering algorithms in 2D tangential velocity space. Conversely, the Crius groups identified here will not extend to very far distances, in part due to our 200\,pc selection cut on the input sample, but also because we only consider the brighter Gaia entries that benefit from heliocentric radial velocities in Gaia~DR2. We might therefore expect that some of the Crius groups discovered here are in fact a spatial extension of previously discovered Theia groups, and we have identified 17 such cases where Crius groups share similar $UVW$ velocities with Theia groups and form a spatial extension in the same plane in the Galactic $Z$ coordinate. These cases are listed in Table~\ref{tab:extensions} and the individual distributions of Galactic positions and space velocities of these candidate extensions are shown in the Appendix. Nine of these newly discovered extensions do not match a known spatially localized core that would make it possible to categorize them as coronae; these cases therefore simply correspond to new spatial extensions of spatially loose groups of stars.

The cases of Crius~168 and Crius~227 (displayed in the Appendix) are less straightforward because of a spatial gap between them and their corresponding Theia groups, but Gaia~DR3 will likely determine without ambiguity whether they are physical extensions by bridging the gap with the radial velocities for fainter and more distant stars.

\subsubsection{The MELANGE--1 Moving Group}\label{sec:mel1}

\cite{2021AJ....161..171T} recently identified MELANGE--1, a new moving group at $\approx 100$\,pc from the Sun. This new association was identified following the serendipitous discovery that the young exoplanet host star HD~110082 was co-moving with a large number of other seemingly young stars. The kinematics of this group of stars, which include HD~110082, did not clearly match any previously known young association, suggesting that MELANGE--1 was likely a newly discovered moving group. \cite{2021AJ....161..171T} identified a total of 133 members of MELANGE--1 (including HD~110082) and combined available lithium measurements and rotation periods to determine an age of $250_{-70}^{+50}$\,Myr for the newly discovered MELANGE--1 ensemble. They further identified that this moving group would be a compelling extension to Theia~786, given its consistent color magnitude diagram-based age ($280_{-60}^{+80}$\,Myr), similar kinematics and adjacent Galactic positions.

\begin{figure*}
	\centering
	\subfigure[Crius~170, Theia~786 and Theia~94 in $XY$ space]{\includegraphics[width=0.49\textwidth]{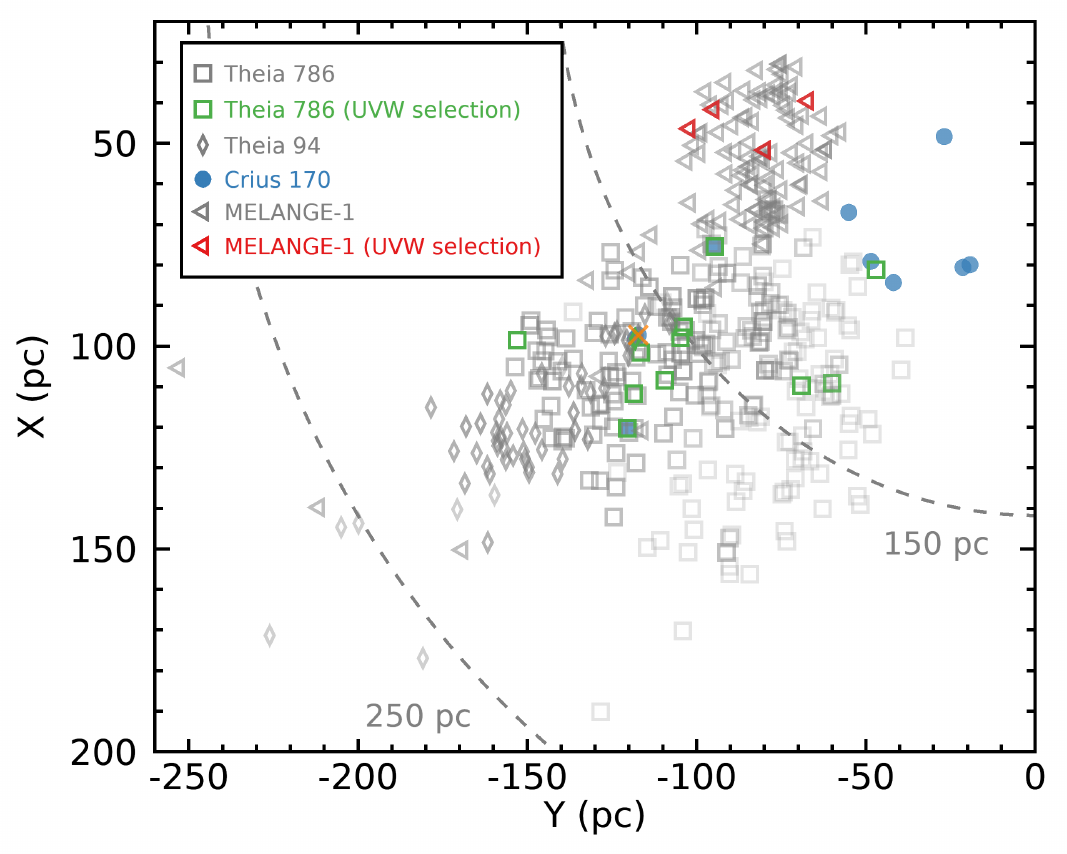}\label{fig:crius170_xy}}
	\subfigure[Crius~170, Theia~786 and Theia~94 in $UV$ space]{\includegraphics[width=0.49\textwidth]{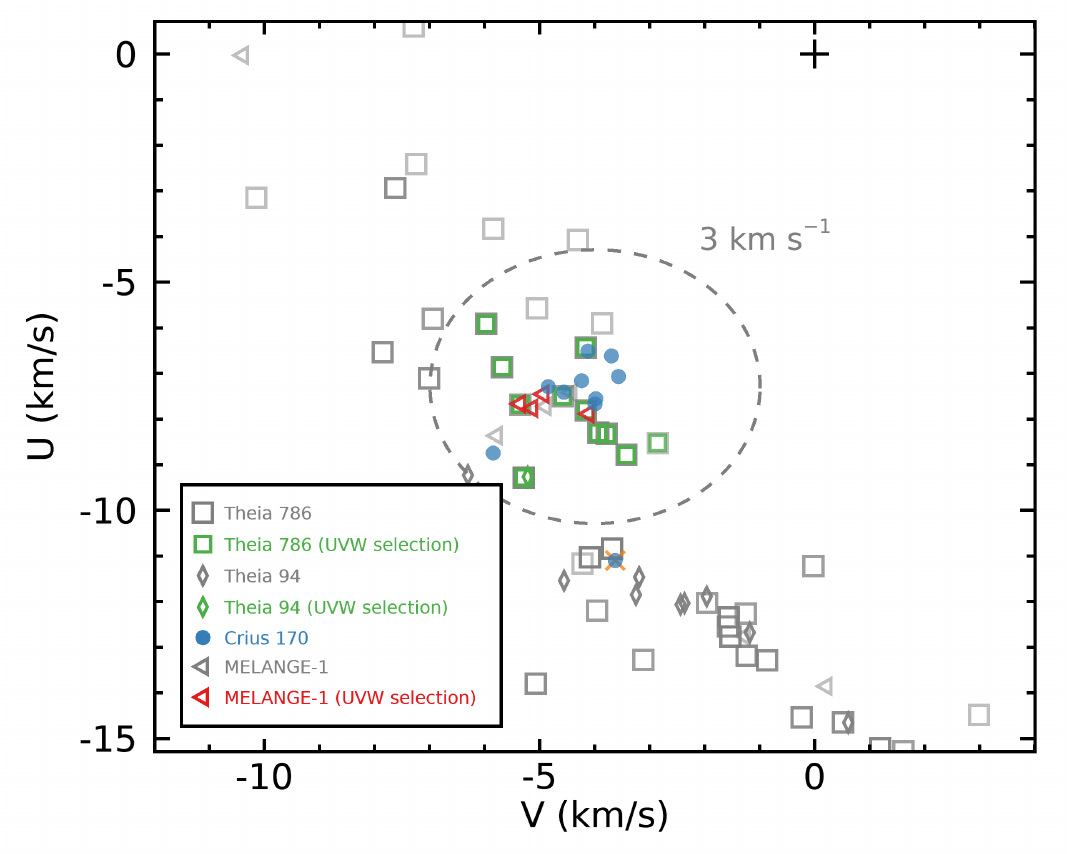}\label{fig:crius170_uv}}
	\caption{Galactic positions and space velocities of Crius~170 which we have identified as a putative extension of Theia~786. Members of Theia~786 and MELANGE--1 with UVW velocities located within 3\,\kms\ from the median bulk velocity of Crius~170 are marked shown with colored symbols, and the single Crius~170 member with outlier velocities is marked with an orange cross. Both Crius~170 and Theia~786 are also likely related with MELANGE--1, as noted by \cite{2021AJ....161..171T}. Theia~94 seems unrelated but causes slight contamination due to its similar space velocities and Galactic positions. The position and velocity of the Sun are marked with a black crosshair, when inside the figure range.}
	\label{fig:melange1}
\end{figure*}

Although we have not recovered any known member of MELANGE--1 in our Crius groups, Crius~170 (10 members) shares similar space velocities and is adjacent to MELANGE--1 in Galactic positions, as shown in Figure~\ref{fig:melange1}. We have also independently identified Crius~170 as an extension of Theia~786 before identifying the similarity in $UVW$ with MELANGE--1, and we also show Theia~94 in these figures because it has similar but slightly discrepant kinematics and it is located in the neighborhood of Theia~786 spatially. It is unclear at this point whether Theia~94 is related to Theia~786, Crius~170 and MELANGE--1, but it is likely to be a distinct sub-population of a larger group at best. For now, we consider it likely that Crius~170, Theia~786 and MELANGE--1 are different spatial segments of the same coeval population of co-moving stars, and it will be interesting to investigate whether additional radial velocity measurements in Gaia~DR3 will allow us to recover the full spatial extent of this population at once in a clustering analysis, and how Theia~94 fits in that picture.

\subsubsection{Older Moving Groups Associated with the Taurus-Auriga Star-Forming Region}\label{sec:tau}

\begin{figure*}
	\centering
	\subfigure[Crius~216 and Taurus-Auriga in $XY$ space]{\includegraphics[width=0.49\textwidth]{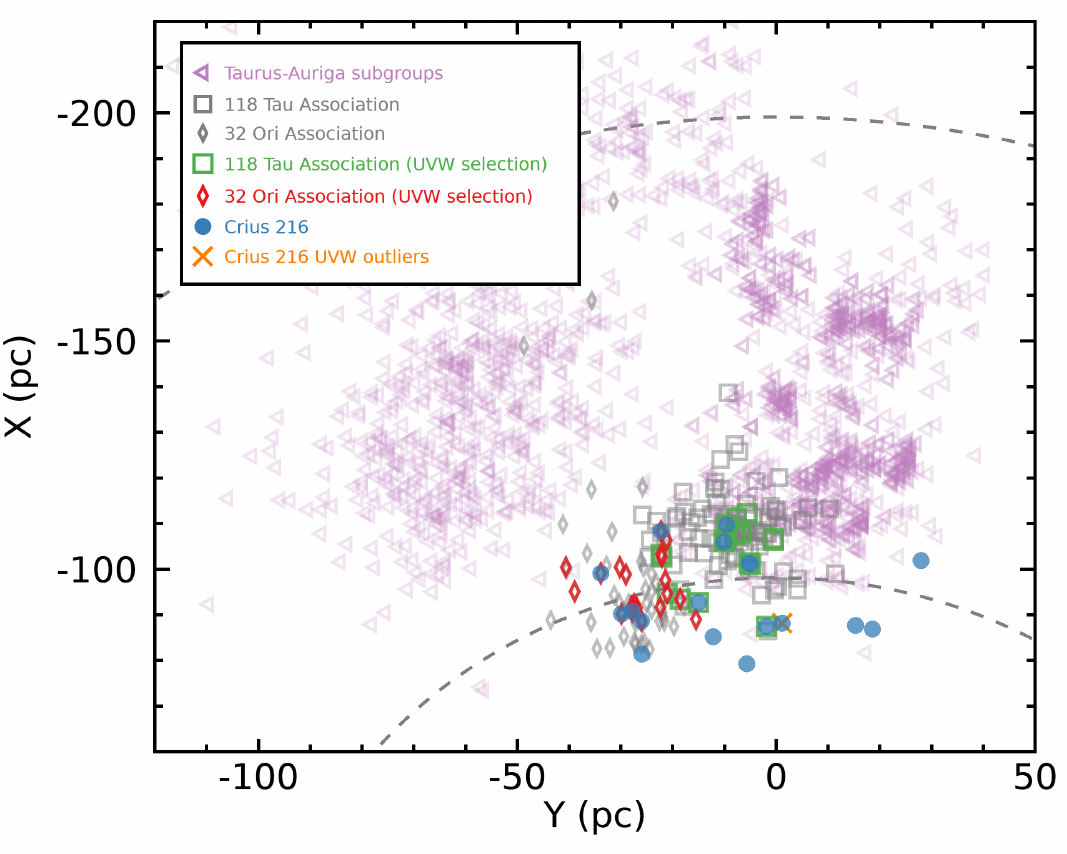}\label{fig:tau_xy}}
	\subfigure[Crius~216 and Taurus-Auriga in $UV$ space]{\includegraphics[width=0.49\textwidth]{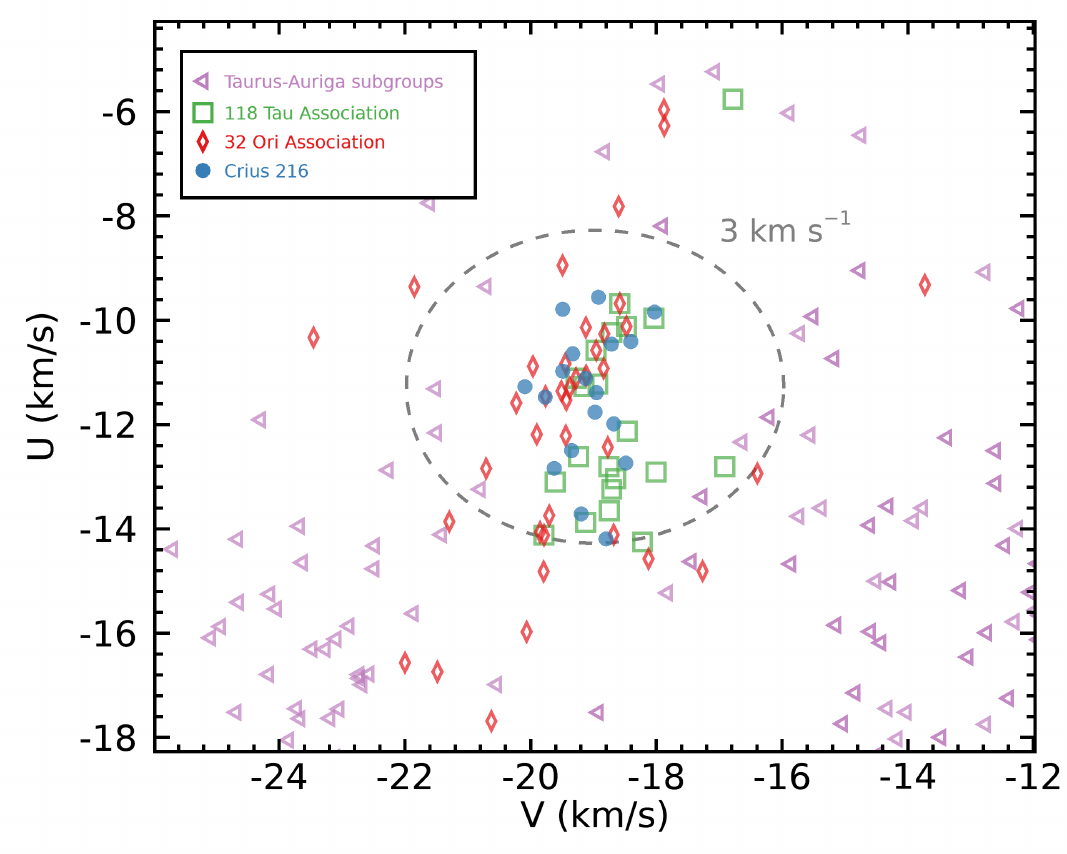}\label{fig:tau_uv}}
	\caption{Galactic positions and space velocities of Crius~216 and various subgroups associated with the Taurus-Auriga star-forming region \citep{2021ApJS..254...20L,2021ApJ...917...23K}. The subgroup corresponding to $\mu$~Tau (Greater Taurus Subgroup~1 of \citealp{2021ApJ...917...23K}) is not shown, and the subgroups corresponding to the 118~Tau association (Greater Taurus Subgroup~6 of \citealp{2021ApJ...917...23K}, and Group~11 of \citealp{2021ApJS..254...20L}) and the 32~Ori association (Greater Taurus Subgroup~7 of \citealp{2021ApJ...917...23K}) are with separate symbols. Crius~216 includes members of both the 118~Tau and 32~Ori associations, and they have $UVW$ velocities in the vicinity of other Taurus subgroups, consistent with the hypothesis that they represent contiguous star-formation events related with the larger complex of Taurus-Auriga members.}
	\label{fig:tau}
\end{figure*}

Crius~216 (18 members) includes members of both the relatively poorly characterized 118~Tau association \citep{mamajek118tau} and the $22^{+4}_{-3}$~Myr-old 32~Ori association first discovered by \cite{2007IAUS..237..442M} and further characterized by \cite{2010AAS...21542822S} and \cite{2017MNRAS.468.1198B}. \cite{2017ApJ...836L..15C} estimated that the age of the 118~Tau association seems older than the background Taurus-Auriga star-forming region based on its mid-M stars having \ion{Na}{1} line strengths \citep{2006AJ....131.3016S,2006AJ....132.2665S} comparable to members of the 10--16\,Myr Scorpius Centaurus complex \citep{2016MNRAS.461..794P}, and the lithium lines of its mid K-type stars being larger than those of typical 30--120\,Myr-old stars and comparable with those of members of the $\epsilon$~Cha association (3--8\,Myr, \citealp{2013MNRAS.435.1325M,2021AJ....161...87D}). They have therefore assigned a plausible age range of 3--10\,Myr for the 118~Tau association, slightly younger than the current estimates for the 32~Ori association. \cite{2017ApJ...838..150K} noted that 118~Tau and 32~Ori have very similar kinematics with respect to each other, and are only separated from Taurus-Auriga members by only 9\,\kms\ in $UVW$ space, indicating that these three populations might correspond to individual but contiguous star-formation events that originate from a single extended molecular cloud similar to Scorpius Centaurus. Our independent recovery of members of the 118~Tau and 32~Ori associations within Crius~216, as well as a continuous set of stars that span their spatial and kinematic $XYZUVW$ coordinates (shown in Figure~\ref{fig:tau}), strongly favors the hypothesis of \cite{2017ApJ...838..150K}. This hypothesis can be further tested in the near future by characterizing the ages of stars in Crius~216 with lithium measurements, as well as an isochronal age determination of its missing M-type members that will be recovered with  Gaia~DR3.

It is interesting to note that HDBSCAN has not separated members of 118~Tau from 32~Ori even though we have used the “leaf” method that would encourage breaking such substructure into distinct groups. This could indicate that there are additional subpopulations of intermediate positions, ages and space velocities distributed between the two associations, but a more detailed analysis based on Gaia~DR3 that includes many more stars will be useful to assess this.

\subsection{Coronae}\label{sec:coronae}

Several of the extensions of known associations identified in Section~\ref{sec:extensions} appear to be newly recognized coronae around known open clusters and other young associations. These are listed in Table~\ref{tab:coronae} and maps of their $XY$ Galactic positions and $UV$ space velocities are displayed individually in the Appendix. Figure~\ref{fig:coronae_all} shows the spatial extent of these new coronae compared with those already known in the literature. It is perhaps not surprising that additional known associations also host extended coronae given that most well-studied, nearby and dense open clusters have now been demonstrated to host similar structures \citep{2019AA...621L...3M,2021AA...645A..84M,2019ApJ...877...12T,2019AA...627A...4R} which likely correspond to tidal dissipation tails around the gradually degrading denser core of gravitationally bound stars. The main reason why these coronae were not previously recognized seems to be that they are related to less well studied, sparse open clusters (Platais~3, Alessi~9), or recently discovered associations (Volans~Carina, Groups 23, 51 and 59 of \citealp{2017AJ....153..257O} and RSG~2 of \citealp{2016AA...595A..22R}). We also note that Crius~87 (24 members), which displays a large $Z$ median absolute deviation (29.1\,pc), initially appeared to form a candidate tidal tail around the UPK~612 open cluster of \cite{2020AA...640A...1C}, but a closer inspection of its $UVW$ space velocities appears inconsistent with this hypothesis. However, Theia~216 of \cite{2019AJ....158..122K} appears to form a previously unrecognized corona around UPK~612, with mostly consistent kinematics (A figure is shown in the Appendix).

\begin{figure*}
	\centering
	\subfigure[Volans-Carina corona in $XY$ space]{\includegraphics[width=0.49\textwidth]{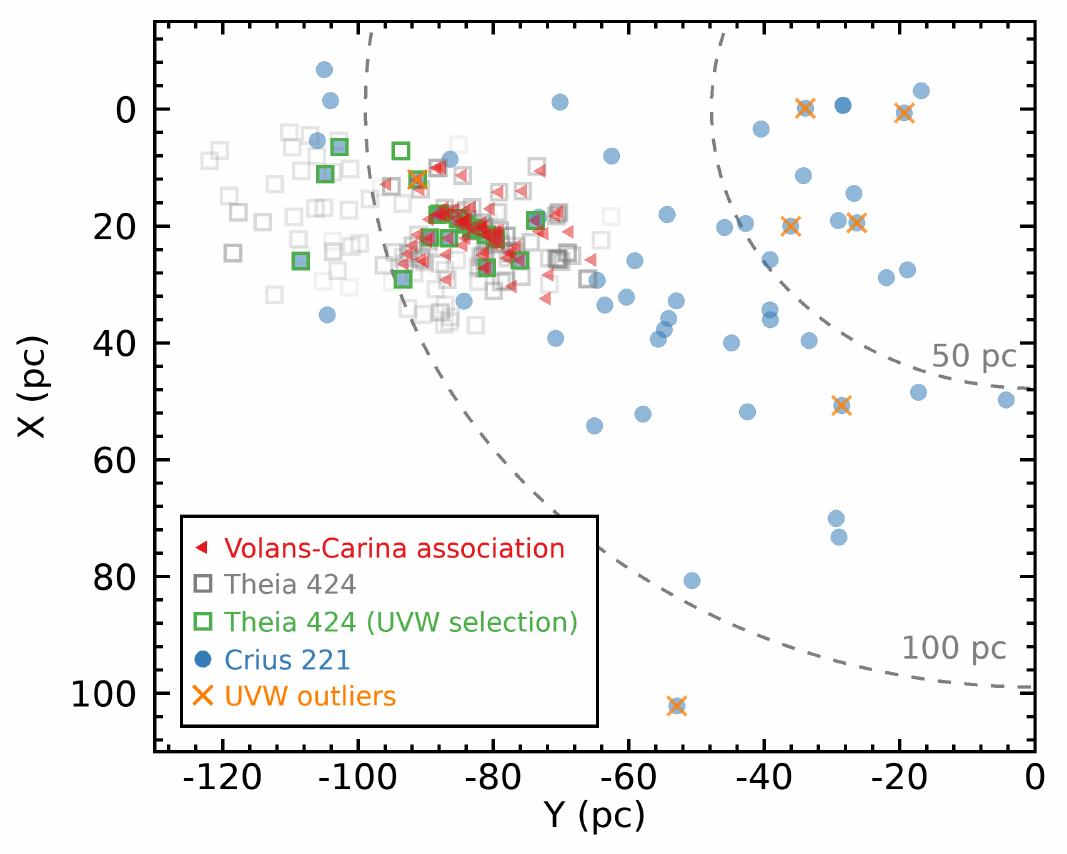}\label{fig:vca_xy}}
	\subfigure[Volans-Carina corona in $UV$ space]{\includegraphics[width=0.49\textwidth]{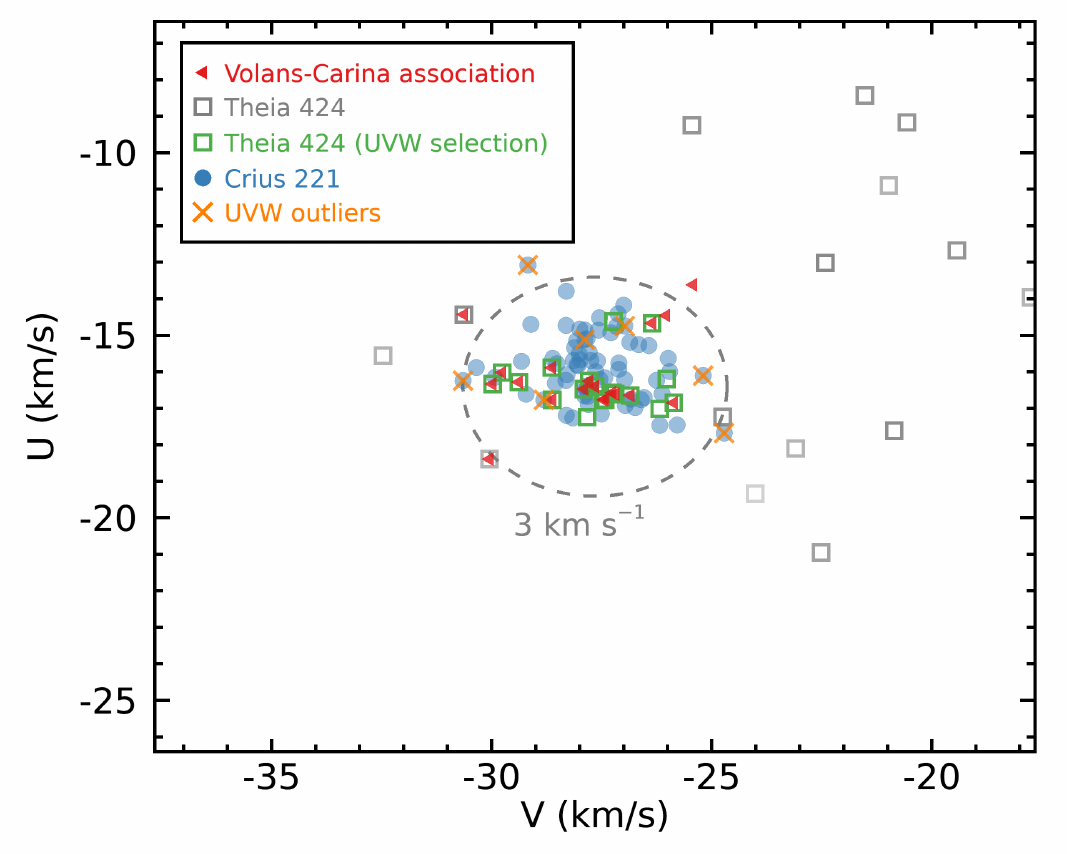}\label{fig:vca_uv}}
	\caption{Galactic positions and space velocities of Crius~221, which likely corresponds to the front-facing tidal dissipation tail (or “corona”) of the 90\,Myr-old Volans-Carina association. Stars with full 3D $UVW$ kinematics consistent with the core members of Volans-Carina within 3\,\kms\ are indicated with distinct symbols; other stars will need heliocentric radial velocity measurements for their membership to be fully confirmed. This young structure that reaches within \nearestvca\,pc of the Sun will be a valuable laboratory to study young exoplanets such as TOI--1224~b as well as substellar objects.}
	\label{fig:vca}
\end{figure*}

The case of Volans-Carina ($89^{+7}_{-6}$\,Myr; \citealp{2018ApJ...865..136G}) is particularly interesting (see Figure~\ref{fig:vca}), because its candidate corona (Crius~221, 67 members) seems to reach within the immediate solar neighborhood, with a member at \nearestvca\,pc from the Sun. This nearest member is the K1V-type, debris disk hosting star V419 Hya, which activity index $\log R^\prime_{HK} = -4.32$ \citep{2010ApJ...725..875I} is consistent with an age of 80--120\,Myr \citep{2008ApJ...687.1264M}, and displays Ca~II and X-ray Emission consistent with an age of $\approx 100$\,Myr \citep{2004ApJ...614L.125S}. It will be particularly interesting to study its substellar population in the near future. In addition to these newly recognized coronae, we note that Crius~178 (55 members) may form a slight extension of the NGC~2451A corona towards the Sun.

\begin{figure*}
	\centering
	\subfigure[$XY$ Galactic positions]{\includegraphics[width=0.49\textwidth]{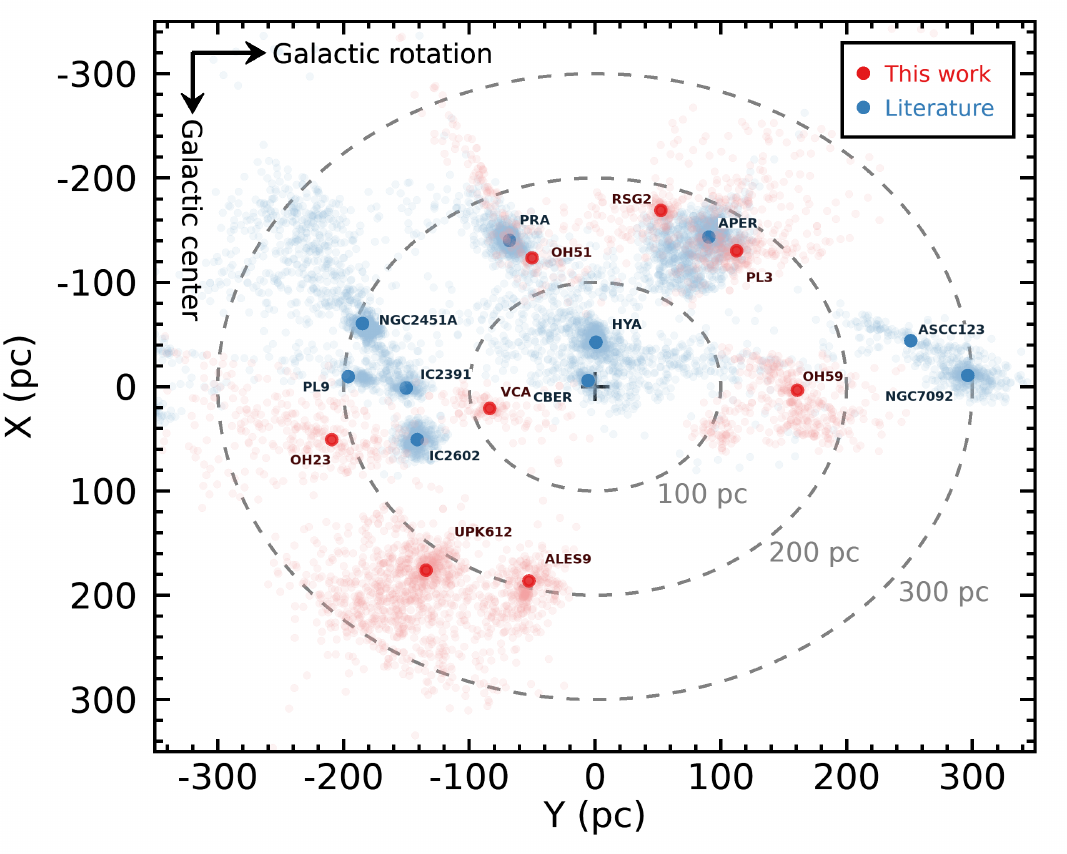}\label{fig:coronae_all_xy}}
	\subfigure[$YZ$ Galactic positions]{\includegraphics[width=0.49\textwidth]{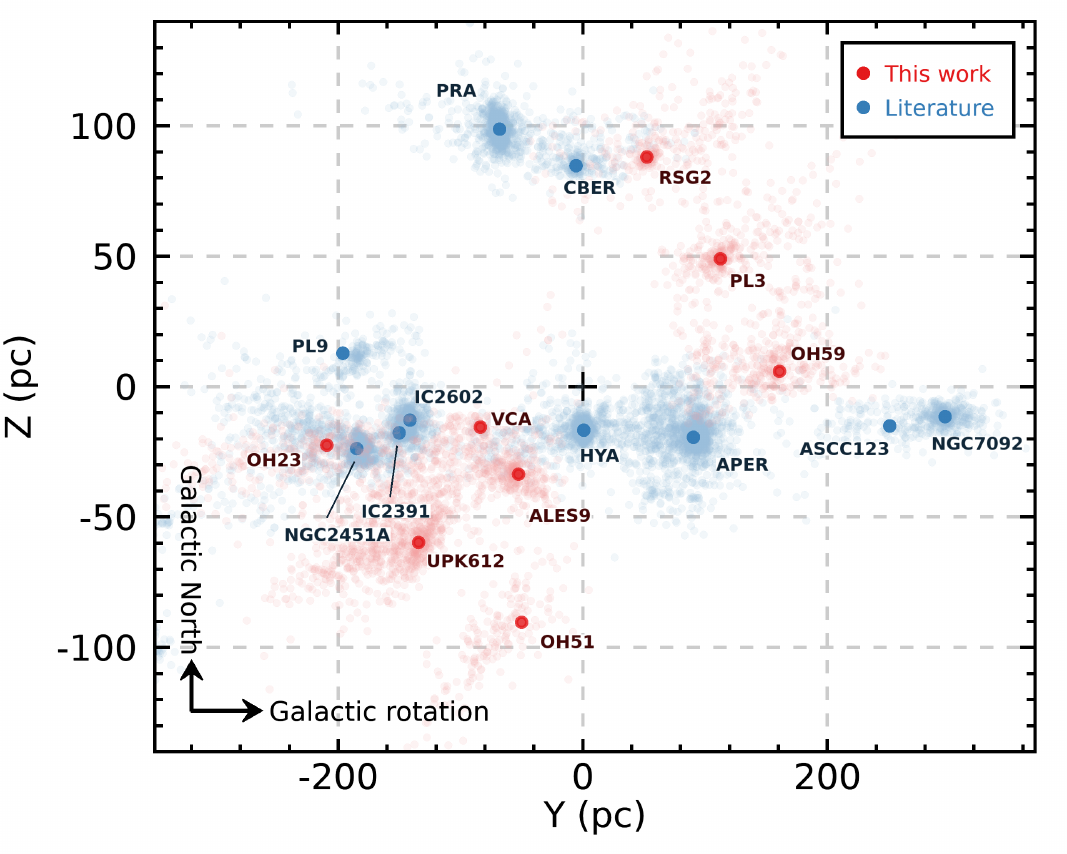}\label{fig:coronae_all_yz}}
	\caption{Galactic positions of the new coronae recognized in this work compared with those already known in the literature. With the exception of Oh~51, Platais~3, Oh~59 and Platais~9, most open clusters near the Sun lie in two distinct planes in $Z$, centered near $Z \approx -25$\,pc and $Z \approx 90$\,pc. The newly discovered corona around the 90\,Myr-old Volans-Carina association is the second closest known corona discovered so far, after that of the 680\,Myr-old Hyades association, and presents a unique opportunity for the search and characterization of its substellar members. Parts of the coronae shown here appear to be distributed along the line-of-sight, which is likely an artifact due to non-negligible parallax measurement errors. As shown by \cite{2021AA...645A..84M}, a proper spatial deconvolution of these measurement errors shows coronae that are flat in the $Z$ direction and not preferentially aligned along the line-of-sight, but distributed along a spatial axis in the $XY$ plane that is unique to each open cluster. The contracted figure tags correspond to the following associations: PL9: Platais~9; ALES9: Alessi~9; PRA: Praesepe; VCA: Volans-Carina association; CBER: Coma~Ber; HYA: Hyades; APER: $\alpha$~Persei; PL3: Platais~3.}
	\label{fig:coronae_all}
\end{figure*}

\subsection{Projection Effects and their Impact on Previous Clustering Analyses}\label{sec:projection}

\begin{figure*}
	\centering
	\subfigure[Spread in sky coordinates]{\includegraphics[width=0.47\textwidth]{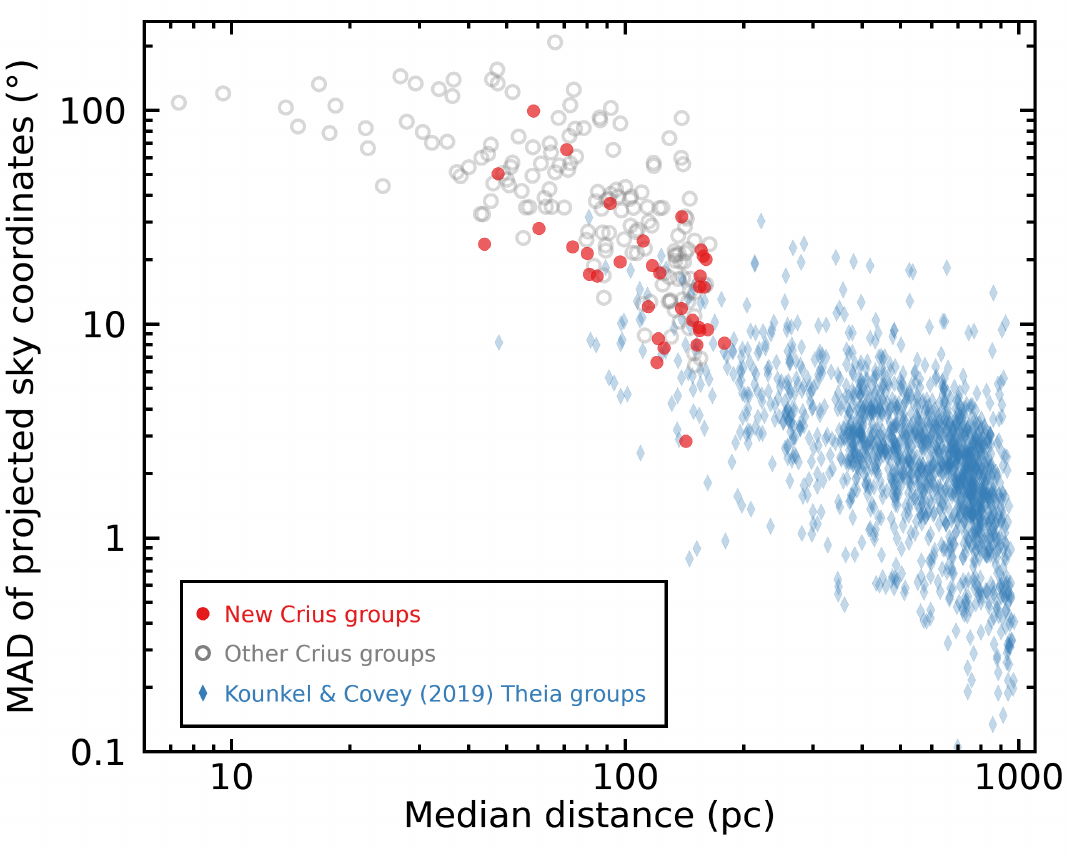}\label{fig:cris_kc_lb}}
	\subfigure[Spread in sky tangential velocities]{\includegraphics[width=0.47\textwidth]{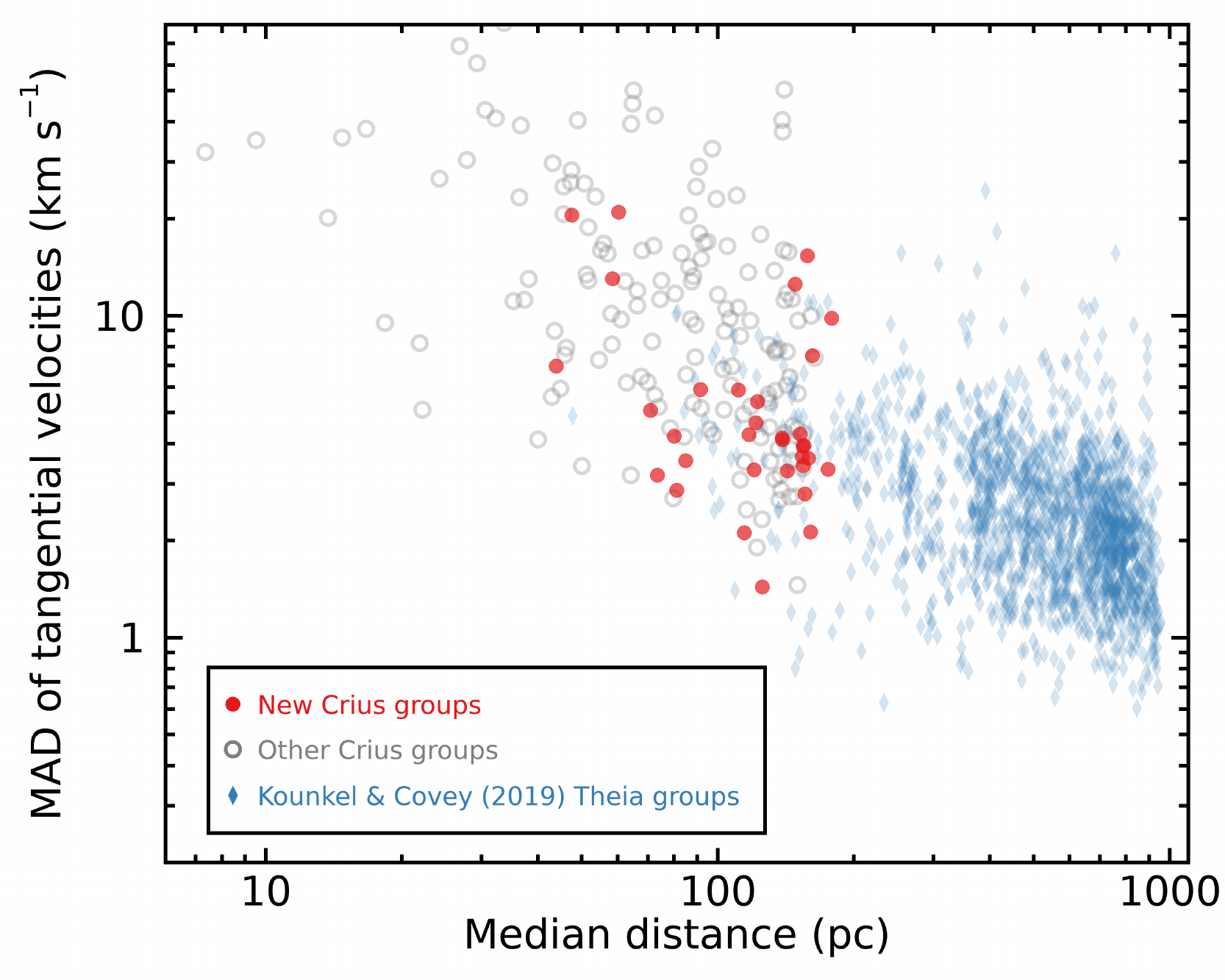}\label{fig:crius_kc_vtan}}
	\subfigure[Spread in $UVW$ velocities]{\includegraphics[width=0.47\textwidth]{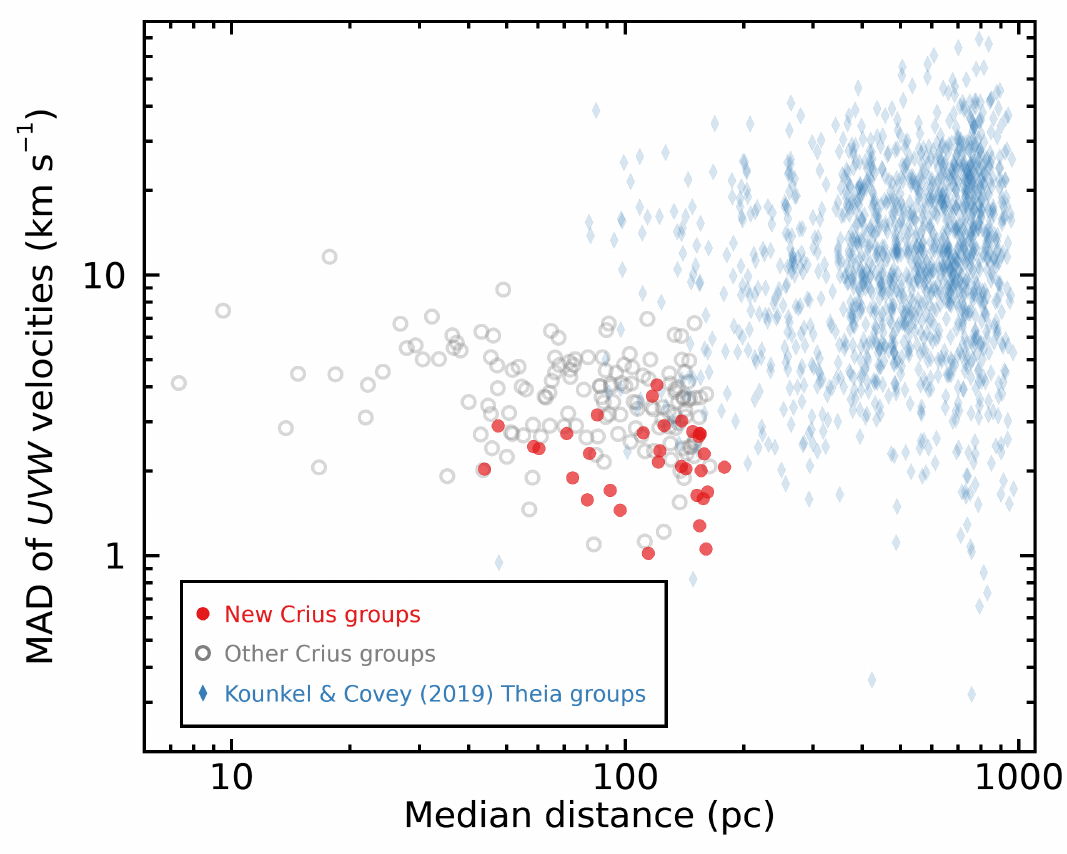}\label{fig:cris_kc_uvw}}
	\subfigure[Spread in $Z$ coordinate]{\includegraphics[width=0.47\textwidth]{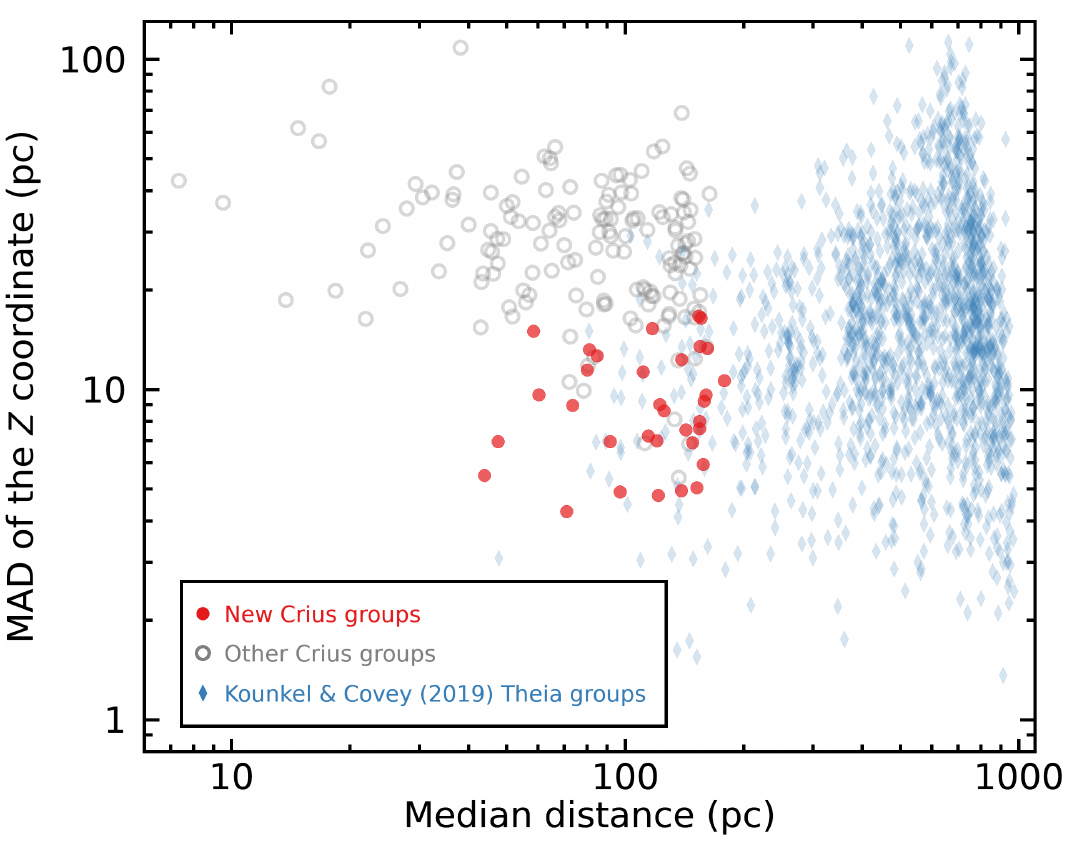}\label{fig:crius_kc_z}}
	\caption{Spread in properties as a function of distance for the new Crius candidate coeval associations presented here (red, filled circles), the other Crius groups which we have rejected based on various quality cuts (gray, open circles) and the Theia groups of \cite{2019AJ....158..122K}. We show the median absolute deviations (MAD) instead of standard deviations to avoid the exaggerated impact of outliers due to low-quality measurements. These distributions outline how the large spread of sky coordinates at nearby distances can drive a larger spread in tangential velocities. Crius groups show much smaller $UVW$ spreads compared with average Theia groups, as described in Section~\ref{sec:projection}. The spreads in the $Z$ coordinate of Crius groups are comparable to those of the Theia groups, as should be expected.}
	\label{fig:spreads}
\end{figure*}

We show in Figure~\ref{fig:spreads} the spreads in properties\footnote{In the case of 2D quantities, we show the quadrature sum of the median absolute deviation of their individual components.} for the Crius groups, compared with the Theia groups of \cite{2019AJ....158..122K} to better demonstrate the impact of projection effects on the ability to recover clusters of stars with coherent $UVW$ velocities when relying on only the projected tangential velocities of their members. The Crius groups show on average tighter $UVW$ distributions compared with Theia groups, even before our kinematic selection criteria are applied. This is probably due at least in part to the fact that \cite{2019AJ....158..122K} did not include Gaia~DR2 heliocentric radial velocities in their analysis, and therefore any star with an available radial velocity that is in reality an interloper that just happens to share a similar 2D tangential velocity with a Theia group will significantly impact our calculation of the Theia group's $UVW$ spread. Despite the Crius groups having small $UVW$ dispersions, the fact that their members are spread on the sky causes them to display larger spreads in tangential velocities, on average. The distribution of the $Z$ median absolute deviations of Crius groups are generally in line with those of Theia groups.

\begin{figure*}
	\centering
	\subfigure[\cite{2017AJ....153..257O} Group 59 corona in $XY$ space]{\includegraphics[width=0.47\textwidth]{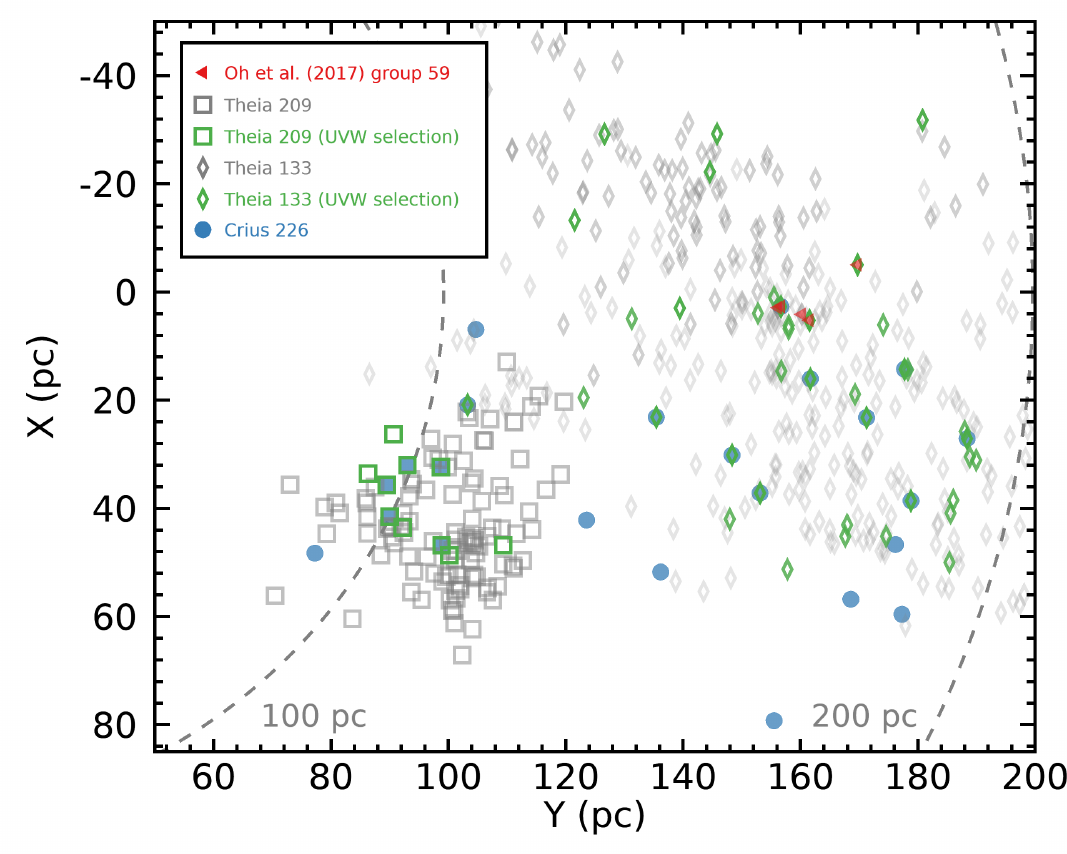}\label{fig:oh59_xy}}
	\subfigure[\cite{2017AJ....153..257O} Group 59 corona in $UV$ space]{\includegraphics[width=0.47\textwidth]{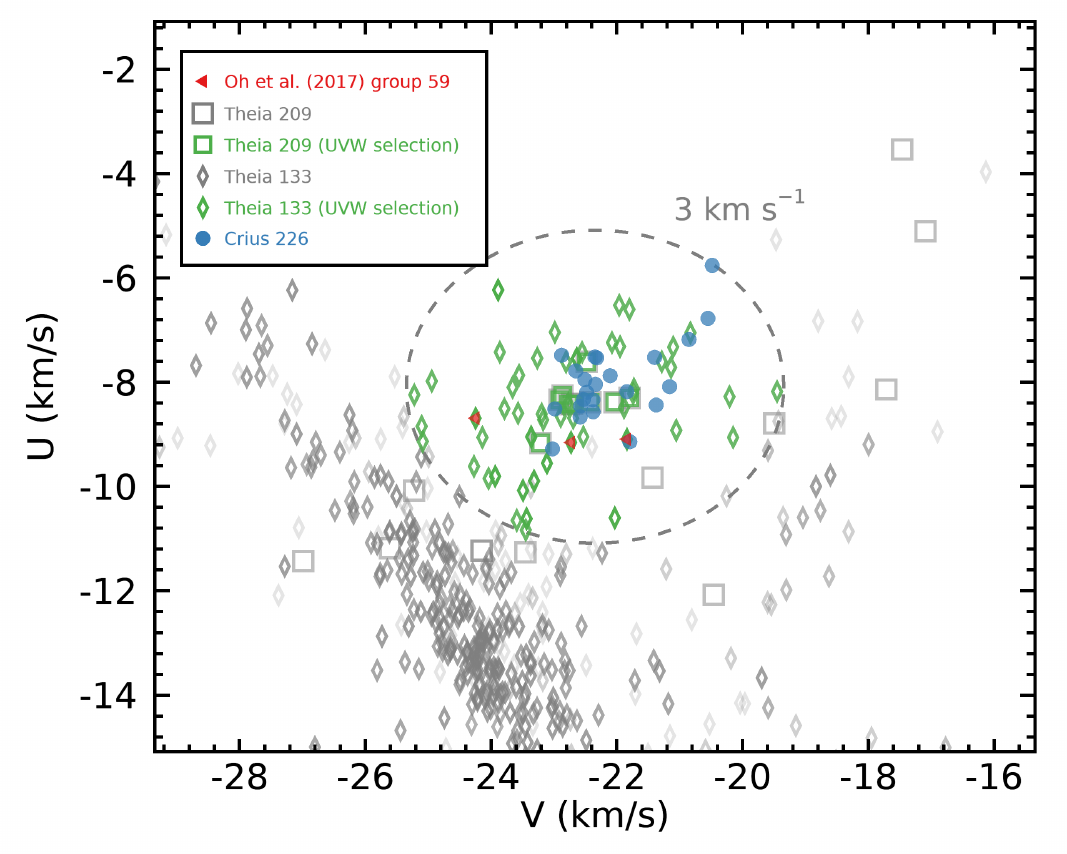}\label{fig:oh59_uv}}
	\caption{Galactic positions and space velocities of Crius~226, which likely corresponds to a corona around Group~59 of \cite{2017AJ....153..257O}. Theia~209 of \cite{2019AJ....158..122K} seems associated with the near-side of Crius~226, and a small fraction of Theia~133 seems associated with its far side, however, Theia~133 also includes a much larger population of stars associated with the $\alpha$~Persei open cluster and its corona, which kinematics are discrepant with Crius~226 and Group~59 of \cite{2017AJ....153..257O}.}
	\label{fig:oh59}
\end{figure*}

\begin{figure*}
	\centering
	\subfigure[$Z$ distance between Group~59 and $\alpha$~Persei]{\includegraphics[width=0.47\textwidth]{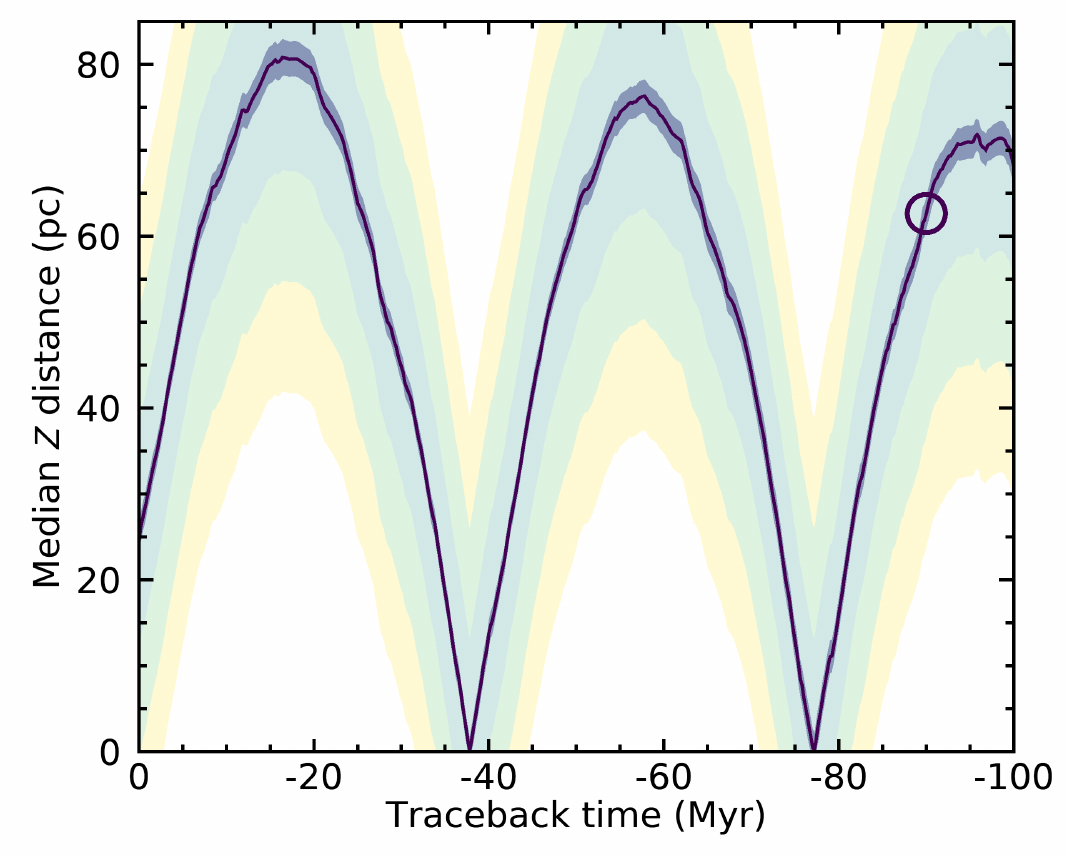}\label{fig:oh59_aper_zdist}}
	\subfigure[$XYZ$ distance between Group~59 and $\alpha$~Persei]{\includegraphics[width=0.47\textwidth]{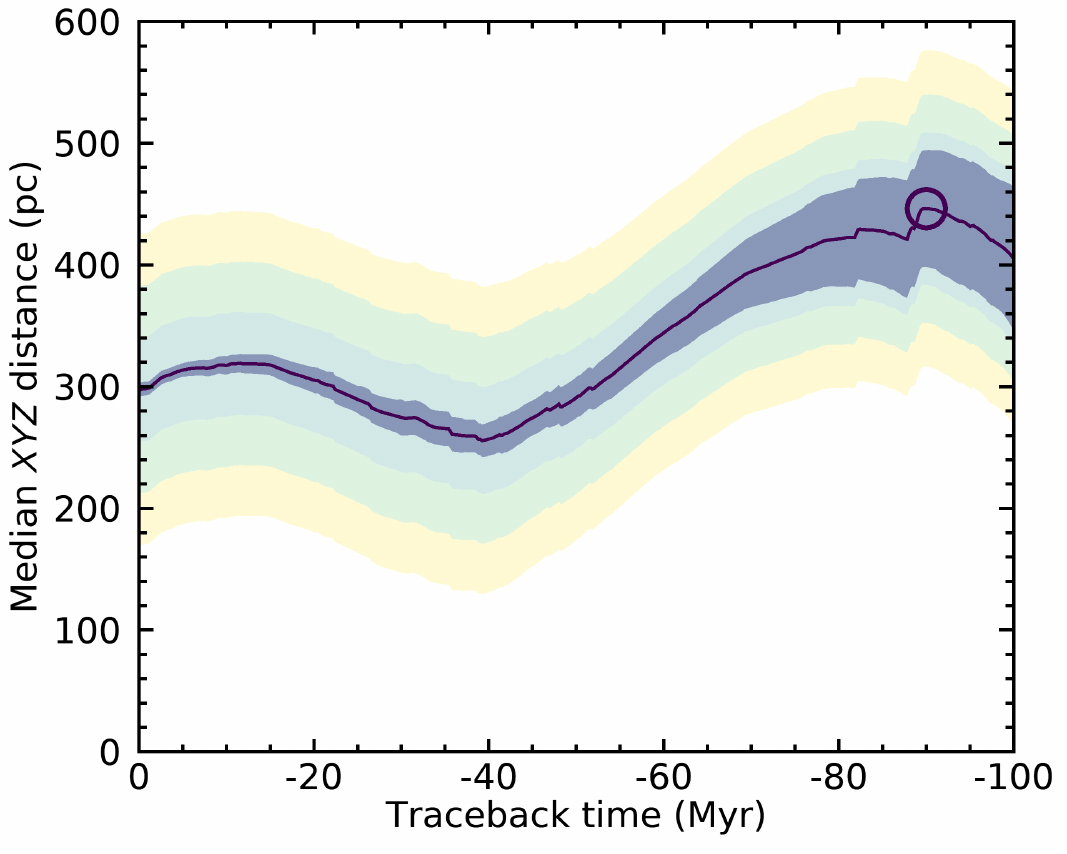}\label{fig:oh59_aper_dist}}
	\caption{Distances between the median position of $\alpha$~Persei and Group~59 \citep{2017AJ....153..257O} members as a function of time in their past trajectories around the Galaxy. The left panel shows the distance along the $Z$ position only, and the right panel shows the 3D distance in $XYZ$ Galactic position. The thick, dark purple line indicate the distances between the median positions, and the associated purple shaded regions around the thick lines indicate the measurement errors on the median distances, which increases over traceback time, especially in the $XY$ directions given that the locations of individual members become uncertain as the initial measurement uncertainties grow over time. The three additional shaded regions indicate areas of overlap at 1, 2 and 3 times the median absolute deviation of the distribution of the $\alpha$~Persei and Group~59 members at the current epoch, which indicates whether the two groups might overlap near the edges of their spatial extent, assuming that the shape of the associations do not change significantly over time. The purple circles indicate the epoch of formation of $\alpha$~Persei. Group~59 of \cite{2017AJ....153..257O} is likely $\approx 160$\,Myr-old based on the age of Theia~209, outside of the traceback range. Although the two associations crossed each other in the $Z$ dimension $\approx 38$ and $\approx 77$\,Myr ago, their 3D distributions do not seem to significantly overlap at any point in the past 100\,Myr.}
	\label{fig:oh59_aper_tracebacks}
\end{figure*}
\subsection{Known Exoplanet Systems}\label{sec:exo}

We cross-matched the members of Crius groups with the Gaia~DR2 source identifiers of exoplanet host stars in the NASA exoplanet archive\footnote{\url{https://exoplanetarchive.ipac.caltech.edu/index.html}} and the NASA list of TESS exoplanet candidates \citep{2019AJ....158..138S} to determine whether additional exoplanet systems had been recovered in our analysis and may potentially benefit from future age constraints based on their host associations. The results of this cross-match are listed in Table~\ref{tab:exo}. We recovered 11 confirmed exoplanet systems in Crius groups that correspond to known associations or open clusters, and which memberships had previously been discussed in the literature, and 5 additional TESS candidate exoplanet systems in Crius groups that we have matched to known young associations, but for which memberships had not been discussed in the literature to our knowledge. 

In addition to these, we have uncovered the two TESS exoplanets host stars TOI--1807 ($204 \pm 50$\,Myr) and TOI--2076 ($180 \pm 40$\,Myr) of \cite{2021AJ....162...54H} in Crius~224. These two exoplanet host stars were firstly noted by \cite{2021AJ....162...54H} as co-moving as well as being coeval despite their unusually large 12.5\textdegree\ separation on the sky. \cite{2017AJ....153..257O} had also recovered these two stars as a likely binary pair based on their similar kinematics using Gaia~DR1 data (in their Group~3914, with only 2 members), however the large separation led \cite{2021AJ....162...54H} to hypothesize that they might be members of an $\approx 200$\,Myr-old stream of co-moving stars which other members might have escaped detection so far. This system also escaped further investigations of likely new coeval associations contained in the \cite{2017AJ....153..257O} groups (e.g., see \citealp{2018ApJ...863...91F}) because this particular group included only two stars. \cite{2022arXiv220603496N} also investigated 24 possible stars co-moving with these two exoplanet hosts, and used lithium equivalent width measurements and TESS rotation periods to estimate an age of $300 \pm 80$\,Myr for the group. We have recovered both stars as members of Crius~224 (28 members), and found that its members are consistent with an age of 100--700\,Myr based on their color-magnitude position. This is yet another scenario similar to MELANGE--1 that is suggestive of the validity of the newly discovered groups presented here, which have likely escaped detection prior to Gaia because of their low spatial density and slightly older age which prevented a discovery based on their X-ray or NUV-bright members or more massive, early B-type stars.

Moreover, we have uncovered one additional confirmed radial velocity exoplanet (HD~103949~b; \citealp{2019ApJS..242...25F}) in the newly discovered 100--700\,Myr candidate association Crius~162 (14 members) and three additional TESS exoplanet candidates (TOI--1598~b, TOI--2481~b and TOI--2133~b, see \citealp{2019AJ....158..138S}) in three additional new candidate associations (Crius~109, $\geq 100$\,Myr with 16 members; Crius~121, $\geq 100$\,Myr with 13 members; and Crius~205, 100--700\,Myr with 12 members, respectively). To our knowledge, no age constraints are currently available for HD~103949~b, TOI--1598~b, TOI--2481~b and TOI--2133~b. It will be particularly interesting to fully characterize Crius~162, 109, 121, and 205, given that exoplanet systems with well-calibrated ages are rare and will provide valuable information about their early evolution.

\subsection{Tracebacks of Oh~59 and $\alpha$~Persei}

Among the likely extensions of known associations discussed in Section~\ref{sec:extensions}, Crius~226 (23 total members) seems related to the $\alpha$~Persei open cluster, Group~59 of \cite{2017AJ....153..257O}, Theia~133 and Theia~209. Figure~\ref{fig:oh59} shows that the kinematics of Crius~226 are consistent with a number of stars in Theia~209 and a small subset of Theia~133 members, whereas they tentatively form a bridge between the two spatially distinct structures in the $XY$ plane. Furthermore, a large fraction of Theia~133 members have kinematics significantly discrepant with Crius~226, and are instead clearly related to the $\alpha$~Persei open cluster and its corona, as noted by \cite{2019AJ....158..122K}. It seems likely that Theia~133 is therefore composed of two populations with distinct kinematics, only one of which may be directly related to Crius~226 and Theia~209, which together may span distances in the range $\approx 100-200$\,pc, partially overlapping with the corona of $\alpha$~Persei. \cite{2019AJ....158..122K} estimated an age of $\approx$\,160\,Myr for Theia~209 (relatively well constrained by the two B8 stars 27~Vul and bet02~Cyg, as well as 7 $\geq$\,M5 members), slightly older than the lithium depletion boundary age of $\alpha$~Persei ($90 \pm 10$; \citealp{1999ApJ...527..219S}). The color-magnitude diagram of Crius~226 is indicative of an age of $\approx 100$\,Myr or older, consistent with the age of Theia~209.

Because of their similar kinematics and spatial overlap, we investigated whether $\alpha$~Persei and Group~59 of \cite{2016AA...595A...1G} might have collided in the recent past, using the dynamical traceback approach of D. Couture et al. (in preparation)\footnote{See \cite{2020AA...642A.179M} for a similar approach.}. Independent backward Galactic orbits were integrated for every member of $\alpha$~Persei and Group~59 with the galpy Python package \citep{Bovy_2015} using the Galactic potential model I from \cite{2013A&A...549A.137I} for 100\,Myr into the past.

The resulting mutual median distances over time between $\alpha$~Persei and Group~59 are shown in Figure~\ref{fig:oh59_aper_tracebacks}. Although the Galactic positions of individual stars become gradually more uncertain as they are projected further in the past, the median positions of the clusters are less affected because of their large number of members. This fact is also much less important in the $Z$ direction, because stars in our sample are gravitationally bound to the Galactic plane. We found that the two groups came at their closest approach $\approx 39$\,Myr ago, while they crossed the same plane in the $Z$ direction twice in the past 100\,Myr, $\approx 38$ and $\approx 77$\,Myr ago. However, their respective median positions have never been closer than $\approx 260$\,pc, which excludes an overlap of their spatial distributions even at three times the median absolute deviation of their current membership distributions, even when including the backward projected measurement errors at this epoch (see Figure~\ref{fig:oh59_aper_tracebacks}). We find that it is therefore unlikely that these two associations are related to each other, unless significant systematics remain in the Galactic potential of \citeauthor{2013AA...549A.137I} (\citeyear{2013AA...549A.137I}; see also \citealp{2015ApJS..216...29B}), or other collision events that would significantly affect the relative positions of Oh~59 and $\alpha$~Persei in the recent past.

\section{CONCLUSIONS}\label{sec:conclusion}

We present a clustering analysis of the Solar neighborhood using Gaia~EDR3 entries in order to reconstruct their full 6D $XYZ$ Galactic positions and $UVW$ space velocities. This allowed us to recover most of the known nearby associations as well as a number of new moving groups and co-moving structures within 200\,pc of the Sun, which was previously hindered by projection effects at such nearby distances. A detailed study of these clustered stars allowed us to identify new groups of stars, including: 

\begin{itemize}
    \item \newgroups\ candidate new coeval associations whose color-magnitude diagrams are consistent with ages older than $\approx 100$\,Myr;
    \item \newcoronae\ newly found coronae of nearby open clusters, including a corona associated with the Volans-Carina associations that reaches within \nearestvca\,pc from the Sun;
    \item \newtheiaextensions\ spatial extensions of Theia goups \citep{2019AJ....158..122K} that do not seem to form coronae around spatially denser open cluster cores;
    \item A potential spatial extension (Crius~170) of the MELANGE--1 association of \cite{2021AJ....161..171T};
    \item The discovery of a moving group (Crius~224) which existence was hypothesized by \cite{2021AJ....162...54H} following the discovery of two exoplanet host stars (TOI--1807 and TOI--2076) with similar velocities but large angular separations;
    \item Additional evidence for the \cite{2017ApJ...838..150K} hypothesis that the 118~Tau and 32~Ori associations may be part of contiguous star-forming events related with the greater Taurus-Auriga star-forming region;
    \item Newly recognized host associations for 3 previously known confirmed exoplanets and 8 TESS candidate exoplanets, makeing it possible to improve the age estimates of these systems.
\end{itemize}

This study hints that the vastly improved radial velocity coverage of Gaia~DR3 will likely allow us to significantly further our understanding of kinematic structures in the Solar neighborhood, and these new co-moving populations will be especially useful to study young, age-calibrated exoplanets and substellar objects.

\facility{Exoplanet Archive}

\acknowledgments

We thank the anonymous reviewer for their extremely useful feedback and thorough comments. This research has made use of the NASA Exoplanet Archive, which is operated by the California Institute of Technology, under contract with the National Aeronautics and Space Administration under the Exoplanet Exploration Program. This research made use of the SIMBAD database and VizieR catalog access tool, operated at the Centre de Donn\'ees astronomiques de Strasbourg, France \citep{2000AAS..143...23O}. This work presents results from the European Space Agency (ESA) space mission Gaia. Gaia data are being processed by the Gaia Data Processing and Analysis Consortium (DPAC). Funding for the DPAC is provided by national institutions, in particular the institutions participating in the Gaia MultiLateral Agreement (MLA). The Gaia mission website is https://www.cosmos.esa.int/gaia. The Gaia archive website is https://archives.esac.esa.int/gaia. This research used the HDBSCAN Clustering Library. The documentation regarding the algorithm can be found at the following website : https://hdbscan.readthedocs.io/en/latest/index.html . J.~F. acknowledges support from NASA TESS award 80NSSC21K0792 as well as support from the Heising Simons Foundation. J.~G. acknowledges the support of the Natural Sciences and Engineering Research Council of Canada (NSERC), funding reference number RGPIN-2021-03121.

\bibliographystyle{apj}
\bibliography{moranta2022_jo,moranta2022_leslie,other_references,Zenodo_Library}
\appendix

\input{table_allstars.tex}
\input{table_allgroups_pass.tex}
\input{table_allgroups_fail.tex}
\input{table_known.tex}
\input{table_new_coronae.tex}
\input{table_new.tex}
\input{table_extensions.tex}
\input{table_exoplanets.tex}

\begin{figure*}[p]
	\centering
	\subfigure[Crius~155 and Theia~1005 in $XY$ space]{\includegraphics[width=0.49\textwidth]{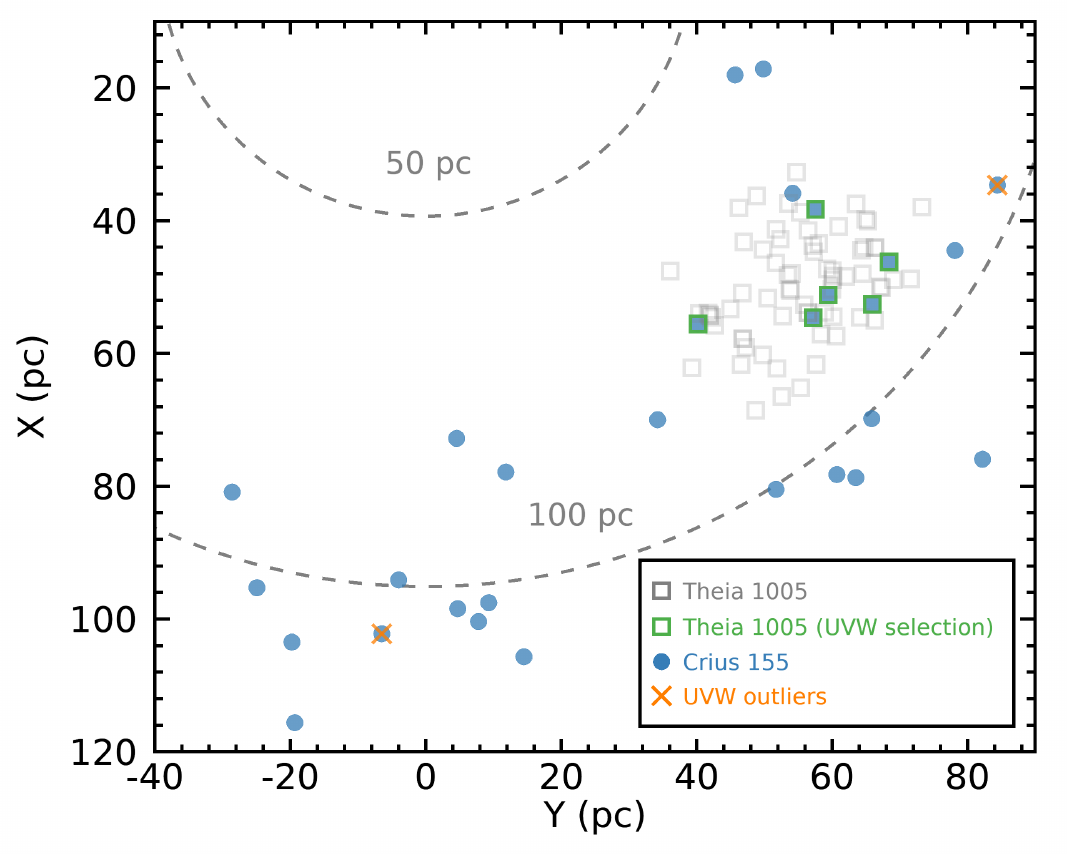}\label{fig:crius155_xy}}
	\subfigure[Crius~155 and Theia~1005 in $UV$ space]{\includegraphics[width=0.49\textwidth]{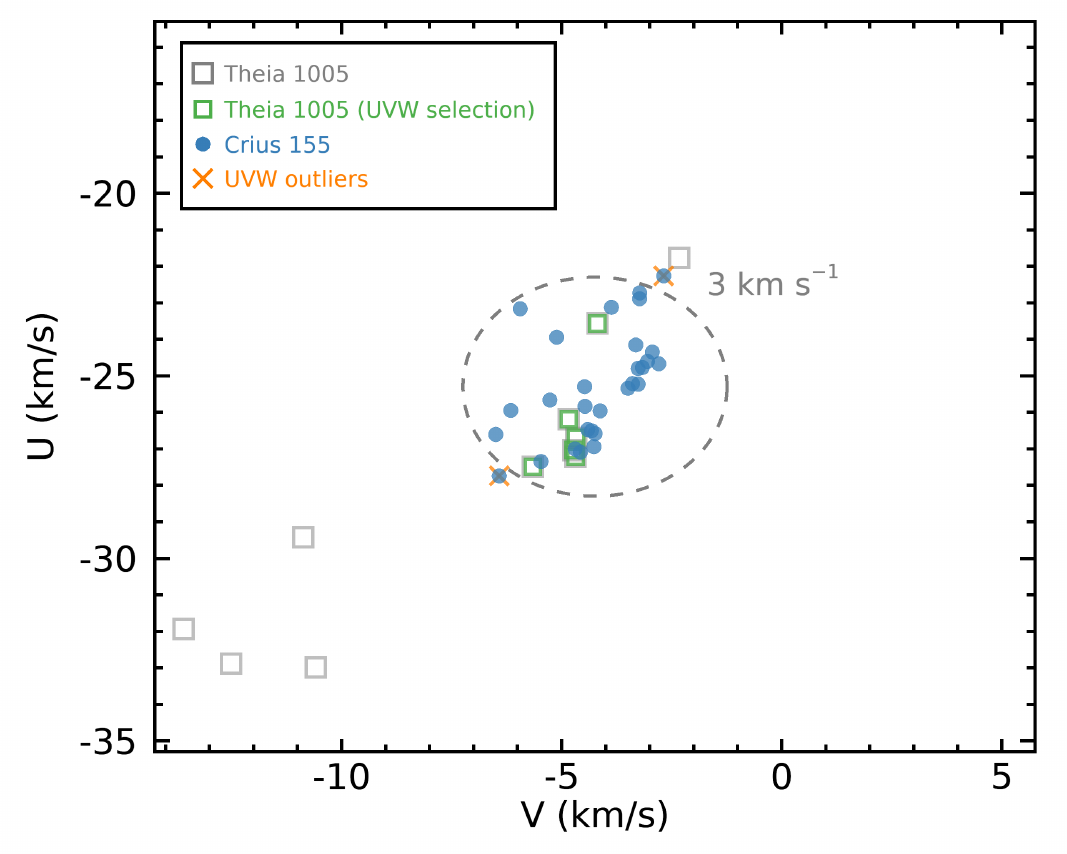}\label{fig:crius155_uv}}
	\subfigure[Crius~151, Crius~207 and Theia~677 in $XY$ space]{\includegraphics[width=0.49\textwidth]{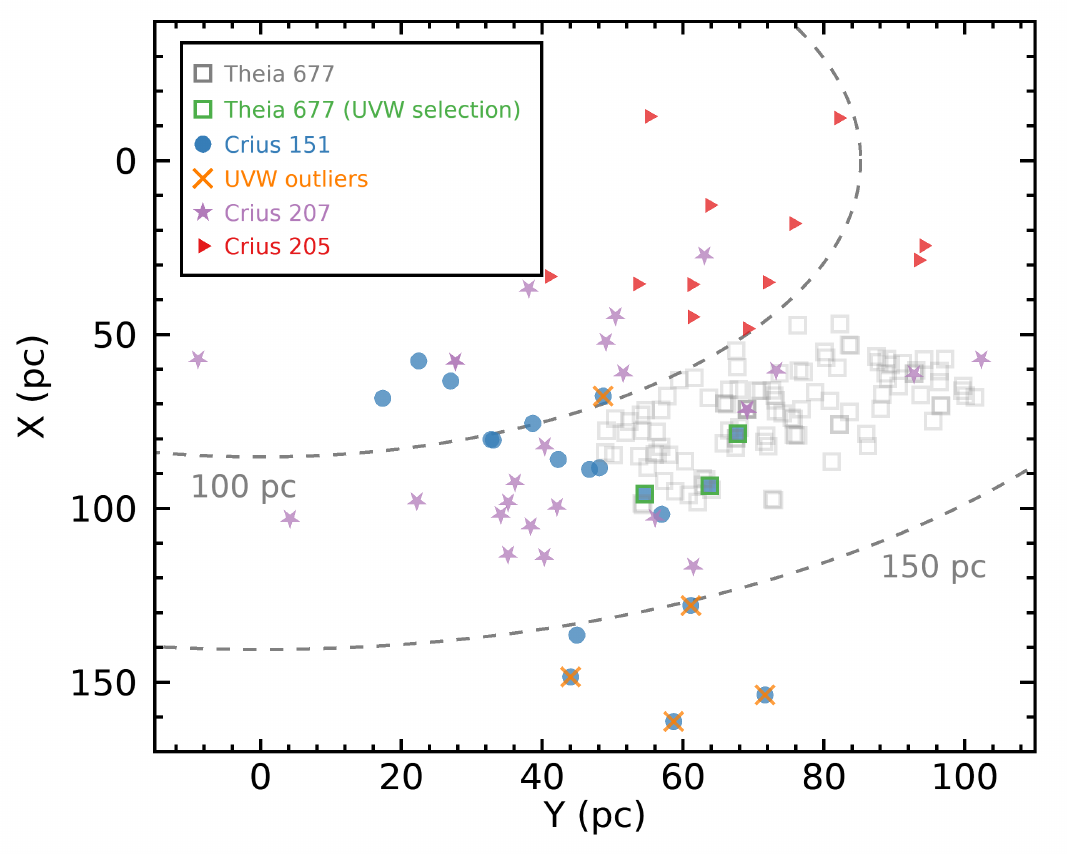}\label{fig:crius151_xy}}
	\subfigure[Crius~151, Crius~207 and Theia~677 in $UV$ space]{\includegraphics[width=0.49\textwidth]{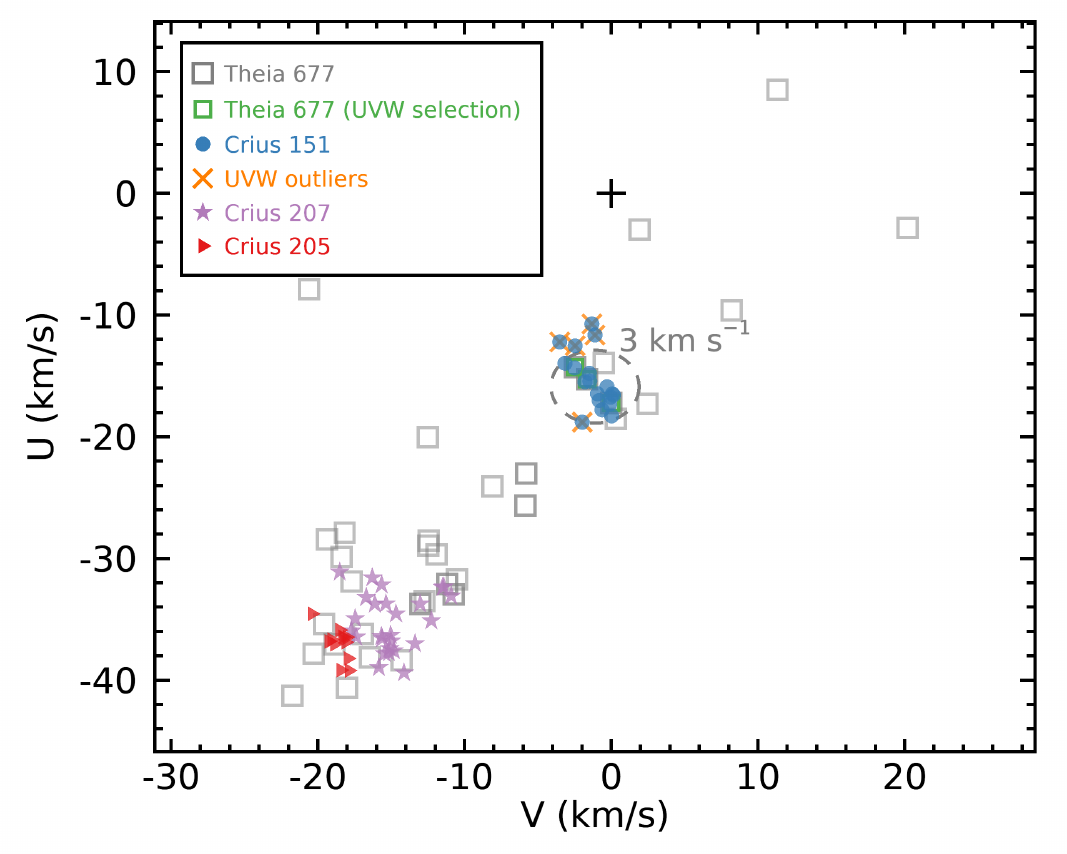}\label{fig:crius151_uv}}
	\caption{Galactic positions and space velocities of Crius groups which we have identified as putative extensions of generally more distant, known associations. Groups which are significantly extended on the sky will show large variations in observed tangential motions, causing clustering algorithms to recover them less efficiently unless they work in 3D $UVW$ space directly. Theia~1005 may be much more spatially extended than previously recognized in the $XY$ plane (top row panels). Both Crius~151, Crius~205 and Crius~207 (bottom row panels) were all associated with Theia~677 of \cite{2019AJ....158..122K}, indicating that Theia~677 may be composed of distinct populations, the largest fraction of which likely corresponds to Crius~205 and Crius~207. These latter two appear as a potentially coeval population with a slightly bimodal density in $XYZUVW$ space, whereas Crius~151 shows significantly different kinematics and may therefore correspond to a distinct population that contaminated Theia~677 in part.}
	\label{fig:extensions1}
\end{figure*}

\begin{figure*}[p]
	\centering
	\subfigure[Crius~215 and Theia~600 in $XY$ space]{\includegraphics[width=0.49\textwidth]{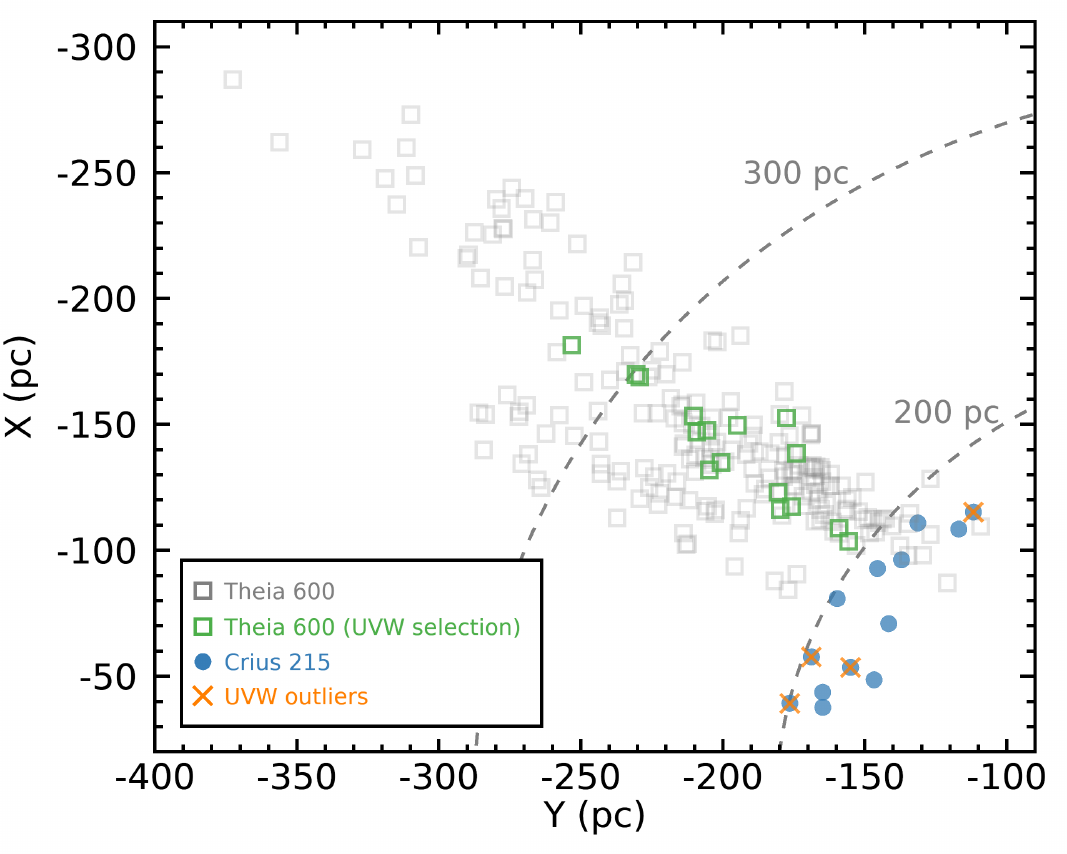}\label{fig:crius215_xy}}
	\subfigure[Crius~215 and Theia~600 in $UV$ space]{\includegraphics[width=0.49\textwidth]{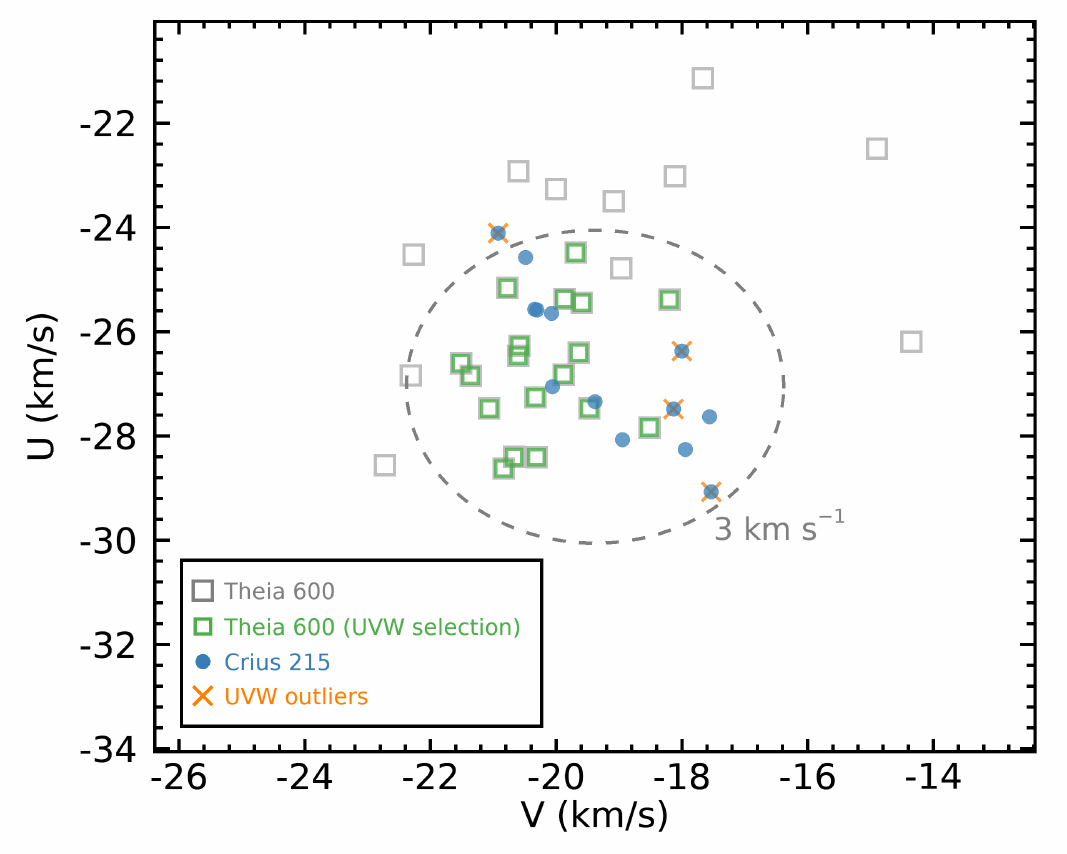}\label{fig:crius215_uv}}
	\caption{Galactic positions and space velocities of Crius groups which we have identified as putative extensions of generally more distant, known associations.  Crius~215 is a possible slight extension of Theia~600, perhaps corresponding to the forefront of the bi-modal structure seen at larger distances.}
	\label{fig:extensions2}
\end{figure*}

\begin{figure*}[p]
	\centering
	\subfigure[Crius~168 and Theia~310 in $XY$ space]{\includegraphics[width=0.49\textwidth]{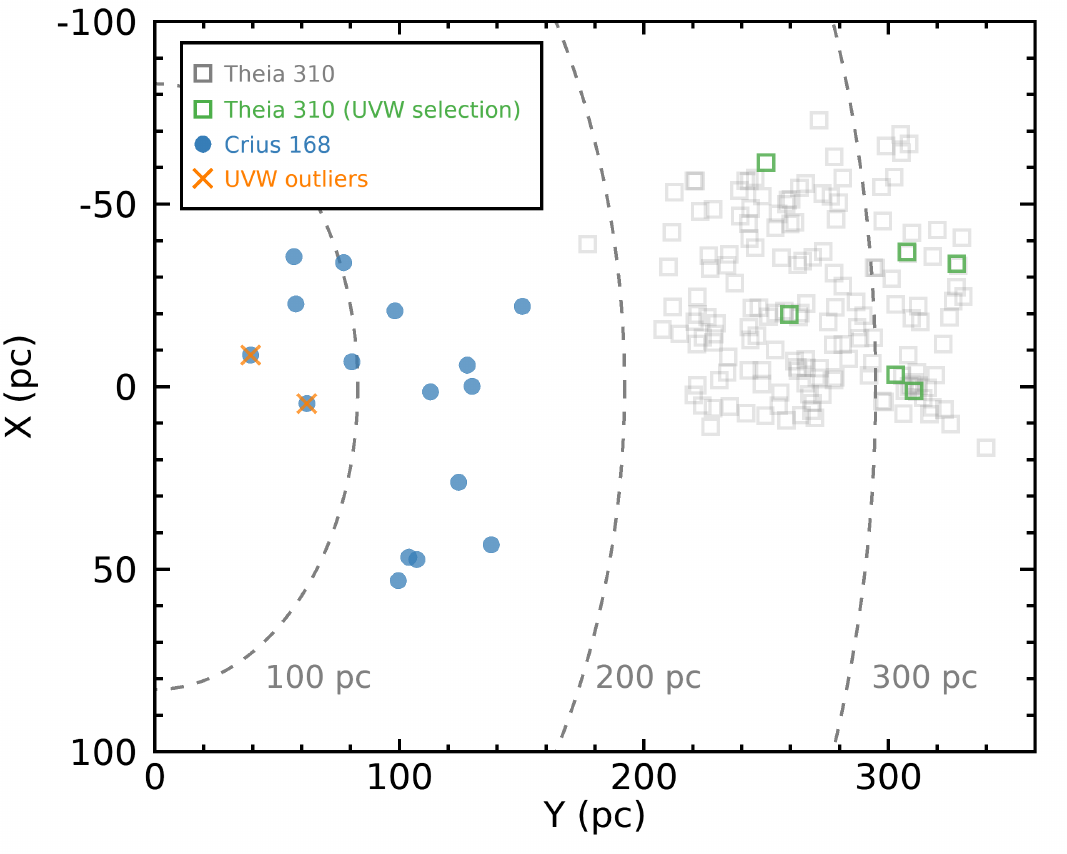}\label{fig:crius168_xy}}
	\subfigure[Crius~168 and Theia~310 in $UV$ space]{\includegraphics[width=0.49\textwidth]{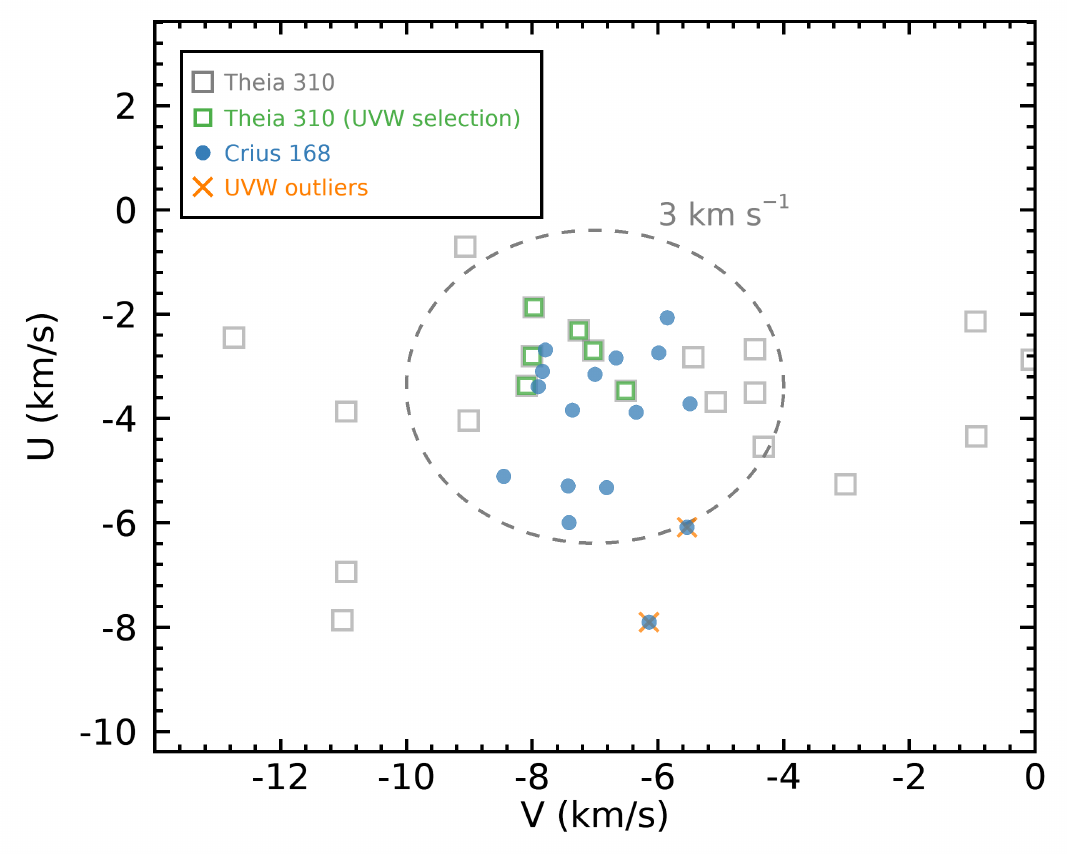}\label{fig:crius168_uv}}
	\subfigure[Crius~227, Theia~209 in $XY$ space]{\includegraphics[width=0.49\textwidth]{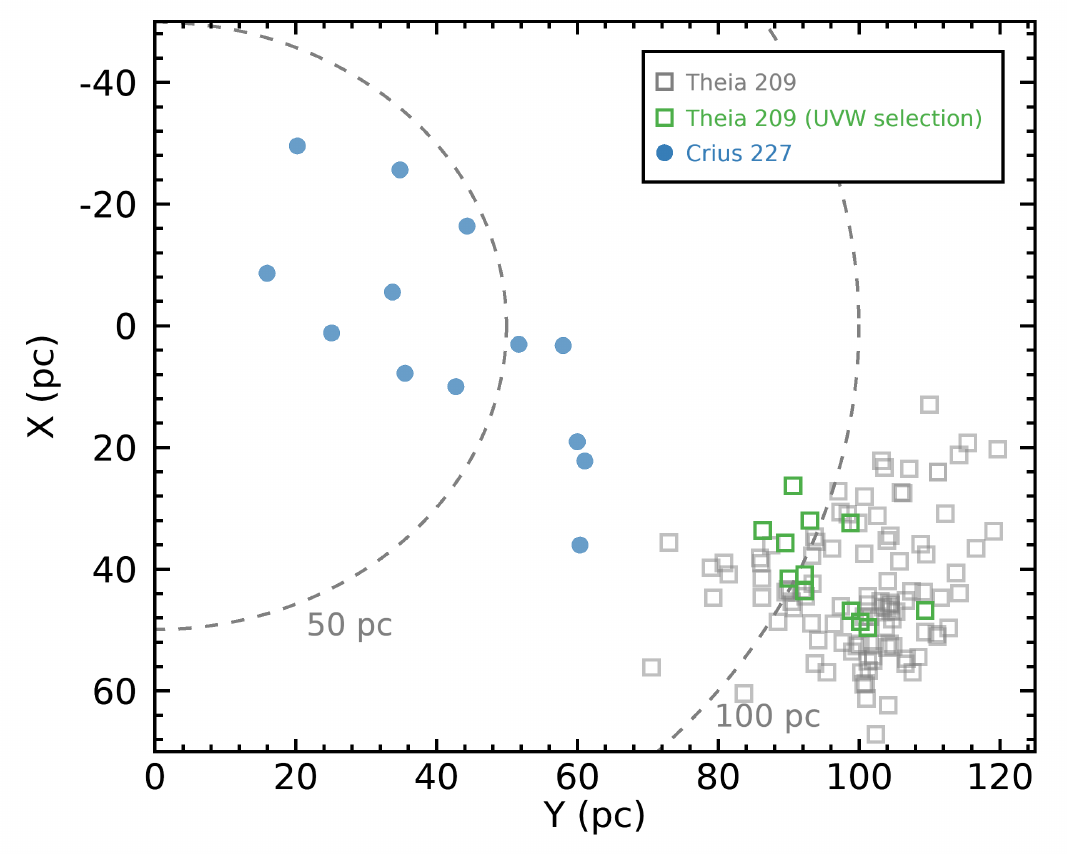}\label{fig:crius227_xy}}
	\subfigure[Crius~227, Theia~209 and Theia~677 in $UV$ space]{\includegraphics[width=0.49\textwidth]{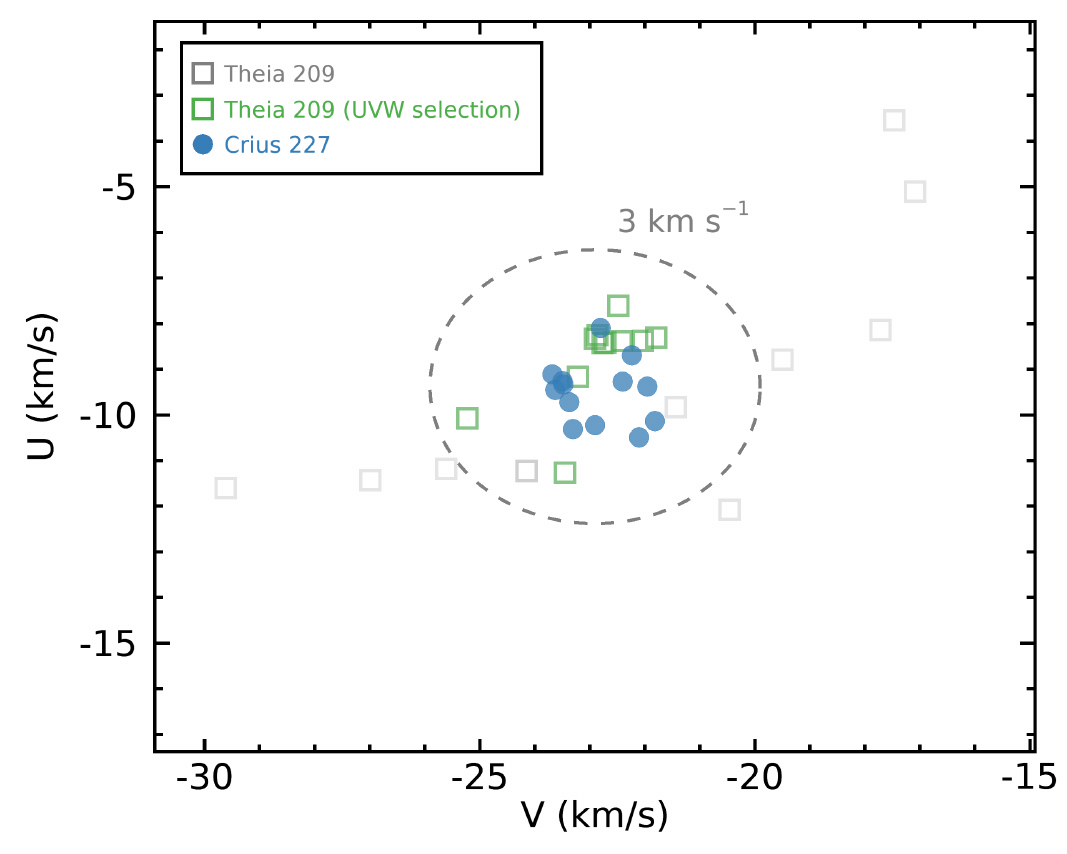}\label{fig:crius227_uv}}
	\caption{Galactic positions and space velocities of Crius groups which we have identified as putative extensions of generally more distant, known associations. Theia~310 of \cite[top row panels]{2019AJ....158..122K} may be spatially more extended than previously recognized along our line of sight, reaching within 100\,pc of the Sun. The gap at 200\,pc may be due to a low recovery rate by \cite{2019AJ....158..122K} caused by projection effects (the larger parallax correlates with larger tangential motions), and a low recovery rate in our study due to the lack of radial velocity measurements in Gaia~DR2 at these farther distances. A similar structure is observed for Crius~227 and Theia~209, although the gap in members is located at more nearby distances ($\approx 80$\,pc), and the putative spatial extension reaches well within the solar neighborhood down to 19\,pc from the Sun, making this ensemble particularly interesting for the study of exoplanets and substellar objects.}
	\label{fig:extensions3}
\end{figure*}

Figures~\ref{fig:extensions1} to \ref{fig:extensions3} show the candidate spatial extensions of distant groups recovered in this analysis, as discussed in Section~\ref{sec:extensions}. $XY$ Galactic positions are shown to illustrate how the Crius groups are displayed such as to potentially form extensions around the known associations; we do not show the $Z$ direction because known groups and newly discovered candidate extensions are in the same $Z$ plane, as would be expected for coeval populations. The $UV$ space velocities are also shown and compared with a 3\,\kms\ circular region around that of the known young association or the new Crius group if the known association displays a large kinematic scatter.

\begin{figure*}
	\centering
	\subfigure[UPK~612 corona in $XY$ space]{\includegraphics[width=0.49\textwidth]{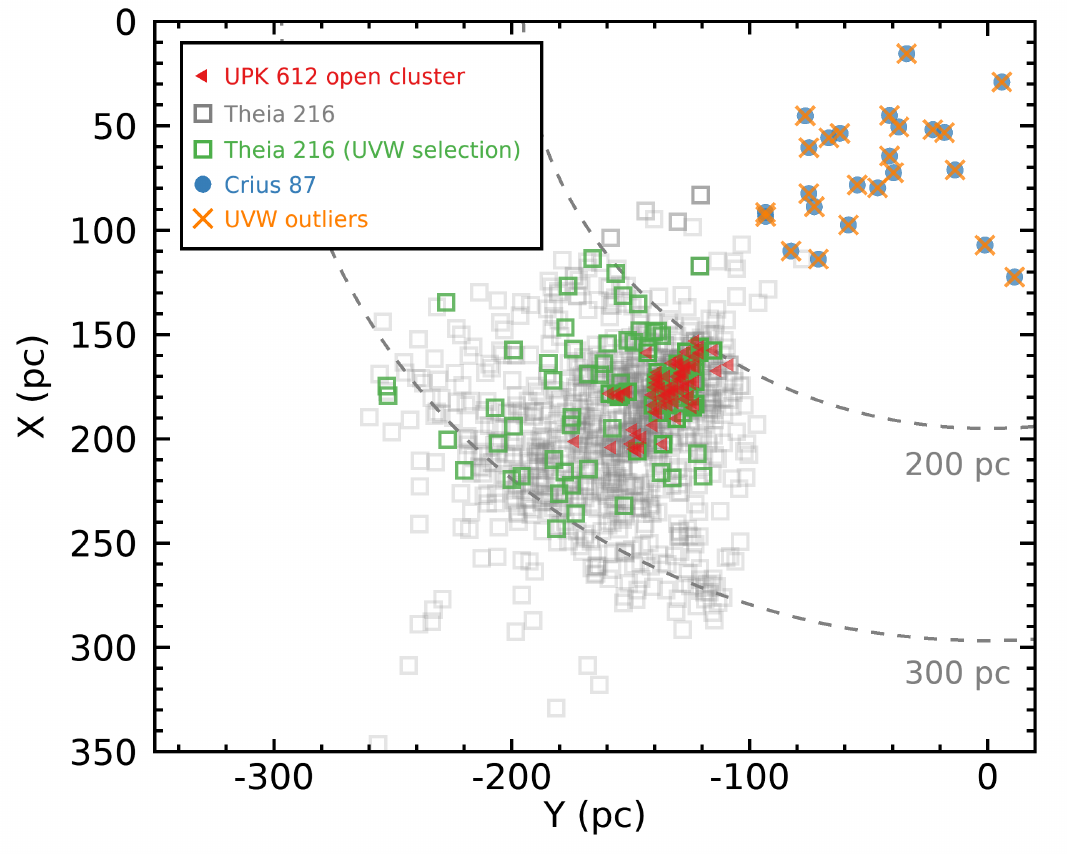}\label{fig:upk612_xy}}
	\subfigure[UPK~612 corona in $UV$ space]{\includegraphics[width=0.49\textwidth]{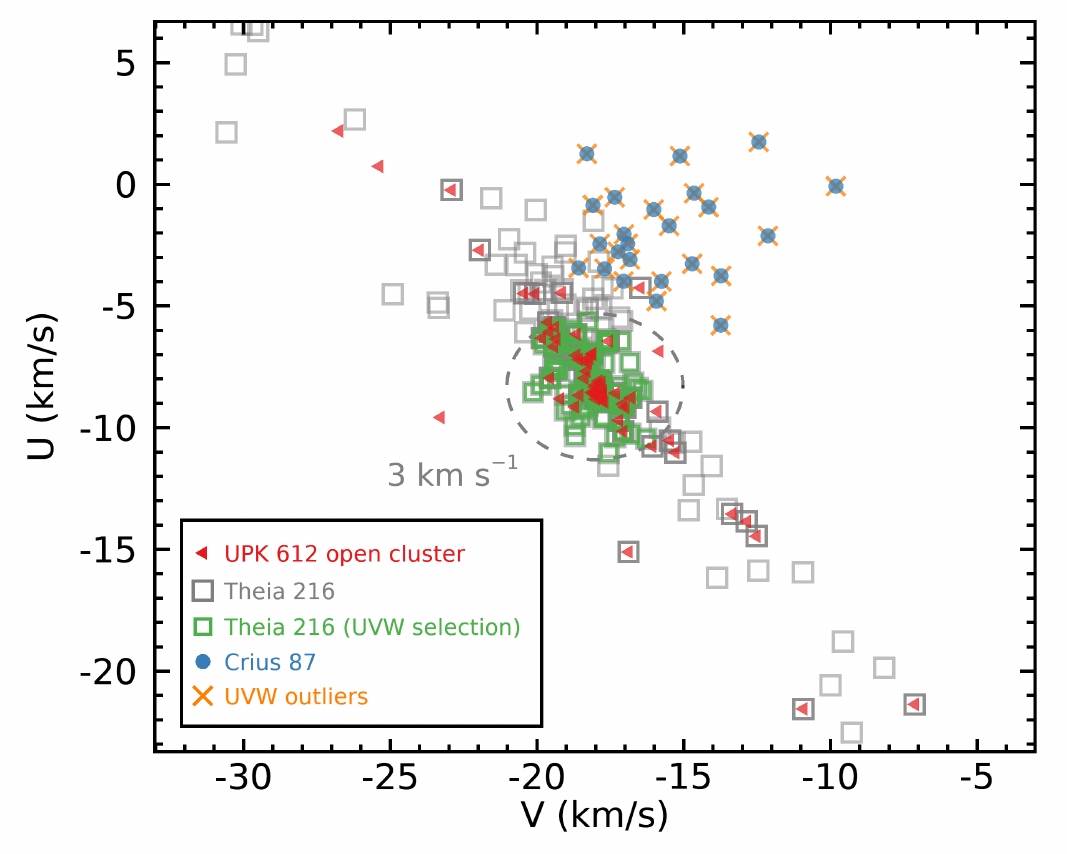}\label{fig:upk612_uv}}
	\caption{In our investigations of known associations possibly related with Crius~87, we have identified a previously unrecognized, likely corona of the UPK~612 open cluster \citep{2020AA...640A...1C} contained in Theia~216 of \cite{2019AJ....158..122K}. Crius~87 appeared as a candidate extension of this corona towards the Sun at a first glance, given its consistent $Z$ coordinate, location in the $XY$ plane, and similar kinematics. However, its median kinematics are discrepant by 8.5\,\kms, enough to call into question whether Crius~87 is related to UPK~612 at all.}
	\label{fig:upk612}
\end{figure*}

\begin{figure*}
	\centering
	\subfigure[\cite{2017AJ....153..257O} Group 51 corona in $XY$ space]{\includegraphics[width=0.49\textwidth]{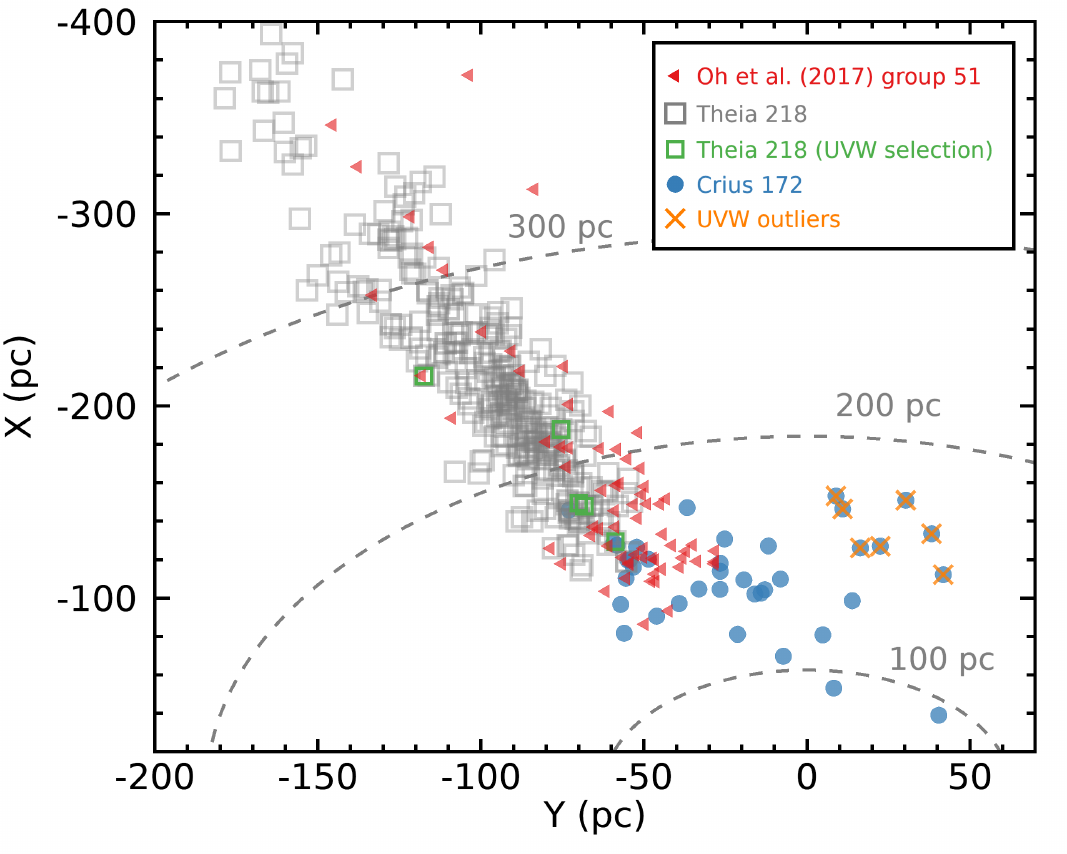}\label{fig:oh51_xy}}
	\subfigure[\cite{2017AJ....153..257O} Group 51 corona in $UV$ space]{\includegraphics[width=0.49\textwidth]{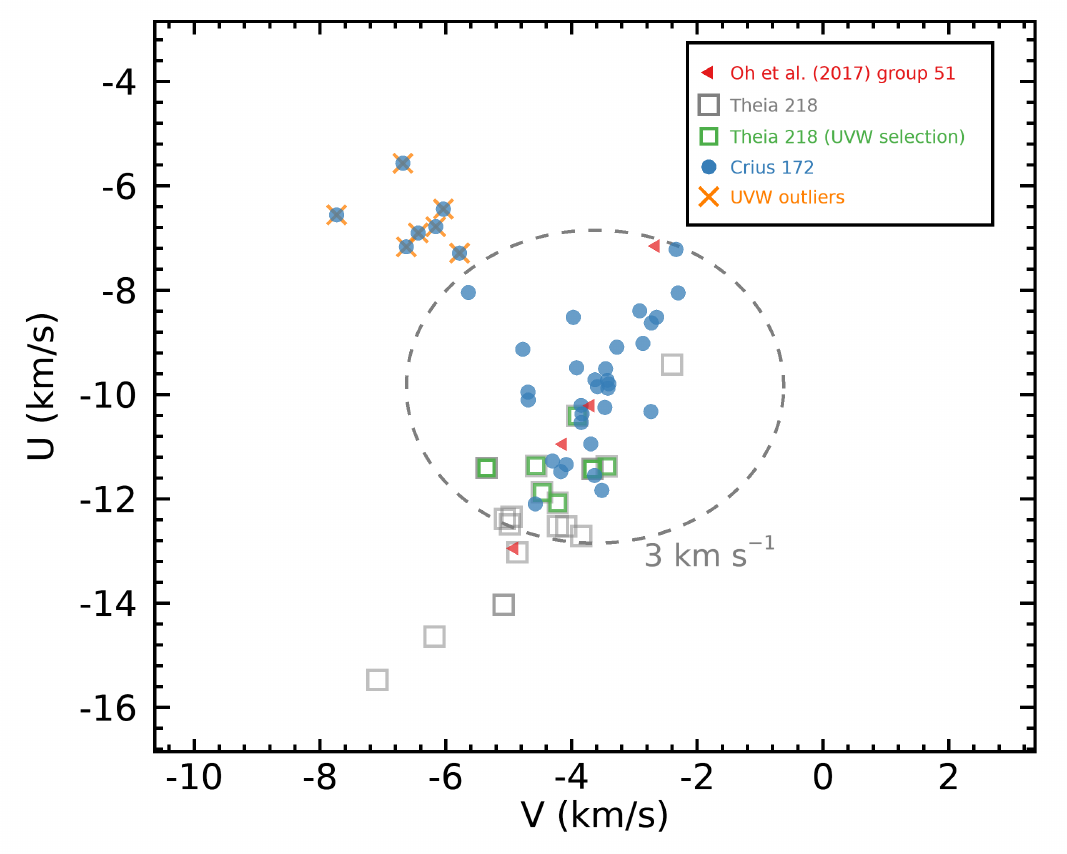}\label{fig:oh51_uv}}
	\caption{Galactic positions and space velocities of Crius~172, which we associate with a likely front-facing tidal tail to the poorly characterized Group~51 of \cite{2017AJ....153..257O}. Theia~218 may correspond to a further tidal tail behind Group~51 given its consistent space velocities and spatial location.}
	\label{fig:oh51}
\end{figure*}

\begin{figure*}
	\centering
	\subfigure[Platais~3 corona in $XY$ space]{\includegraphics[width=0.49\textwidth]{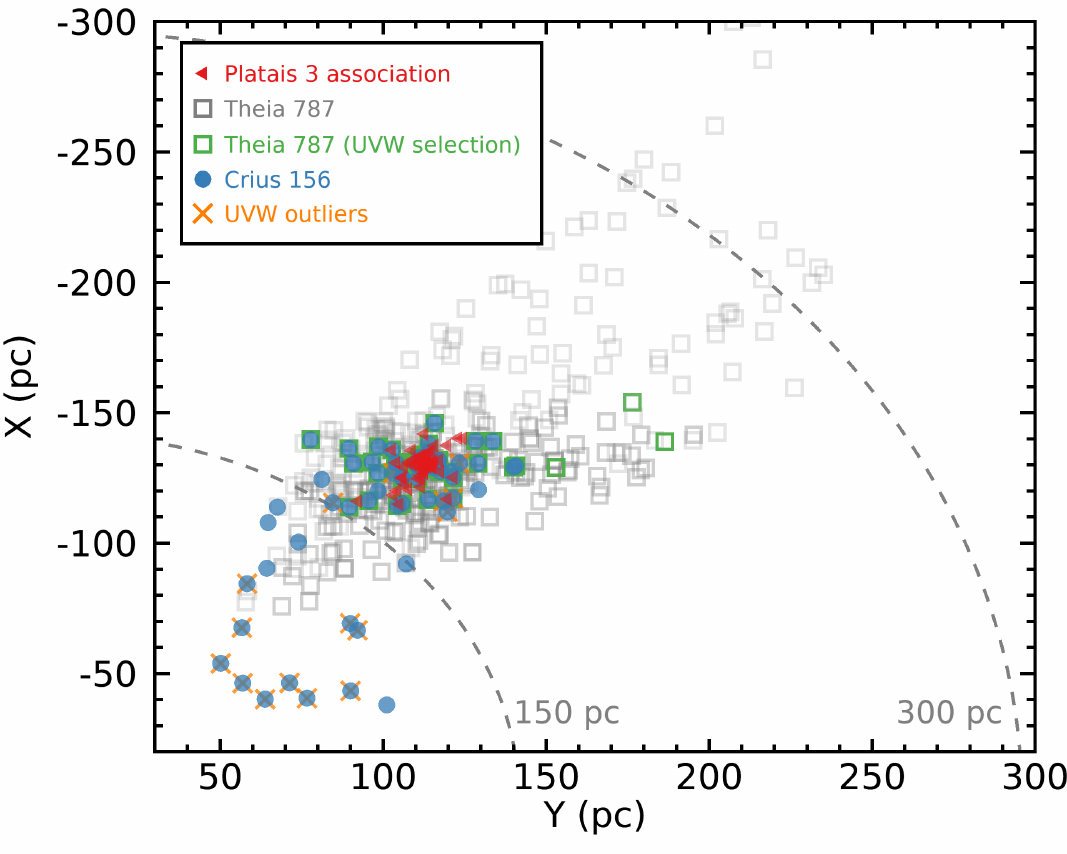}\label{fig:pl3_xy}}
	\subfigure[Platais~3 corona in $UV$ space]{\includegraphics[width=0.49\textwidth]{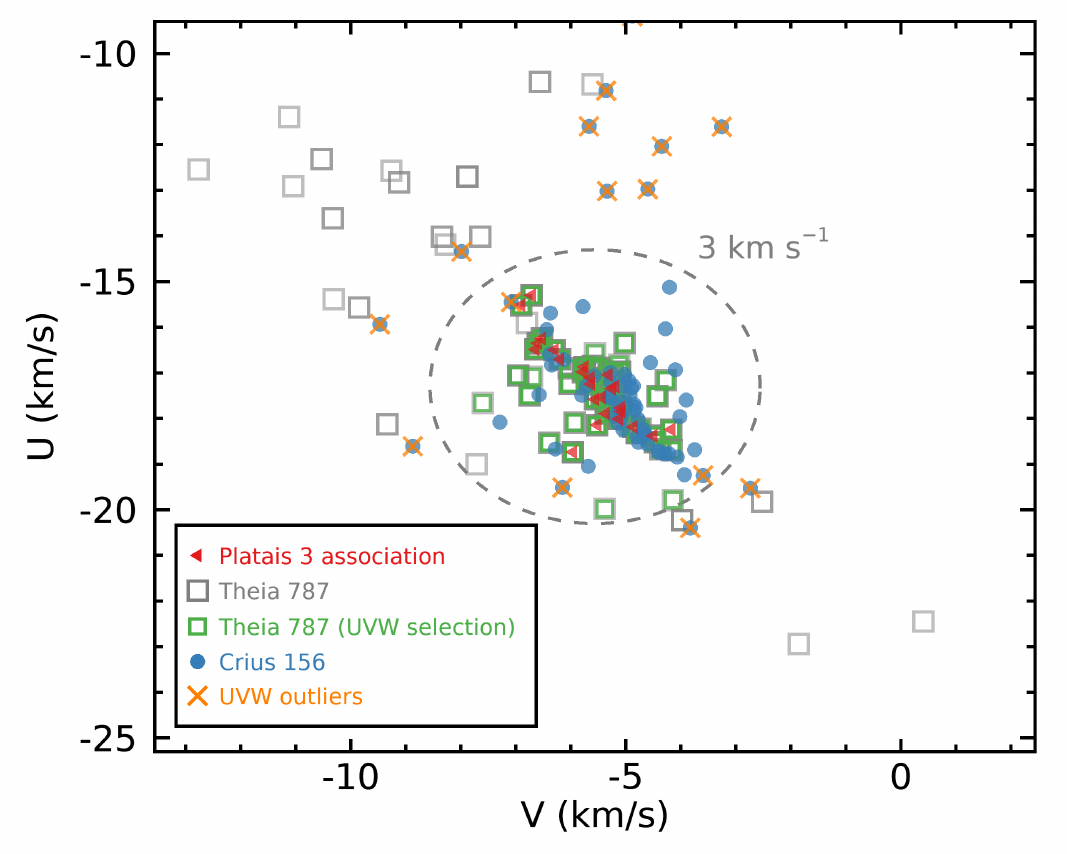}\label{fig:pl3_uv}}
	\subfigure[Alessi~9 corona in $XY$ space]{\includegraphics[width=0.49\textwidth]{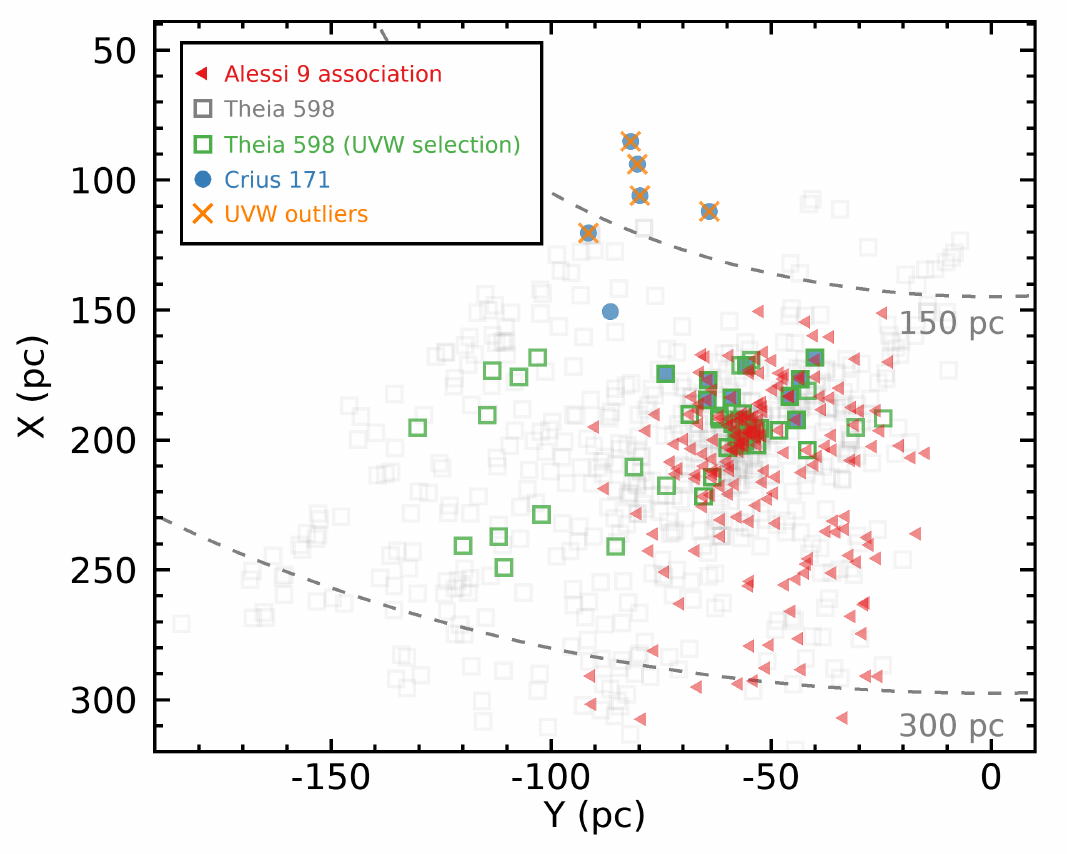}\label{fig:ales9_xy}}
	\subfigure[Alessi~9 corona in $UV$ space]{\includegraphics[width=0.49\textwidth]{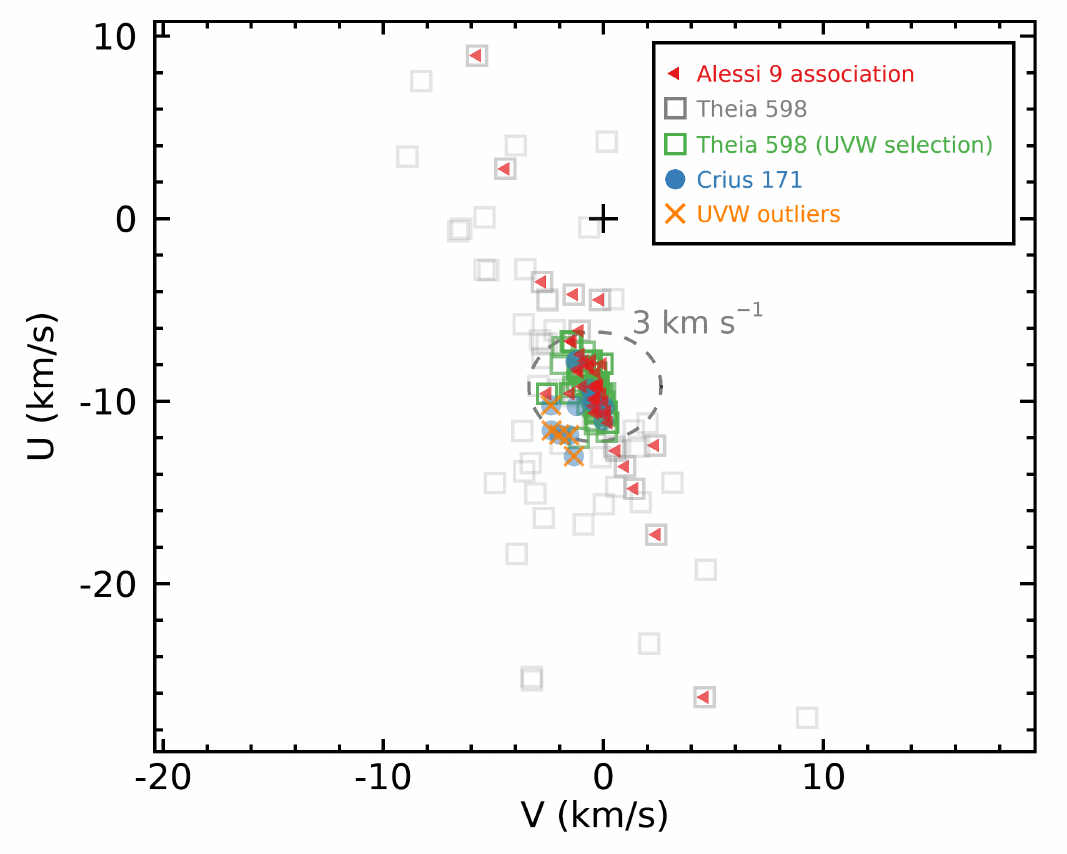}\label{fig:ales9_uv}}
	\caption{Galactic positions and space velocities of Crius~156 and Crius~171, which corresponds to the Platais~3 and Alessi~9 associations, respectively. These associations have been respectively associated with Theia~787 and Theia~598 by \cite{2019AJ....158..122K}. The tentative spatial extensions of both associations towards the Sun show systematically discrepant $UVW$ velocities which may indicate that they are simply caused by contamination. Both Crius~156/Theia~787 and Crius~171/Theia~598 are spatially much more extended than the core of their respective associations, and likely correspond to their extended coronae, which had not been previously discussed in the literature to the extent of our knowledge.}
	\label{fig:pl3_ales9}
\end{figure*}

\begin{figure*}
	\centering
	\subfigure[RSG~2 corona in $XY$ space]{\includegraphics[width=0.49\textwidth]{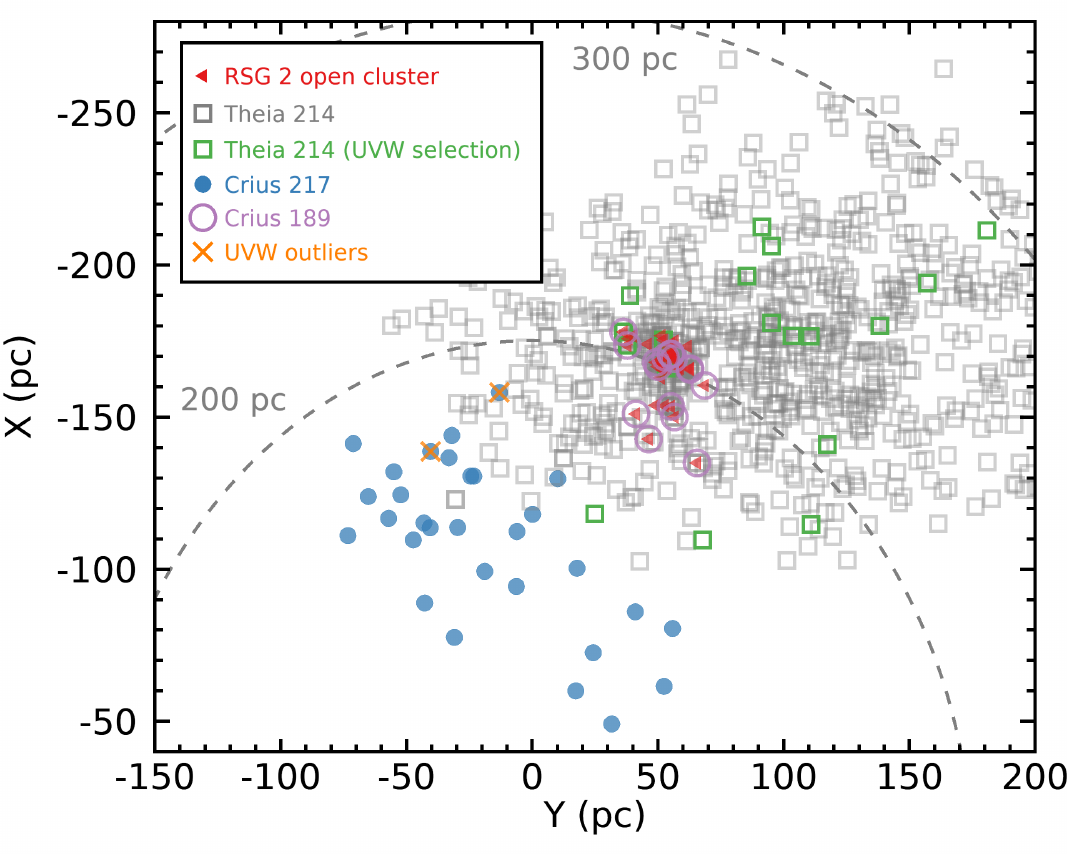}\label{fig:rsg2_xy}}
	\subfigure[RSG~2 corona in $UV$ space]{\includegraphics[width=0.49\textwidth]{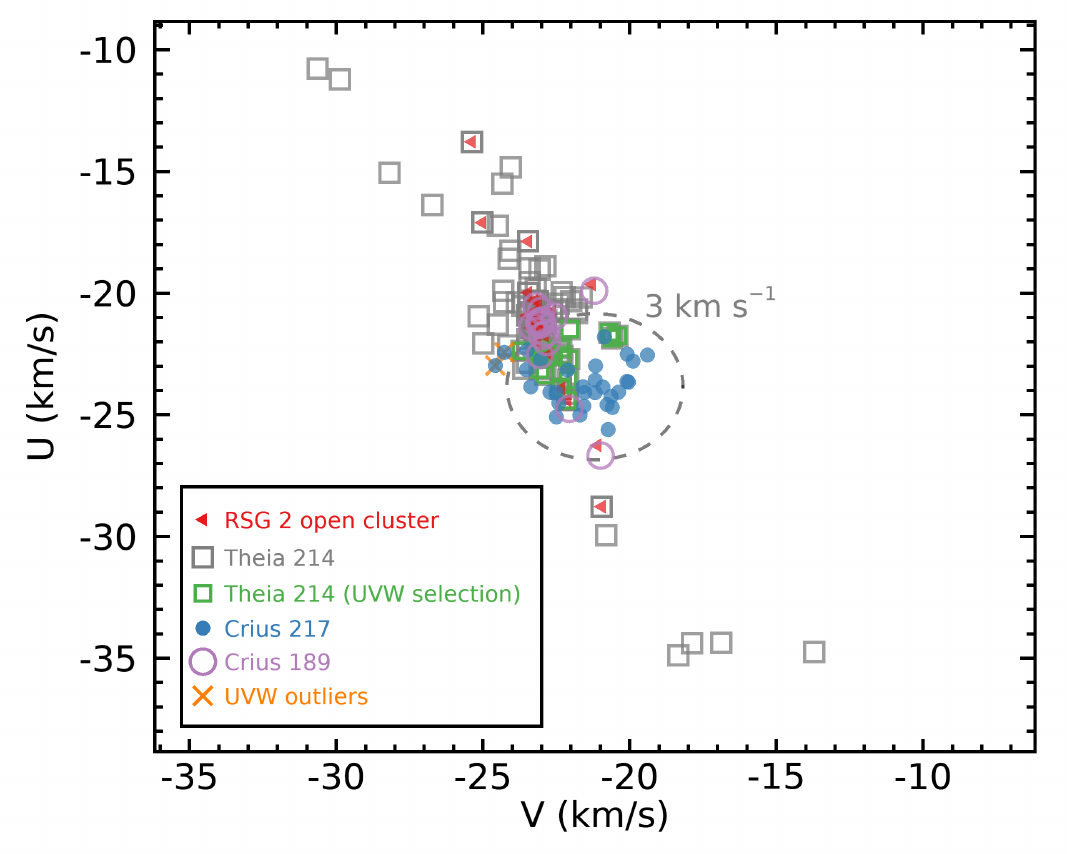}\label{fig:rsg2_uv}}
	\caption{Galactic positions and space velocities of Crius~217 and Theia~214 of \cite{2019AJ....158..122K}, which likely correspond to an extensive corona around the newly discovered RSG~2 open cluster \citep{2016AA...595A..22R}.}
	\label{fig:rsg2}
\end{figure*}

\begin{figure*}
	\centering
	\subfigure[NGC2451A corona in $XY$ space]{\includegraphics[width=0.49\textwidth]{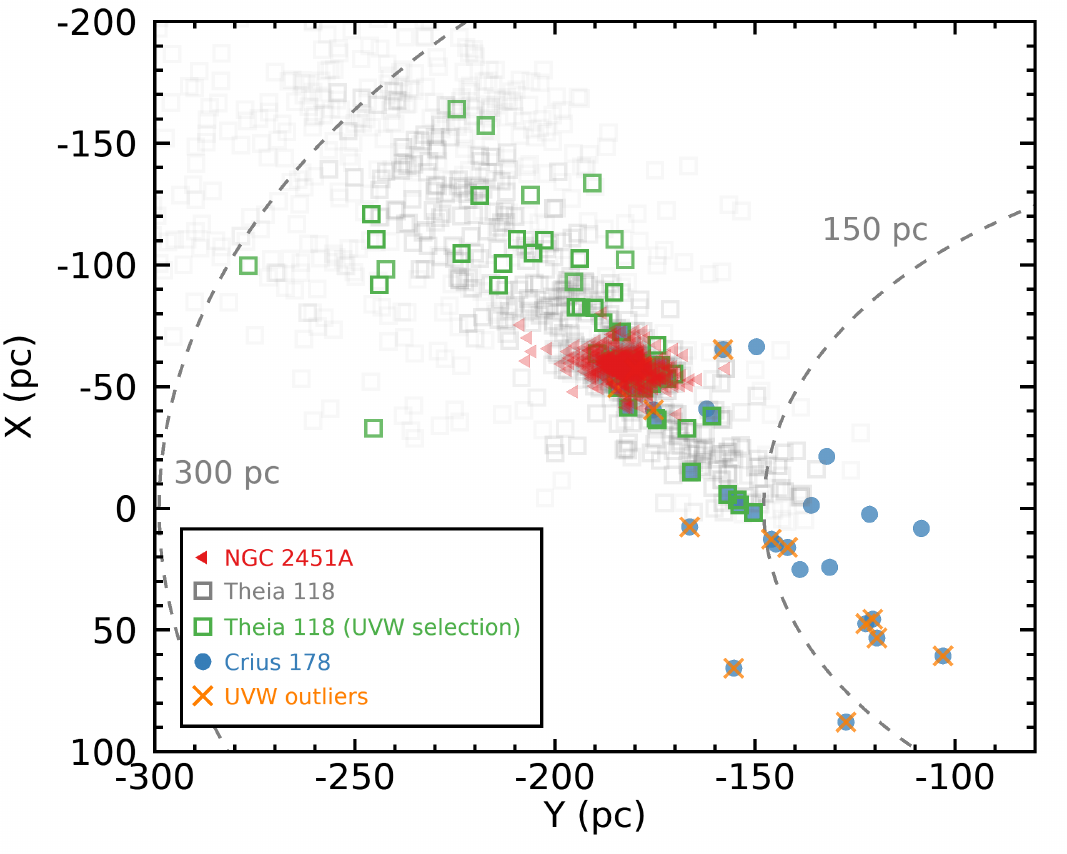}\label{fig:ngc2451a_xy}}
	\subfigure[NGC2451A corona in $UV$ space]{\includegraphics[width=0.49\textwidth]{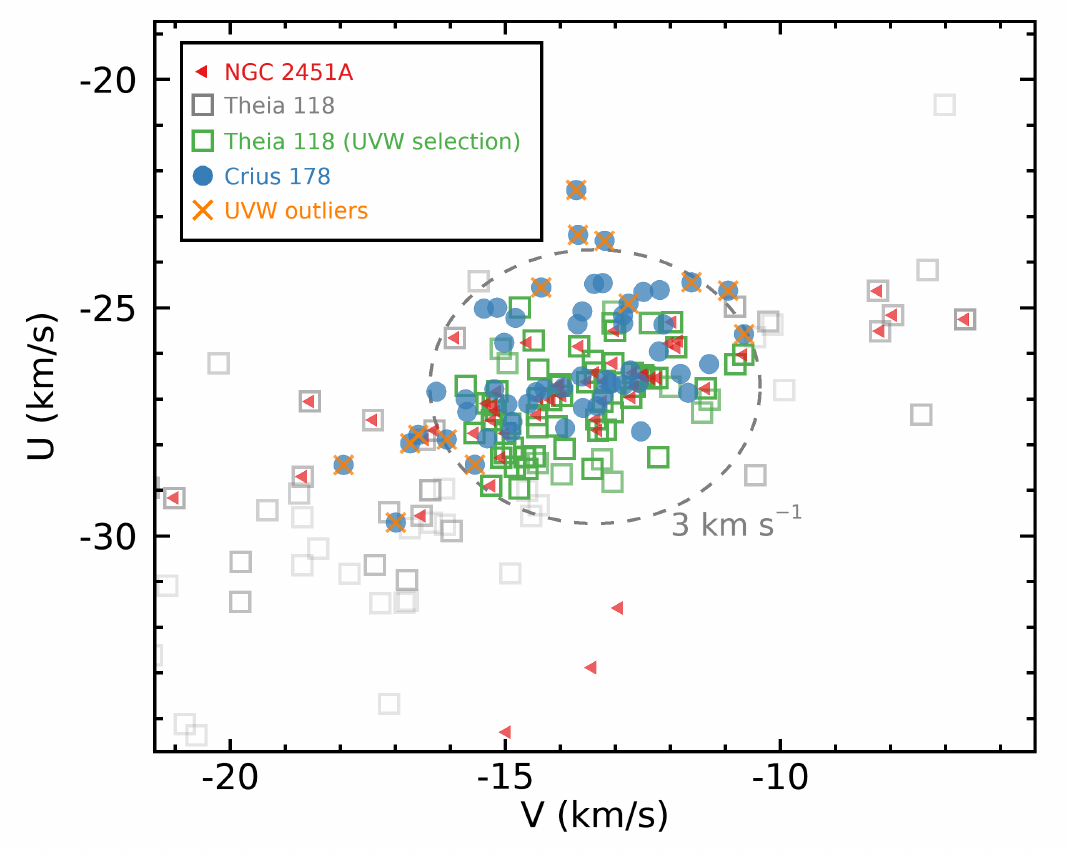}\label{fig:ngc2451a_uv}}
	\caption{The NGC~2451A open cluster with its corona identified by \cite{2019AA...622L..13M}. The spatial position and kinematics of Crius~178 are consistent with it extending the corona slightly towards nearby distances, which was probably not recovered by previous studies because of projection effects. The elongated distribution of NGC~2451A members in $UV$ space is due to larger parallax measurement errors compared with proper motion and heliocentric radial velocity measurement errors in Gaia~EDR3, which is typical of young associations and open clusters beyond 100\,pc.}
	\label{fig:ngc2451a}
\end{figure*}

Figures~\ref{fig:upk612} to \ref{fig:ngc2451a} show the individual distributions of $XY$ Galactic positions and $UV$ space velocities for the newly recognized coronae in this work. These structures also lie in a narrow sheet in the $Z$ direction and have $UVW$ space velocities consistent with their candidate host association.

\end{document}

%% file: table_allstars.tex
\startlongtable
\tabletypesize{\footnotesize}
\tablewidth{0.985\textwidth}
\begin{deluxetable*}{llcccccc}
\tablecolumns{8}
\tablecaption{Individual Gaia~EDR3 entries recovered as clusters by HDBSCAN.\label{tab:allstars}}
\tablehead{\colhead{Crius} & \colhead{Gaia~EDR3} & \colhead{Right ascension} & \colhead{Declination} & \colhead{$\mu_\alpha\cos\delta$} & \colhead{$\mu_\delta$} & \colhead{Parallax} & \colhead{Radial velocity}\\
\colhead{Group ID} & \colhead{source ID} & \colhead{(hh:mm:ss.sss)} & \colhead{dd:mm:ss.ss} & \colhead{(mas/yr)} & \colhead{(mas/yr)} & \colhead{(mas)} & \colhead{(\kms)}}
\startdata
1 &    2491729335519097600 & 01:58:24.266 & -05:09:53.51 & $246.99 \pm 0.04$ & $147.53 \pm 0.04$ & $14.09 \pm 0.04$ & $51.5 \pm 0.4$\\
1 &      83929987451343104 & 02:56:05.817 & 18:53:12.75 & $109.72 \pm 0.02$ & $19.47 \pm 0.01$ & $6.66 \pm 0.02$ & $88.9 \pm 0.8$\\
1 &     459661938487952000 & 02:56:35.793 & 55:26:06.82 & $727.63 \pm 0.02$ & $-455.23 \pm 0.02$ & $49.32 \pm 0.02$ & $76.2 \pm 0.3$\\
1 &      57366301921437952 & 03:38:35.335 & 20:09:45.29 & $95.44 \pm 0.02$ & $16.35 \pm 0.01$ & $8.41 \pm 0.02$ & $102 \pm 1$\\
1 &    3441580727627952512 & 05:45:22.715 & 27:38:13.04 & $-5.22 \pm 0.02$ & $-31.27 \pm 0.01$ & $15.63 \pm 0.02$ & $108 \pm 1$\\
1 &    3813898626334257664 & 11:21:30.469 & 05:17:40.32 & $-275.02 \pm 0.02$ & $60.96 \pm 0.01$ & $11.41 \pm 0.01$ & $12.8 \pm 0.5$\\
1 &    3813323276811078656 & 11:32:39.157 & 05:13:43.76 & $-301.36 \pm 0.02$ & $72.59 \pm 0.01$ & $12.03 \pm 0.02$ & $15.6 \pm 0.4$\\
1 &    3930922058356114432 & 12:47:13.039 & 14:42:16.43 & $-354.72 \pm 0.02$ & $149.31 \pm 0.02$ & $16.50 \pm 0.02$ & $-24 \pm 1$\\
1 &    6197890597023337344 & 15:00:38.072 & -38:41:07.83 & $-210.94 \pm 0.02$ & $-92.50 \pm 0.01$ & $14.09 \pm 0.02$ & $-86 \pm 1$\\
1 &    4595150639153952000 & 17:27:35.020 & 27:01:38.22 & $-9.09 \pm 0.01$ & $370.94 \pm 0.02$ & $20.34 \pm 0.01$ & $-72.1 \pm 0.3$\\
\enddata
\tablecomments{Only a portion of the full table is shown here. The complete table is available as a machine-readable table in the online-only material.}
\end{deluxetable*}

%% file: table_allgroups_pass.tex
\startlongtable
\tabletypesize{\scriptsize}
\tablewidth{0.985\textwidth}
\begin{deluxetable*}{lccccccccccccccccc}
\tablecolumns{18}
\tablecaption{Median properties of individual groups recovered by HDBSCAN which passed all selection criteria.\label{tab:allgroups_pass}}
\tablehead{\colhead{Crius} & \colhead{Number of} & \colhead{Number with} & \colhead{} & \multicolumn{7}{c}{Median (pc, \kms)} & \colhead{} & \multicolumn{4}{c}{MAD$\dagger$ (pc, \kms)} & \colhead{}\\
\cline{5-11}
\cline{13-16}
\colhead{ID} & \colhead{members} & \colhead{RUWE $>$ 1.4} & \colhead{} & \colhead{distance} & \colhead{$X$} & \colhead{$Y$} & \colhead{$Z$} & \colhead{$U$} & \colhead{$V$} & \colhead{$W$} & \colhead{} & \colhead{$Z$} & \colhead{$U$} & \colhead{$V$} & \colhead{$W$} & \colhead{}}
\startdata
77 & 8 & 0 &  & $148.4$ & $133.1$ & $29.3$ & $-58.6$ & $-36.2$ & $-29.3$ & $-10.4$ &  & $6.9$ & $1.5$ & $1.1$ & $2.1$ & \\
85 & 12 & 4 &  & $154.5$ & $6.4$ & $-118.0$ & $99.6$ & $-15.0$ & $-6.4$ & $-19.3$ &  & $8.0$ & $1.7$ & $2.1$ & $0.3$ & \\
109 & 16 & 2 &  & $158.7$ & $-100.8$ & $86.3$ & $-87.1$ & $2.6$ & $1.5$ & $-16.3$ &  & $9.2$ & $1.6$ & $0.8$ & $1.4$ & \\
110 & 8 & 1 &  & $142.6$ & $119.5$ & $-2.6$ & $-77.8$ & $5.8$ & $7.4$ & $-15.7$ &  & $7.5$ & $0.8$ & $1.1$ & $1.5$ & \\
113 & 10 & 1 &  & $120.4$ & $98.3$ & $-68.7$ & $10.6$ & $-0.5$ & $2.4$ & $-12.8$ &  & $7.0$ & $2.3$ & $1.7$ & $2.8$ & \\
116 & 11 & 0 &  & $114.5$ & $62.7$ & $22.6$ & $-93.1$ & $-11.8$ & $-5.9$ & $5.7$ &  & $7.2$ & $0.4$ & $0.6$ & $0.7$ & \\
119 & 11 & 2 &  & $155.1$ & $108.5$ & $-104.9$ & $35.7$ & $-14.7$ & $-5.9$ & $-2.2$ &  & $13.5$ & $1.6$ & $1.7$ & $1.5$ & \\
124 & 15 & 0 &  & $160.4$ & $-48.1$ & $-92.4$ & $-122.0$ & $-0.7$ & $8.5$ & $-16.7$ &  & $9.6$ & $0.6$ & $0.7$ & $0.5$ & \\
127 & 10 & 0 &  & $121.4$ & $-70.3$ & $74.2$ & $-65.5$ & $23.7$ & $-2.9$ & $-3.6$ &  & $4.8$ & $0.9$ & $0.8$ & $1.8$ & \\
131 & 40 & 2 &  & $112.8$ & $10.9$ & $-100.5$ & $-49.9$ & $-0.2$ & $0.5$ & $-1.7$ &  & $10.0$ & $1.1$ & $1.1$ & $1.3$ & \\
134 & 24 & 2 &  & $111.1$ & $-107.4$ & $26.4$ & $10.3$ & $20.5$ & $-5.7$ & $-13.6$ &  & $11.3$ & $1.6$ & $1.2$ & $1.9$ & \\
135 & 13 & 2 &  & $84.9$ & $-17.5$ & $-75.1$ & $-35.4$ & $15.7$ & $-1.2$ & $-3.6$ &  & $12.6$ & $1.4$ & $1.1$ & $2.6$ & \\
137 & 13 & 0 &  & $97.1$ & $-75.9$ & $30.8$ & $-52.1$ & $5.1$ & $-1.9$ & $-7.3$ &  & $4.9$ & $1.1$ & $0.7$ & $0.7$ & \\
144 & 10 & 2 &  & $161.9$ & $-105.2$ & $114.5$ & $-45.4$ & $-42.1$ & $-22.7$ & $-9.3$ &  & $13.3$ & $0.9$ & $1.0$ & $1.0$ & \\
147 & 44 & 3 &  & $100.0$ & $-7.2$ & $56.7$ & $82.0$ & $-3.4$ & $-9.0$ & $-1.7$ &  & $2.6$ & $0.7$ & $0.6$ & $0.8$ & \\
148 & 62 & 9 &  & $85.7$ & $-6.3$ & $-5.6$ & $85.3$ & $-2.4$ & $-5.6$ & $-0.5$ &  & $2.6$ & $0.5$ & $0.4$ & $0.7$ & \\
149 & 16 & 2 &  & $81.1$ & $25.7$ & $7.6$ & $76.6$ & $-10.4$ & $-20.7$ & $5.6$ &  & $13.2$ & $1.1$ & $1.2$ & $1.6$ & \\
150 & 26 & 2 &  & $152.3$ & $149.2$ & $29.7$ & $-7.7$ & $-7.9$ & $-17.2$ & $-13.6$ &  & $5.3$ & $1.2$ & $0.7$ & $0.8$ & \\
151 & 19 & 3 &  & $112.8$ & $88.3$ & $46.7$ & $52.3$ & $-15.9$ & $-1.1$ & $-2.5$ &  & $7.9$ & $2.2$ & $1.3$ & $1.2$ & \\
152 & 11 & 1 &  & $122.4$ & $-81.6$ & $-60.3$ & $68.5$ & $-33.4$ & $-13.9$ & $-16.7$ &  & $9.0$ & $1.9$ & $1.4$ & $0.5$ & \\
153 & 22 & 3 &  & $169.4$ & $-55.6$ & $135.5$ & $-85.2$ & $-12.1$ & $-0.8$ & $-7.1$ &  & $6.3$ & $1.4$ & $0.7$ & $0.5$ & \\
154 & 10 & 3 &  & $178.7$ & $35.9$ & $-172.3$ & $-30.8$ & $-35.4$ & $-18.4$ & $-7.4$ &  & $10.6$ & $1.4$ & $1.4$ & $0.5$ & \\
155 & 29 & 4 &  & $93.8$ & $75.9$ & $45.7$ & $30.9$ & $-25.3$ & $-4.2$ & $-5.0$ &  & $8.7$ & $1.7$ & $1.4$ & $0.9$ & \\
156 & 70 & 8 &  & $175.4$ & $-127.3$ & $110.3$ & $48.8$ & $-17.5$ & $-5.0$ & $-6.3$ &  & $5.2$ & $1.3$ & $0.9$ & $1.0$ & \\
158 & 12 & 0 &  & $139.1$ & $43.8$ & $116.1$ & $63.0$ & $-12.2$ & $-4.0$ & $-9.7$ &  & $12.3$ & $2.0$ & $1.9$ & $1.2$ & \\
159 & 13 & 1 &  & $125.5$ & $22.3$ & $122.3$ & $17.6$ & $-14.1$ & $-9.4$ & $-9.9$ &  & $8.6$ & $0.9$ & $1.3$ & $2.4$ & \\
162 & 14 & 1 &  & $73.5$ & $27.7$ & $-62.1$ & $28.0$ & $-17.7$ & $-3.7$ & $-11.9$ &  & $8.9$ & $1.1$ & $1.0$ & $1.1$ & \\
163 & 18 & 1 &  & $170.9$ & $-48.1$ & $-160.5$ & $-33.6$ & $-10.5$ & $-8.3$ & $-9.8$ &  & $11.1$ & $2.1$ & $1.7$ & $0.5$ & \\
166 & 26 & 3 &  & $136.8$ & $135.1$ & $-15.6$ & $14.9$ & $-1.0$ & $-28.9$ & $-14.0$ &  & $5.4$ & $2.9$ & $0.5$ & $0.5$ & \\
167 & 13 & 0 &  & $191.3$ & $-188.9$ & $-25.3$ & $16.0$ & $-35.6$ & $-19.8$ & $1.4$ &  & $14.7$ & $1.6$ & $1.0$ & $0.5$ & \\
168 & 16 & 0 &  & $116.3$ & $-3.0$ & $101.7$ & $-56.3$ & $-3.8$ & $-6.9$ & $-9.9$ &  & $12.3$ & $1.6$ & $1.2$ & $1.2$ & \\
169 & 60 & 11 &  & $126.2$ & $-117.8$ & $5.8$ & $-44.7$ & $-12.7$ & $-6.7$ & $-9.1$ &  & $13.4$ & $1.9$ & $1.2$ & $1.4$ & \\
170 & 10 & 1 &  & $107.8$ & $80.3$ & $-51.8$ & $-49.9$ & $-7.3$ & $-4.1$ & $-8.9$ &  & $3.3$ & $0.4$ & $0.6$ & $0.4$ & \\
171 & 15 & 2 &  & $187.0$ & $171.2$ & $-64.3$ & $-38.9$ & $-10.3$ & $-1.2$ & $-11.2$ &  & $10.5$ & $1.2$ & $1.1$ & $1.1$ & \\
172 & 38 & 5 &  & $139.5$ & $-113.0$ & $-23.3$ & $-78.3$ & $-9.6$ & $-3.8$ & $-11.4$ &  & $14.1$ & $1.6$ & $1.0$ & $0.9$ & \\
174 & 28 & 0 &  & $123.7$ & $-8.9$ & $65.8$ & $-104.3$ & $-7.3$ & $-4.1$ & $-18.5$ &  & $5.7$ & $1.0$ & $0.8$ & $1.3$ & \\
175 & 13 & 1 &  & $114.8$ & $-57.7$ & $8.3$ & $-98.9$ & $-9.2$ & $-4.2$ & $-18.4$ &  & $6.2$ & $0.5$ & $0.3$ & $0.7$ & \\
176 & 12 & 1 &  & $162.3$ & $-121.2$ & $-55.8$ & $-92.5$ & $-11.5$ & $-5.1$ & $-18.3$ &  & $2.0$ & $0.7$ & $0.6$ & $0.2$ & \\
178 & 55 & 8 &  & $182.8$ & $-49.5$ & $-174.1$ & $-25.4$ & $-26.6$ & $-13.7$ & $-13.1$ &  & $4.6$ & $1.7$ & $1.8$ & $0.8$ & \\
182 & 11 & 0 &  & $152.1$ & $-60.2$ & $85.8$ & $110.3$ & $-14.2$ & $-21.6$ & $-3.6$ &  & $5.0$ & $1.0$ & $1.0$ & $0.8$ & \\
183 & 20 & 3 &  & $160.7$ & $80.4$ & $-136.3$ & $28.1$ & $-34.5$ & $-17.4$ & $-10.9$ &  & $13.5$ & $1.3$ & $2.0$ & $2.9$ & \\
184 & 11 & 1 &  & $138.9$ & $-91.8$ & $-89.5$ & $53.4$ & $-29.3$ & $-19.3$ & $2.3$ &  & $4.9$ & $0.7$ & $0.5$ & $1.9$ & \\
187 & 232 & 33 &  & $184.2$ & $-140.0$ & $-67.9$ & $98.6$ & $-42.8$ & $-20.4$ & $-9.5$ &  & $3.3$ & $1.1$ & $0.8$ & $0.7$ & \\
188 & 14 & 0 &  & $154.5$ & $-107.9$ & $-69.3$ & $-86.1$ & $-14.3$ & $-29.3$ & $-4.8$ &  & $7.6$ & $1.0$ & $0.4$ & $0.8$ & \\
189 & 16 & 4 &  & $195.9$ & $-166.2$ & $55.2$ & $87.7$ & $-21.3$ & $-23.0$ & $-2.1$ &  & $2.0$ & $0.6$ & $0.2$ & $0.5$ & \\
190 & 10 & 0 &  & $182.1$ & $53.1$ & $-172.4$ & $-24.8$ & $-34.4$ & $-16.8$ & $-0.7$ &  & $6.4$ & $0.7$ & $0.5$ & $0.9$ & \\
191 & 18 & 2 &  & $125.4$ & $38.1$ & $-103.0$ & $60.6$ & $-26.3$ & $-7.2$ & $-10.9$ &  & $3.7$ & $0.5$ & $1.1$ & $0.7$ & \\
192 & 14 & 1 &  & $71.0$ & $1.8$ & $-14.5$ & $69.5$ & $-24.6$ & $-8.4$ & $-8.6$ &  & $4.3$ & $1.9$ & $1.1$ & $1.5$ & \\
193 & 14 & 1 &  & $154.8$ & $90.8$ & $-123.0$ & $-24.5$ & $-22.6$ & $-13.5$ & $-6.1$ &  & $6.1$ & $0.4$ & $0.9$ & $0.8$ & \\
195 & 20 & 0 &  & $164.0$ & $-5.6$ & $-163.3$ & $13.4$ & $-23.6$ & $-15.8$ & $-5.9$ &  & $4.9$ & $0.5$ & $0.8$ & $0.5$ & \\
196 & 38 & 2 &  & $150.3$ & $0.8$ & $-149.1$ & $-18.3$ & $-23.8$ & $-14.5$ & $-5.7$ &  & $1.4$ & $0.4$ & $1.3$ & $0.3$ & \\
198 & 11 & 0 &  & $47.6$ & $-27.1$ & $36.8$ & $13.1$ & $-34.3$ & $-12.9$ & $-11.0$ &  & $6.9$ & $1.5$ & $2.1$ & $1.4$ & \\
199 & 21 & 2 &  & $157.9$ & $106.5$ & $-89.3$ & $-74.8$ & $-37.0$ & $-11.8$ & $-4.1$ &  & $5.9$ & $1.0$ & $1.0$ & $0.7$ & \\
202 & 20 & 2 &  & $104.0$ & $13.3$ & $-95.2$ & $-39.6$ & $-33.1$ & $-16.3$ & $0.9$ &  & $15.0$ & $1.7$ & $1.0$ & $0.5$ & \\
204 & 17 & 2 &  & $120.7$ & $100.7$ & $56.6$ & $-35.0$ & $-26.5$ & $-13.7$ & $2.7$ &  & $11.5$ & $2.0$ & $0.9$ & $0.5$ & \\
205 & 12 & 0 &  & $83.2$ & $31.0$ & $66.4$ & $39.3$ & $-36.8$ & $-18.4$ & $-8.1$ &  & $12.5$ & $0.5$ & $0.6$ & $0.8$ & \\
207 & 26 & 2 &  & $90.8$ & $76.8$ & $40.3$ & $27.0$ & $-35.6$ & $-15.3$ & $-7.5$ &  & $9.9$ & $2.7$ & $1.4$ & $2.8$ & \\
208 & 13 & 5 &  & $112.1$ & $79.7$ & $75.3$ & $23.1$ & $-26.2$ & $-14.1$ & $0.8$ &  & $6.9$ & $1.0$ & $0.6$ & $0.1$ & \\
210 & 18 & 1 &  & $91.6$ & $48.7$ & $-55.7$ & $54.0$ & $-32.5$ & $-14.0$ & $0.5$ &  & $6.9$ & $0.8$ & $1.3$ & $0.7$ & \\
213 & 11 & 0 &  & $60.4$ & $-46.3$ & $38.6$ & $3.9$ & $-41.4$ & $-19.3$ & $-10.7$ &  & $9.6$ & $1.4$ & $0.9$ & $1.8$ & \\
214 & 293 & 28 &  & $45.0$ & $-41.8$ & $1.1$ & $-16.9$ & $-42.2$ & $-19.1$ & $-1.4$ &  & $4.0$ & $1.0$ & $0.5$ & $0.5$ & \\
215 & 13 & 2 &  & $183.8$ & $-70.9$ & $-146.7$ & $85.0$ & $-27.1$ & $-19.4$ & $-1.7$ &  & $9.8$ & $1.8$ & $1.6$ & $1.4$ & \\
217 & 30 & 6 &  & $151.4$ & $-113.7$ & $-27.0$ & $96.3$ & $-23.8$ & $-21.4$ & $-3.2$ &  & $9.3$ & $1.1$ & $1.5$ & $1.0$ & \\
220 & 23 & 0 &  & $47.7$ & $42.6$ & $1.7$ & $-21.4$ & $-9.1$ & $-15.3$ & $-8.4$ &  & $10.1$ & $1.2$ & $0.6$ & $1.1$ & \\
221 & 67 & 9 &  & $63.7$ & $22.1$ & $-57.9$ & $-14.7$ & $-15.9$ & $-27.8$ & $-0.8$ &  & $7.0$ & $1.0$ & $0.9$ & $0.7$ & \\
222 & 14 & 1 &  & $80.1$ & $-31.2$ & $9.4$ & $73.2$ & $-18.9$ & $-18.4$ & $-7.5$ &  & $11.4$ & $1.2$ & $0.5$ & $0.9$ & \\
224 & 28 & 3 &  & $43.9$ & $10.8$ & $14.3$ & $40.1$ & $-16.4$ & $-20.6$ & $-3.7$ &  & $5.5$ & $1.1$ & $1.5$ & $0.9$ & \\
225 & 15 & 3 &  & $171.1$ & $-100.7$ & $-77.2$ & $-114.7$ & $-15.9$ & $-22.4$ & $-4.3$ &  & $4.7$ & $0.7$ & $0.5$ & $0.5$ & \\
226 & 23 & 0 &  & $153.1$ & $35.7$ & $148.3$ & $12.7$ & $-8.1$ & $-22.3$ & $-6.1$ &  & $7.5$ & $0.6$ & $0.5$ & $0.7$ & \\
227 & 13 & 0 &  & $42.9$ & $3.1$ & $42.7$ & $-2.7$ & $-9.4$ & $-22.9$ & $-5.4$ &  & $7.9$ & $0.5$ & $0.9$ & $0.5$ & \\
228 & 24 & 2 &  & $153.3$ & $-44.8$ & $-132.5$ & $-62.8$ & $-7.5$ & $-24.7$ & $-14.1$ &  & $7.8$ & $0.7$ & $1.1$ & $0.7$ & \\
229 & 142 & 6 &  & $166.2$ & $-138.5$ & $89.9$ & $-18.8$ & $-13.5$ & $-23.8$ & $-6.7$ &  & $8.7$ & $1.2$ & $0.7$ & $0.4$ & \\
230 & 13 & 1 &  & $147.3$ & $-124.4$ & $2.1$ & $-78.8$ & $-14.4$ & $-24.1$ & $-7.0$ &  & $4.7$ & $0.4$ & $0.5$ & $0.4$ & \\
231 & 30 & 2 &  & $82.5$ & $-51.2$ & $-49.5$ & $-41.7$ & $-6.1$ & $-27.9$ & $-14.2$ &  & $6.1$ & $0.5$ & $0.3$ & $0.9$ & \\
232 & 211 & 23 &  & $135.4$ & $-120.4$ & $29.1$ & $-54.6$ & $-6.8$ & $-28.4$ & $-14.3$ &  & $3.0$ & $1.1$ & $0.5$ & $0.7$ & \\
233 & 45 & 5 &  & $44.7$ & $7.3$ & $-25.1$ & $-36.3$ & $-9.8$ & $-20.8$ & $-0.5$ &  & $4.2$ & $0.6$ & $0.3$ & $0.8$ & \\
234 & 15 & 0 &  & $147.3$ & $142.5$ & $-3.6$ & $-37.3$ & $-4.7$ & $-17.5$ & $-7.9$ &  & $4.9$ & $1.3$ & $0.3$ & $1.1$ & \\
235 & 46 & 3 &  & $149.9$ & $49.7$ & $-140.9$ & $-12.0$ & $-8.4$ & $-21.0$ & $-0.9$ &  & $4.6$ & $0.7$ & $1.3$ & $0.6$ & \\
236 & 31 & 0 &  & $175.6$ & $167.6$ & $-50.5$ & $14.0$ & $-4.1$ & $-19.6$ & $-4.2$ &  & $3.1$ & $1.2$ & $0.6$ & $0.4$ & \\
237 & 11 & 0 &  & $73.6$ & $-26.4$ & $-49.8$ & $-47.4$ & $-12.9$ & $-21.8$ & $-5.0$ &  & $11.9$ & $0.5$ & $0.4$ & $0.6$ & \\
238 & 57 & 8 &  & $130.6$ & $11.2$ & $-129.1$ & $-16.6$ & $-11.2$ & $-22.8$ & $-4.1$ &  & $10.4$ & $0.5$ & $1.2$ & $0.4$ & \\
239 & 14 & 1 &  & $144.3$ & $140.2$ & $-24.5$ & $24.0$ & $-2.4$ & $-16.7$ & $-6.8$ &  & $12.2$ & $0.8$ & $0.4$ & $0.5$ & \\
240 & 60 & 12 &  & $143.3$ & $131.5$ & $-19.0$ & $53.6$ & $-5.9$ & $-15.9$ & $-8.0$ &  & $6.5$ & $1.6$ & $0.9$ & $1.4$ & \\
241 & 126 & 9 &  & $111.2$ & $57.8$ & $-93.4$ & $17.3$ & $-8.7$ & $-20.5$ & $-6.6$ &  & $12.9$ & $1.0$ & $1.0$ & $0.9$ &
\enddata
\tablenotetext{}{$^\dagger$\ Median absolute deviation.}
\end{deluxetable*}

%% file: table_allgroups_fail.tex
\startlongtable
\tabletypesize{\scriptsize}
\tablewidth{0.985\textwidth}
\begin{deluxetable*}{lcccccccccccccccc}
\tablecolumns{17}
\tablecaption{Median properties of individual groups recovered by HDBSCAN which failed at least one selection criterion.\label{tab:allgroups_fail}}
\tablehead{\colhead{Crius} & \colhead{Number of} & \colhead{} & \multicolumn{7}{c}{Median (pc, \kms)} & \colhead{} & \multicolumn{4}{c}{MAD$^\dagger$ (pc, \kms)} & \colhead{} & \colhead{Selection} \\
\cline{4-10}
\cline{12-15}
\colhead{ID} & \colhead{members} & \colhead{} & \colhead{distance} & \colhead{$X$} & \colhead{$Y$} & \colhead{$Z$} & \colhead{$U$} & \colhead{$V$} & \colhead{$W$} & \colhead{} & \colhead{$Z$} & \colhead{$U$} & \colhead{$V$} & \colhead{$W$} & \colhead{} & \colhead{criteria$^\ddagger$}}
\startdata
1 & 10 &  & $17.7$ & $-13.7$ & $4.4$ & $10.3$ & $-111.9$ & $-14.7$ & $-10.8$ &  & $82.6$ & $2.9$ & $9.5$ & $6.1$ &  & FFFF\\
2 & 11 &  & $32.3$ & $7.0$ & $12.1$ & $29.1$ & $-17.2$ & $-90.1$ & $-12.9$ &  & $39.5$ & $3.8$ & $5.5$ & $2.5$ &  & FPFF\\
3 & 10 &  & $86.4$ & $25.8$ & $-23.3$ & $-79.1$ & $-67.1$ & $-66.3$ & $1.1$ &  & $33.7$ & $2.8$ & $2.7$ & $1.0$ &  & FPPF\\
4 & 26 &  & $64.8$ & $-37.9$ & $-46.4$ & $24.6$ & $47.7$ & $-2.2$ & $-39.6$ &  & $48.4$ & $4.5$ & $2.2$ & $3.9$ &  & PPFF\\
5 & 15 &  & $90.8$ & $4.1$ & $90.1$ & $10.0$ & $8.6$ & $-65.7$ & $3.7$ &  & $38.8$ & $4.3$ & $3.3$ & $3.9$ &  & FPFF\\
6 & 14 &  & $74.0$ & $16.6$ & $-50.8$ & $51.1$ & $-7.3$ & $27.1$ & $6.3$ &  & $34.3$ & $1.1$ & $2.2$ & $4.1$ &  & PFFF\\
7 & 7 &  & $140.5$ & $-33.8$ & $-120.3$ & $64.1$ & $-55.9$ & $3.8$ & $2.6$ &  & $25.9$ & $3.3$ & $1.1$ & $1.0$ &  & PPFF\\
8 & 10 &  & $72.5$ & $43.0$ & $49.4$ & $31.0$ & $-71.3$ & $-11.6$ & $-20.4$ &  & $41.0$ & $3.5$ & $1.9$ & $1.7$ &  & PPFF\\
9 & 11 &  & $91.9$ & $23.1$ & $-40.2$ & $79.4$ & $-89.1$ & $-49.4$ & $3.0$ &  & $32.8$ & $2.0$ & $1.9$ & $2.9$ &  & FPPF\\
10 & 11 &  & $88.2$ & $-70.7$ & $50.1$ & $-16.1$ & $26.4$ & $-27.4$ & $-42.6$ &  & $18.6$ & $1.8$ & $2.2$ & $2.0$ &  & PPPF
\enddata
\tablecomments{Only a portion of the full table is shown here. The complete table is available as a machine-readable table in the online-only material.}
\tablenotetext{}{$^\dagger$\ Median absolute deviation.}
\tablenotetext{}{$^\ddagger$\ Pass (P) or fail (F) flags that indicate whether a given Crius group has passed the selection criteria for further study in this work. The respective character positions refer to the following criteria: (1) rejection based on the Hercules stream; (2) rejection based on other high-velocity groups in the $UV$ plane ($U < -100$\,\kms, $U > 60$\,\kms, or $V > 20$\,\kms); (3) rejection based on individual $UVW$ MADs above 3\,\kms; (4) rejection based on $Z$ MAD above 15\,pc.}
\end{deluxetable*}

%% file: table_known.tex
\startlongtable
\tabletypesize{\footnotesize}
\tablewidth{0.985\textwidth}
\begin{deluxetable*}{lcccl}
\tablecolumns{5}
\tablecaption{Known associations recovered in this work.\label{tab:known}}
\tablehead{\colhead{Crius} & \colhead{Approx.} & \colhead{Age} & \colhead{Age} & \colhead{Association}\\
\colhead{ID} & \colhead{distance (pc)} & \colhead{(Myr)} & \colhead{ref.} & \colhead{Name}}
\startdata
141$^\mathsection$ & 25 & 414 & 1 & Ursa Major association\\
220 & 50 & 26 & 2 & $\beta$ Pic moving group$^\dagger$\\
233 & 50 & 40 & 3 & Tucana-Horologium\\
214 & 50 & 676 & 4 & Hyades + \cite{2019AA...621L...3M} corona\\
237 & 70 & 42 & 5 & Columba$^\dagger$\\
155 & 80 & 400 & 6 & Theia 1005\\
148 & 90 & 560 & 7 & Coma Berenices + \cite{2019ApJ...877...12T} corona\\
221 & 90 & 90 & 8 & Volans-Carina (\citealp{2017AJ....153..257O} Group 30) + corona (Theia 424)\\
216 & 100 & 24 & 9 & \cite{mamajek118tau} 118~Tau, \cite{2017MNRAS.468.1198B} 32~Ori\\
170 & 100 & 250 & 10 & \cite{2021AJ....161..171T} MELANGE--1, Theia 786\\
147 & 100 & 400 & 11 & \cite{2017AJ....153..257O} Group 10\\
151 & 110 & 300 & 6 & Theia 677$^\dagger$\\
204 & 110 & 500 & 6 & Theia 595$^\dagger$ (low-$Z$ part)\\
205 & 110 & 300 & 6 & Theia 677$^\dagger$\\
207 & 110 & 300 & 6 & Theia 677$^\dagger$\\
208 & 110 & 500 & 6 & Theia 595$^\dagger$ (high-$Z$ part)\\
227 & 110 & 160 & 6 & Theia 209\\
169 & 120 & 30--38 & 12 & \cite{2021ApJ...917...23K} Greater Taurus Subgroup 4\\
174 & 130 & 120 & 13 & \cite{2019AA...622L..13M} Pisces-Eridanus stream$^\ddagger$\\
175 & 130 & 120 & 13 & \cite{2019AA...622L..13M} Pisces-Eridanus stream$^\ddagger$\\
176 & 130 & 120 & 13 & \cite{2019AA...622L..13M} Pisces-Eridanus stream$^\ddagger$\\
238 & 130 & 60 & 14 & Platais 8 + \cite{2021ApJ...915L..29G} candidate corona (Theia 113, Theia 92$^\dagger$) \\
241 & 130 & 8--23 & 12 & \cite{2021ApJ...917...23K} Greater Scorpius-Centaurus Subgroup 27, Upper Centaurus Lupus$^\dagger$\\
164$^\mathsection$ & 140 & 35 & 15 & Octans$^\dagger$\\
166 & 140 & 170 & 6 & Theia~368\\
232 & 140 & 112 & 16 & Pleiades\\
239 & 140 & 19 & 12 & \cite{2021ApJ...917...23K} Greater Scorpius-Centaurus Subgroups 16$^\dagger$, 17$^\dagger$\\
240 & 140 & 11 & 12 & \cite{2021ApJ...917...23K} Greater Scorpius-Centaurus Subgroup 17\\
170 & 150 & 280 & 6 & Theia 786\\
196 & 150 & 51 & 17 & IC~2391 + \cite{2021AA...645A..84M} corona\\
230 & 150 & 60 & 18 & \cite{2020ApJ...903...96G} $\mu$ Tau\\
234 & 150 & 5 & 19 & Corona Australis\\
235 & 150 & 46 & 20 & IC~2602 + \cite{2021AA...645A..84M} corona (Theia 92$^\dagger$)\\
226 & 160 & 160 & 6 & \cite{2017AJ....153..257O} Group 59 + corona (Theia 209, Theia 133$^\dagger$)\\
172 & 160 & 90 & 6 & \cite{2017AJ....153..257O} Group 51 + corona (Theia 218)\\
191 & 160 & 180 & 6 & Theia 430\\
225 & 160 & 50 & 21 & \cite{1956ApJ...123..408B} Cas-Tau\\
228 & 160 & 200 & 6 & Theia 301$^\dagger$\\
231 & 160 & 200 & 6 & Theia 301$^\dagger$\\
150 & 170 & 230 & 6 & Theia 431\\
229 & 170 & 90 & 22 & $\alpha$ Persei + \cite{2021AA...645A..84M} corona\\
187 & 180 & 617 & 4 & Praesepe + \cite{2019AA...627A...4R} corona (Theia 1184)\\
193 & 180 & 45 & 6 & Theia 115$^\dagger$\\
195 & 180 & 50 & 23 & Platais 9 + \cite{2021AA...645A..84M} corona (Theia 134, Theia 508)\\
156 & 180 & 180 & 23 & Platais 3 + corona (Theia 787)\\
236 & 180 & 19 & 12 & \cite{2021ApJ...917...23K} Greater Scorpius-Centaurus Subgroup 15\\
131 & 190 & 210 & 6 & Theia 371\\
153 & 190 & 220 & 6 & Theia 372\\
163 & 200 & 320 & 6 & Theia 788\\
178 & 200 & 36 & 24 & NGC 2451A + \cite{2021AA...645A..84M} corona (Theia 118)\\
189 & 200 & 126 & 25 & Core of \cite{2016AA...595A..22R} RSG~2 open cluster (\citealp{2017AJ....153..257O} Group 16)\\
217 & 200 & 126 & 25 & Corona of \cite{2016AA...595A..22R} RSG~2 (Theia 214)\\
171 & 210 & 280 & 24 & Alessi 9 + corona (Theia 598)\\
167 & 210 & 340 & 6 & Theia 792\\
190 & 220 & 300 & 6 & \cite{2017AJ....153..257O} Group 23 + corona (Theia 599)\\
202 & 220 & 300 & 6 & \cite{2017AJ....153..257O} Group 23 + corona (Theia 599)\\
215 & 260 & 270 & 6 & Theia 600\\
168 & 270 & 190 & 6 & Theia 310
\enddata
\tablenotetext{}{$^\mathsection$\ This Crius group was initially rejected from our list of interesting groups because of selection cuts in the median absolute deviations of $ZUVW$ or further selection cuts in the $UV$ plane.}
\tablenotetext{}{$^\dagger$\ Only a subset of the members of these associations correspond to this Crius group.}
\tablenotetext{}{$^\ddagger$\ The Pisces-Eridanus stream was previously noted to contain local over-densities \citep{2019AA...622L..13M}, and we have recovered them as distinct groups.}
\tablerefs{(1)~\citealt{2015AAS...22511203J}; (2)~\citealt{2014ApJ...792...37M}; (3)~\citealt{2014AJ....147..146K}; (4)~\citealt{2018ApJ...863...67G}; (5)~\citealt{2015MNRAS.454..593B}; (6)~\citealt{2019AJ....158..122K}; (7)~\citealt{2014AA...566A.132S}; (8)~\citealt{2018ApJ...865..136G}; (9)~\citealt{2017MNRAS.468.1198B}; (10)~\citealt{2021AJ....161..171T}; (11)~\citealt{2019ApJ...877...12T}; (12)~\citealt{2021ApJ...917...23K}; (13)~\citealt{2019AJ....158...77C};
(14)~\citealt{1998AJ....116.2423P};
(15)~\citealt{2015MNRAS.447.1267M};
(16)~\citealt{2015ApJ...813..108D};
(17)~\citealt{2005MNRAS.358...13J};
(18)~\citealt{2020ApJ...903...96G};
(19)~\citealt{2012MNRAS.420..986G};
(20)~\citealt{2010MNRAS.409.1002D};
(21)~\citealt{1999AJ....117..354D};
(22)~\citealt{1999ApJ...527..219S};
(23)~\citealt{2021AA...647A..19T};
(24)~\citealt{2020AA...640A...1C};
(25)~\citealt{2016AA...595A..22R}.}
\end{deluxetable*}

%% file: table_new_coronae.tex
\startlongtable
\tabletypesize{\footnotesize}
\tablewidth{0.985\textwidth}
\begin{deluxetable*}{lccccccccl}
\tablecolumns{10}
\tablecaption{Newly recognized coronae.\label{tab:coronae}}
\tablehead{\colhead{Name} & \colhead{Age} & \colhead{} & \multicolumn{2}{c}{Number of stars in} & \colhead{} & \multicolumn{2}{c}{Distances (pc)} & \colhead{} & \colhead{Other}\\
\cline{4-5}
\cline{7-8}
\colhead{} & \colhead{(Myr)} & \colhead{} & \colhead{core} & \colhead{corona} & \colhead{} & \colhead{Core} & \colhead{Corona} & \colhead{} & \colhead{names}}
\startdata
Volans-Carina & $89_{-6}^{+7}$ & & 48 & 93 & & 87 & 26--124 & & Oh~30$^\mathsection$, Theia~424, Crius~221\\%Not recognized in Kounkel et al
Oh~59$^\mathsection$ & $\approx 160$ & & 33 & 376 & & 160 & 82--249 & & Theia~209, Theia~133$^\dagger$, Crius~226\\%Not recognized in Kounkel et al
Oh~51$^\mathsection$ & $\approx 90$ & & 11 & 170 & & 161 & 93--431 & & Theia~218, Crius~172\\
Platais 3 & $\approx 180$ & & 85 & 241 & & 179 & 108--518 & & Theia 787, Crius~156\\%Recognized as PL3 in Kounkel et al
RSG~2 & $\approx 125$ & & 60 & 276 & & 198 & 123--329 & & Oh~16$^\mathsection$, Theia~214, Crius~189, Crius~217\\%Not recognized in Kounkel et al
Alessi 9 & $\approx 280$ & & 85 & 418 & & 208 & 119--762 & & Theia 598, Crius~171\\%Theia 598 was matched with Alessi 9 by Kounkel and Covey
UPK~612 & $\approx 100$ & & 60 & 276 & & 216 & 154--476 & & Theia~216\\%Not recognized in Kounkel et al
Oh~23$^\mathsection$ & $\approx 300$ & & 17 & 347 & & 218 & 96--493 & & Theia~599, Crius~190, Crius~202%Not recognized in Kounkel et al
\enddata
\tablecomments{Ages are from \citet[Volans-Carina]{2018ApJ...865..136G}, \citet[Oh~23, Oh~51 and Oh~59]{2019AJ....158..122K}, \citet[Platais~3]{2021AA...647A..19T}, \citet[Alessi~9]{2020AA...640A...1C}, \citet[RSG~2]{2016AA...595A..22R}.}
\tablenotetext{}{$^\dagger$\ Only a small part of Theia~133 is likely associated with the Oh~59 corona, the rest being mainly associated with the corona of the $\alpha$~Per open cluster (e.g., see \citealp{2021AA...645A..84M}). Further studies based on Gaia~DR3 will help to properly disentangle the coronae of $\alpha$~Per and Oh~59 given their similar kinematics.}
\tablenotetext{}{$^\mathsection$\ \cite{2017AJ....153..257O} groups.}
\end{deluxetable*}

%% file: table_new.tex
\startlongtable
\tabletypesize{\tiny}
\tablewidth{0.985\textwidth}
\begin{deluxetable*}{llcccccccccccc}
\tablecolumns{14}
\tablecaption{New candidate associations discovered in this work.\label{tab:new}}
\tablehead{\colhead{Crius} & \colhead{Distance$^\dagger$} & \colhead{CMD} & \colhead{Number} & \colhead{M.A.D.} & \colhead{M.A.D.} & \colhead{X$^\dagger$} & \colhead{Y$^\dagger$} & \colhead{Z$^\dagger$} & \colhead{U$^\dagger$} & \colhead{V$^\dagger$} & \colhead{W$^\dagger$} & \colhead{Nearest} & \colhead{$\Delta UVW$}\\
\colhead{ID} & \colhead{(pc)} & \colhead{age (Gyr)} & \colhead{of stars} & \colhead{$UVW$ (\kms)} & \colhead{$Z$ (pc)} & \colhead{(pc)} & \colhead{(pc)} & \colhead{(pc)} & \colhead{(\kms)} & \colhead{(\kms)} & \colhead{(\kms)} & \colhead{Asso.} & \colhead{(\kms)}}
\startdata
85 & 152 & $\geq 0.1$ & 12 & 2.7 & 8.0 & 7.9 & -116.6 & 97.9 & -15.1 & -6.4 & -19.5           & UPK535 & 5.8\\
109 & 157 & $\geq 0.1$ & 17 & 2.3 & 7.1 & -102.0 & 83.0 & -86.3 & 2.6 & 1.6 & -16.2           & UBC8 & 8.0\\
110 & 144 & $\approx$\,0.7 & 10 & 2.1 & 10.5 & 117.8 & -2.8 & -83.7 & 5.9 & 7.2 & -15.4       & COINGA13 & 10.8\\
113 & 122 & 0.7--1.2 & 10 & 4.1 & 7.0 & 99.4 & -69.3 & 10.1 & -0.4 & 2.2 & -12.8              & UBC8 & 7.2\\
116 & 118 & 0.1--3 & 11 & 1.0 & 7.2 & 66.4 & 26.8 & -93.2 & -11.8 & -6.1 & 5.3                & COL350 & 5.2\\
119 & 154 & $\geq 0.1$ & 11 & 2.7 & 13.5 & 112.8 & -98.8 & 36.8 & -14.6 & -5.3 & -2.0         & ASCC21 & 4.1\\
121 & 123 & $\geq 0.1$ & 14 & 4.0 & 14.2 & -102.9 & -37.4 & -57.0 & 1.1 & 3.8 & -11.6         & TURN5 & 7.7\\
124 & 158 & 0.1--1 & 17 & 1.1 & 8.3 & -44.9 & -89.6 & -121.6 & -0.3 & 8.5 & -16.6             & RUP98 & 13.4\\
127 & 124 & 0.1--2 & 10 & 2.2 & 4.8 & -75.2 & 79.3 & -59.4 & 23.9 & -2.9 & -4.0               & NGC7092 & 9.9\\
134 & 104 & 0.1--2 & 26 & 2.9 & 13.4 & -101.3 & 24.4 & 5.4 & 20.4 & -5.6 & -13.6              & CNGC7092 & 6.7\\
135 & 83 & 0.1--0.7 & 13 & 3.2 & 12.6 & -13.5 & -75.1 & -32.7 & 15.8 & -1.0 & -3.6            & STOC10 & 3.0\\
137 & 91 & $\geq 1$ & 13 & 1.5 & 4.9 & -71.7 & 25.5 & -50.6 & 5.0 & -1.8 & -7.4               & TURN5 & 6.4\\
144 & 158 & $\geq 0.1$ & 10 & 1.7 & 13.3 & -107.3 & 105.9 & -47.2 & -41.9 & -23.0 & -9.4      & CPRA & 2.8\\
149 & 83 & $\geq 0.1$ & 16 & 2.3 & 13.2 & 34.5 & 13.1 & 73.8 & -10.3 & -20.5 & 5.6            & UPK612 & 4.2\\
152 & 122 & $\geq 0.1$ & 11 & 2.4 & 9.0 & -85.5 & -55.8 & 66.0 & -33.4 & -14.5 & -16.3        & STOC2 & 4.2\\
154 & 168 & $\geq 0.1$ & 11 & 2.0 & 12.6 & 30.7 & -162.9 & -28.9 & -35.1 & -18.5 & -7.5       & LUPIII & 5.1\\
157 & 64 & $\geq 0.1$ & 18 & 2.4 & 15.0 & 47.4 & -12.9 & -31.7 & -19.4 & -3.4 & -15.5         & COL135 & 5.6\\
158 & 126 & $\geq 1$ & 12 & 3.0 & 12.3 & 23.4 & 107.7 & 61.7 & -12.7 & -3.6 & -9.5            & OCTN & 1.5\\
159 & 125 & 0.1--0.7 & 13 & 2.9 & 8.6 & 24.0 & 121.3 & 20.2 & -13.0 & -8.9 & -9.5             & STOC12 & 1.3\\
162 & 70 & 0.1--0.7 & 14 & 1.9 & 8.9 & 23.4 & -59.0 & 28.9 & -17.7 & -3.8 & -12.3             & SPL5 & 2.8\\
181 & 155 & $\geq 0.1$ & 11 & 2.6 & 13.4 & -103.7 & -111.5 & 27.0 & -32.9 & -17.9 & -8.2      & UPK552 & 4.0\\
182 & 149 & 0.1--2 & 12 & 1.7 & 4.3 & -53.9 & 84.3 & 110.4 & -14.5 & -22.3 & -3.4             & TAUMGLIU21 & 1.5\\
184 & 137 & 0.7--1.2 & 11 & 2.1 & 4.9 & -95.5 & -82.6 & 52.5 & -29.4 & -19.2 & 2.2            & CARN & 5.9\\
188 & 155 & 0.1--2 & 14 & 1.3 & 7.6 & -111.5 & -62.3 & -87.8 & -14.1 & -29.1 & -4.8           & MOPH & 4.6\\
192 & 71 & $\geq 0.1$ & 14 & 2.7 & 4.3 & -0.5 & -15.0 & 69.1 & -24.3 & -8.4 & -8.1            & COL69 & 3.5\\
198 & 47 & 0.1--0.7 & 11 & 2.9 & 6.9 & -27.3 & 36.9 & 10.3 & -34.5 & -13.6 & -10.7            & UPK552 & 3.7\\
199 & 151 & $\geq 0.1$ & 21 & 1.6 & 5.9 & 100.4 & -84.2 & -75.0 & -36.9 & -11.5 & -3.8        & PL10 & 4.1\\
200 & 160 & $\geq 0.1$ & 14 & 2.0 & 12.7 & 0.3 & -129.7 & 94.3 & -28.1 & -16.2 & -5.2         & BH23 & 2.0\\
210 & 89 & 1--3 & 18 & 1.7 & 6.9 & 45.8 & -55.3 & 52.6 & -32.5 & -13.7 & 0.5                  & UBC12 & 6.3\\
213 & 59 & 0.1--0.7 & 11 & 2.4 & 9.6 & -41.0 & 41.8 & 6.5 & -41.2 & -19.3 & -10.5             & CPRA & 1.7\\
222 & 80 & 0.1--0.7 & 14 & 1.6 & 11.4 & -34.1 & 13.8 & 70.6 & -19.1 & -18.5 & -7.6            & NGC6633 & 2.2\\
224$^\ddagger$ & 47 & 0.1--0.7 & 28 & 2.0 & 5.5 & 9.6 & 17.3 & 42.4 & -16.2 & -20.6 & -3.9    & UTAU & 1.0
\enddata
\tablecomments{The last two columns indicate the nearest known association in $UVW$ space with the bulk $UVW$ separation in \kms. The association names are shortened as described here: UPK 535 open cluster (UPK535); UBC 8 open cluster (UBC8); COIN-Gaia 13 open cluster (COINGA13); Collinder 350 open cluster (COL350); ASCC 21 open cluster (ASCC21); Turner 5 open cluster (TURN5); Ruprecht 98 open cluster (RUP98); NGC 7092 open cluster (NGC7092); NGC 7092 corona (CNGC7092); Stock 10 open cluster (STOC10); Turner 5 open cluster (TURN5); Praesepe corona (CPRA); UPK 612 open cluster (UPK612); Stock 2 open cluster (STOC2); Lupus III subgroup (LUPIII); Collinder 135 open cluster (COL135); Octans-Near association (OCTN); Stock 12 open cluster (STOC12); Platais 5 system (SPL5); UPK 552 open cluster (UPK552); Taurus Moving Group Jiaming Liu 21 (TAUMGLIU21); Carina-Near moving group (CARN); $\mu$ Oph association (MOPH); Collinder 69 open cluster (COL69); UPK 552 open cluster (UPK552); Platais 10 open cluster (PL10); BH 23 open cluster (BH23); UBC 12 open cluster (UBC12); Praesepe corona (CPRA); NGC 6633 open cluster (NGC6633); u Tau association (UTAU).}
\tablenotetext{}{$^\dagger$\ Median values are reported here.}
\tablenotetext{}{$^\ddagger$\ We recovered two co-moving exoplanet-hosting stars (TOI--2076 and TOI--1807) in Crius~224, which were discussed by \cite{2021AJ....162...54H}, who suggested that they may be part of a small $\approx$200\,Myr-old moving group.}
\end{deluxetable*}

%% file: table_extensions.tex
\startlongtable
\tabletypesize{\footnotesize}
\tablewidth{0.985\textwidth}
\begin{deluxetable*}{lcclc}
\tablecolumns{5}
\tablecaption{Newly discovered extensions of known associations.\label{tab:extensions}}
\tablehead{\multicolumn{2}{c}{Crius} & \colhead{} & \multicolumn{2}{c}{Known association}\\
\cline{1-2}
\cline{4-5}
\colhead{ID} & \colhead{Distances (pc)} & \colhead{} & \colhead{Name} & \colhead{Distances (pc)}}
\startdata
%Total extensions = 17 - 8 coronae
\sidehead{\textbf{Extensions within 60\,pc}}
155 & 50$-$121 & & Theia 1005 & 65$-$92\\
168 & 51$-$161 & & Theia 310 & 187$-$343\\
207 & 59$-$135 & & Theia 677$^\dagger$ & 93$-$127\\
%Not counted in total of extensions (corona)
221 & 22$-$116 & & Volans-Carina & 71$-$98\\
227 & 19$-$71 & & Theia 209 & 82$-$124\\
\sidehead{\textbf{Other extensions}}
131 & 80$-$191 & & Theia 371 & 118$-$330\\
151 & 80$-$182 & & Theia 677$^\dagger$ & 93$-$127\\
170 & 77$-$177 & & \cite{2021AJ....161..171T} MELANGE--1, Theia 786 & 100$-$242\\
%Not counted in total of extensions (corona)
172 & 90$-$188 & & \cite{2017AJ....153..257O} Group 51, Theia~218 & 131$-$205\\
%Not counted in total of extensions (corona)
178 & 114$-$198 & & NGC 2451A, Theia 118 & 122$-$797\\
%Not counted in total of extensions (corona)
190 & 171$-$191 & & \cite{2017AJ....153..257O} Group 23, Theia~599, Crius~202 & 123$-$497\\
%Not counted in total of extensions (corona)
202 & 75$-$174 & & \cite{2017AJ....153..257O} Group 23, Theia~599, Crius~190 & 123$-$496\\
215 & 170$-$199 & & Theia 600 & 162$-$518\\
%Not counted in total of extensions (corona)
217 & 104$-$189 & & RSG2, \cite{2017AJ....153..257O} Group 16, Theia 214 & 167$-$327\\
%Not counted in total of extensions (corona)
226 & 93$-$191 & & \cite{2017AJ....153..257O} Group 59, Theia 209, Theia 133$\ddagger$ & 82$-$170\\
\sidehead{\textbf{Candidate extensions}}
77 & 131$-$168 & & Theia 1186 & 187$-$330\\
156 & 84$-$199 & & Platais 3, Theia 787 & 102$-$518
\enddata
\tablecomments{We consider Crius~77 and Crius~156 candidate extensions of Theia~1186, Platais~3 and Theia~787 for now, given that their velocities are similar but slightly discrepant. Please see Section~\ref{sec:extensions} for more details on extensions and candidate extensions listed here.}
\tablenotetext{}{$^\dagger$\ Crius~151 only matches a subset of Theia~677, which seems to contain at least two distinct populations. Crius~207 matches the largest locus of Theia~677 members.}
\tablenotetext{}{$\ddagger$\ Crius~226 only matches a subset of Theia~133, which seems to contain at least two distinct populations.}
\end{deluxetable*}

%% file: table_exoplanets.tex
\startlongtable
\tabletypesize{\footnotesize}
\tablewidth{0.985\textwidth}
\begin{deluxetable*}{llccl}
\tablecolumns{5}
\tablecaption{Known exoplanet host stars recovered in Crius groups.\label{tab:exo}}

\tablehead{\colhead{NASA} & \colhead{Simbad} & \colhead{} & \colhead{Crius} & \colhead{}\\
\colhead{exoplanet IDs} & \colhead{host ID} & \colhead{Reference} & \colhead{ID} & \colhead{Notes}}
\startdata
\sidehead{\textbf{Confirmed exoplanets in new Crius groups}}
HD 103949 b & HD 103949 & 1 & 162 & New 100--700 Myr Crius group\\
TOI--1807 b & BD+39 2643 & 2 & 224 & Previously suspected 180--200 Myr moving group\\
TOI--2076 b, c, d & BD+40 2790 & 2 & 224 & Previously suspected 180--200 Myr moving group\\
\sidehead{\textbf{Exoplanet candidates in new Crius groups}}
TOI--1598 b & BD+36 344 & 3 & 109 & New $\geq 100$\,Myr Crius group\\
TOI--2481 b & TYC 103--445--1 & 3 & 121 & New $\geq 100$\,Myr Crius group\\
TOI--2133 b & UCAC4 614--055633 & 3 & 205 & New 100--700 Myr Crius group\\
\sidehead{\textbf{Exoplanet candidates with newly recognized host associations}}
TOI--2048 b & TYC 3496--1082--1 & 3 & 147 & \cite{2017AJ....153..257O} Group 10\\
TOI--447 b & HD 33512 & 3 & 164 & Octans association\\
TOI--4364 b & PM J05202--0414 & 3 & 214 & Corona of the Hyades\\
TOI--1224 b & UCAC4 046--001384 & 3 & 221 & Corona of Volans-Carina\\
TOI--1990 b & TYC 8613--1781--1 & 3 & 235 & Corona of IC~2602\\
\sidehead{\textbf{Confirmed exoplanets which membership is already known}}
HD 63433 b, c & HD 63433 & 4 & 141 & Ursa Major "moving group"\\
K2--100 b & K2--100 & 5 & 187 & Praesepe open cluster\\
K2--102 b & K2--102 & 5 & 187 & Praesepe open cluster\\
Pr0201 b & BD+20 2184 & 6 & 187 & Praesepe open cluster\\
Pr0211 b, c & 2MASS~J08421149+1916373 & 6,12 & 187 & Praesepe open cluster\\
K2--101 b & K2--101 & 5 & 187 & Praesepe open cluster\\
K2--136 b, c, d & K2--136 & 7 & 214 & Hyades association\\
HD 285507 b & HD 285507 & 8 & 214 & Hyades association\\
AU Mic b, c & AU Mic & 9,13 & 220 & $\beta$ Pic moving group\\
DS Tuc A b & HD 222259A & 10 & 233 & Tuacana-Horologium association\\
TOI--837 b & CPD--63 1435 & 11 & 235 & IC~2602
\enddata
\tablerefs{(1)~\citealt{2019ApJS..242...25F}; (2)~\citealt{2021AJ....162...54H}; (3)~\citealt{2019AJ....158..138S}; (4)~\citealt{2020AJ....160..179M}; (5)~\citealt{2017AJ....153...64M}; (6)~\citealt{2012ApJ...756L..33Q}; (7)~\citealt{2018AJ....155....4M}; (8)~\citealt{2014ApJ...787...27Q}; (9)~\citealt{2020Natur.582..497P}; (10)~\citealt{2019ApJ...880L..17N}; (11)~\citealt{2020AJ....160..239B}; (12)~\citealt{2016AA...588A.118M}; (13)~\citealt{2021AA...649A.177M}.}
\end{deluxetable*}